\documentclass[a4paper,11pt]{article}
\usepackage{jheppub} 
\usepackage{todonotes}
\usepackage{cleveref}
\crefformat{footnote}{#2\footnotemark[#1]#3}

\DeclareMathOperator{\sech}{sech}
\usepackage{tablefootnote}
\usepackage{multirow}
\usepackage{booktabs}

\title{\boldmath New asymptotically (Anti)-de Sitter black holes in (super)gravity}

\author[a,b, c, d]{Jay Armas,}
\author[a]{and Gianbattista-Piero Nicosia}
\affiliation[a]{Institute for Theoretical Physics, University of Amsterdam, 1090 GL Amsterdam, The Netherlands}
\affiliation[b]{Dutch Institute for Emergent Phenomena, 1090 GL Amsterdam, The Netherlands}
\affiliation[c]{Institute for Advanced Study, University of Amsterdam, Oude Turfmarkt 147, 1012 GC
	Amsterdam, The Netherlands}
\affiliation[d]{Niels Bohr International Academy, The Niels Bohr Institute, University of Copenhagen,
	Blegdamsvej 17, DK-2100 Copenhagen \O{}, Denmark}

\emailAdd{j.armas@uva.nl, g.nicosia@uva.nl}

\abstract{We use the duality between gravitational dynamics and fluids living on dynamical surfaces carrying multiple charges, known as the blackfold approach, to perturbatively construct new asymptotically global (Anti)-de Sitter multi-spinning, non-extremal, multi-charged black holes in theories of higher-dimensional gravity minimally coupled to a dilaton and higher-form gauge fields in spacetime dimensions $D\ge5$, and new asymptotically AdS$_{l}\times S^{m}$ black holes in type IIB and eleven-dimensional supergravity. These solutions include the generalisation of the Kerr-Newman solution to (A)dS carrying either electric or string charge, generalisations of black rings to higher-dimensions with $\mathbb{S}^p\times \mathbb{S}^{n+1}$ horizon topology, static de Sitter solutions carrying arbitrary $q$-brane charge, as well as various asymptotically AdS$_{l}\times S^{m}$ multi-charged and multi-spinning black hole solutions, some of which correspond to novel thermal states in $\mathcal{N}=4$ Super-Yang-Mills theory.}

\begin{document}
\maketitle
\flushbottom

\section{Introduction}\label{introduction}
In the last decade, the study of black holes in higher-dimensional gravity has been an intense topic of investigation, motivated by gravitational theories with extra dimensions, in particular string theory, and more generally by quantum gravity, the AdS/CFT correspondence and its various extensions.  New black hole solutions, phenomenological and potential astrophysical signatures of higher-dimensional black holes, de Sitter compactifications in string theory, the search for black hole microstates, properties of the string theory landscape and the swampland, as well as the study of new features of gravity including non-spherical topologies, generalisations of rigidity theorems, and horizon instabilities, have been at the core of this line of research. 

Of particular importance is the search for asymptotically anti-de Sitter (AdS) black holes, which provide clues to the physics and properties of (confining) gauge theories at finite temperature. The best understood case is the AdS/CFT correspondence \cite{Maldacena:1997re}, where asymptotically AdS$_5 \times S^5$ black holes in type IIB supergravity are dual to thermal states in $\mathcal{N}=4$ SU(N) Super-Yang-Mills theory on $R \times S^3$ in the large N limit. However, even in this widely studied case, we still lack a complete picture of the phase space of the gauge theory at finite temperature. This is not surprising: gravity is more difficult in higher dimensions than in four \cite{Emparan:2008eg}.

Exact solutions of asymptotically global (Anti)-de Sitter black holes, or (A)dS black holes, in theories of gravity with a cosmological constant, (higher-form) gauge fields and a dilaton, are not abundant. In fact, apart from the uncharged Schwarzschild (A)dS black hole \cite{witten1998antide}, its rotating extension, the Kerr (A)dS black hole \cite{Hawking:1998kw,Gibbons:2004uw}, and the Reissner-Nordstrom-(A)dS black hole \cite{Konoplya:2007jv}, there is no other exact and regular analytic black hole solution in $D\ge5$. Additional information about the phase space of charged (A)dS black holes comes from numerical constructions such as the AdS black ring in $D=5$ \cite{Figueras:2014dta} and various perturbative constructions which have revealed the existence of new black hole topologies that generalise black ring solutions in $D=5$, as well as black hole spacetimes with disconnected horizons \cite{Caldarelli:2008pz, Armas:2010hz, Armas:2015kra, Armas:2015qsv}.

In the case of asymptotically global AdS$_{l}\times S^{m}$, with $(l,m)=\{(4,7),\ (5,5),\ (7,4)\}$, black holes in type IIB and M-theory the situation is similar. Asymptotically AdS solutions of pure gravity with a cosmological constant can be uplifted to type IIB/M-theory and thus the above neutral solutions for pure gravity with a cosmological constant can be converted into uncharged solutions of type IIB and M-theory \cite{Chamblin:1999tk, Cvetic:2000nc}. Exact solutions of spherically-symmetric black holes spinning in the $S^m$ \cite{Cvetic:2004hs,Chong:2004na,Chong:2005hr,Chong:2004dy,Chong:2005da,Kunduri:2006ek,Bobev:2023bxl} have also been found by means of consistent truncations of type IIB and eleven dimensional supergravity down to five- and four- or seven-dimensional gauged supergravity respectively \cite{Cvetic:2000nc,Cvetic:1999un,Nastase:1999cb,Nastase:1999kf,deWit:1986oxb,Cvetic:1999xp,Bhattacharyya:2010yg,Liu:1999ai}. Numerical techniques have also been used to construct various types of non-uniform solutions \cite{Dias:2015pda, Dias:2016eto, Cardona:2020unx, Dias:2022eyq} in type IIB and M-theory and using perturbative methods, thermal giant graviton solutions were constructed, which can simultaneously rotate in the AdS$_l$ and the $S^m$ parts \cite{Armas:2012bk, Armas:2013ota}.

In this paper we make progress towards understanding the phase space of asymptotically (A)dS higher-dimensional compact black holes in theories of gravity minimally coupled to higher-form gauge fields and a dilaton as well as the phase space of asymptotically AdS$_{l}\times S^{m}$ in type IIB/M-theory using the blackfold approach, which exploits the duality between gravitational dynamics and fluid branes carrying multiple higher-form charges. This approach is an effective theory describing the long-wavelength perturbations of asymptotically flat (charged) black p-branes and includes modulations of the horizon radius $r_0$ or the breaking of spherical symmetry of the transverse space by wrapping the p-branes on world volumes with characteristic curvature scales $\mathcal{R}$. At leading order in $r_0/\mathcal{R}$ we can treat the p-brane as a probe brane moving in a given background (the required asymptotic spacetime for the black hole). This approach has been used in the search for asymptotic flat black holes in the context of (super)gravity \cite{Emparan:2011br,Emparan:2011hg,Caldarelli:2010xz,Grignani:2010xm,Grignani:2011mr,Armas:2015kra, Armas:2015nea, Armas:2017myl} as well as in (A)dS backgrounds \cite{Armas:2010hz,Caldarelli:2008pz}, AdS$_{l}\times S^{m}$ \cite{Grignani:2012iw, Armas:2012bk,Armas:2013ota}, Schrödinger spacetimes, plane-wave spacetimes \cite{Armas:2015kra}, mass-deformed AdS \cite{Armas:2022bkh}, Klebanov-Strassler \cite{Armas:2018rsy} and CGLP backgrounds \cite{Armas:2019asf}.

We use the blackfold approach to find new black hole solutions in backgrounds that are either (A)dS or AdS$_{l}\times S^{m}$ with flux $F_l$. In the former case we find new multi-spinning non-extremal black holes carrying $q$-brane charges. The simplest cases are black hole solutions with horizon topology $\prod_a \mathbb{S}^{p_a}\times \mathbb{S}^{n+1}$ where $p_a$ is odd and $D=n+p+1$ where $p=\sum_a p_a$. When $p=1$ in $D=5$ this includes electrically charged ($q=0$) and dipole charged ($q=1$) black rings with horizon topology $\mathbb{S}^{1}\times \mathbb{S}^{2}$. We also construct a family of static $dS$ solutions carrying arbitrary $q$-brane charge as well as generalisations of the Kerr-Newman solution to (A)dS with $q=0,1$-brane charges. In the latter case of asymptotically AdS$_{l}\times S^{m}$ black holes in type IIB/M-theory we construct various multi-spinning non-extremal black holes with horizon topology $\prod_a \mathbb{S}^{p_a}\times \mathbb{S}^{n+1}$ carrying multiple q-brane charges. We also study in detail the different types of extremal limits that these black holes admit. These solutions provide new insights into the possible thermal states of $\mathcal{N}=4$ Super-Yang-Mills theory.

The structure of the paper is as follows. In section \ref{Essentialsoftheblackfoldapproach}, we review the blackfold approach in the presence of background fluxes and describe the main features of perfect fluids carrying (multiple) $q$-brane current(s). We give the necessary ingredients to obtain stationary black holes solutions with $q$-brane charge. In section \ref{Higher-dimensionalgravity}, we find new asymptotically (A)dS black holes in higher-dimensional gravity minimally coupled to higher-form gauge fields and a dilaton, and study their extremal limits. In section \ref{TypeII/M-theory}, we obtain new asymptotically AdS$_{l}\times S^{m}$ black holes in type II/M-theory with a single non-zero $F_l$ flux. These black holes are non-extremal, multi-spinning, and can carry $q=1,\ p,\ (1,p)$-brane charges. Finally, in section \ref{Discussion and Outlook}, we discuss the regime of validity of our approach and future directions. We also provide appendix \ref{staticsolutions} with details of static solutions in dS with arbitrary $q$-brane charge while in appendix \ref{physicalparameterstot}, we give additional details on the thermodynamic properties of several solutions that we construct.

\paragraph{Notation.} In the following, we use the notation of \cite{Emparan:2009at, Caldarelli:2010xz, Armas:2012bk} which we briefly introduce. For a p-brane in a $D$-dimensional spacetime we have that the transverse directions to the black brane are given by $n=D-p-3\geq1$ and 
\begin{itemize}
    \item Spacetime (background) tensors carry Greek indices $\mu,\nu,... = 0,...,D-1$. Spacetime coordinates are denoted by $x^{\mu}$. The spacetime metric is $g_{\mu\nu}$ and its associated covariant derivative given by $\nabla_{\mu}$. The tensor $h_{\mu\nu}$ is the first fundamental form of the worldvolume $\mathcal{W}_{p+1}$ and $\perp_{\mu}^{\ \nu}=g_{\mu}^{\ \nu}-h_{\mu}^{\ \nu}$ is the orthogonal projector to $\mathcal{W}_{p+1}$.
    \item  Wordlvolume tensors carry Latin indices $a, b,... = 0,...,p$. Worldvolume coordinates are denoted by $\sigma^{a}$. The induced metric on the worldvolume is given by $\gamma_{ab}=g_{\mu\nu}\partial_{a}X^{\mu}\partial_{b}X^{\nu}$ and its associated worldvolume covariant derivative $D_a$. The set of functions $X^{\mu}(\sigma)$ is the embedding map for the p-brane in the background spacetime. The tensor $\hat{h}_{ab}$ is the first fundamental form of the current worldline/sheet $\mathcal{C}_{q+1}\subset\mathcal{W}_{p+1}$ and $\hat{\perp}_{a}^{\ b}=\gamma_{a}^{\ b}-\hat{h}_{a}^{\ b}$ is the orthgonal projector to $\mathcal{C}_{q+1}$.
\end{itemize}

\section{Essentials of the blackfold approach}\label{Essentialsoftheblackfoldapproach}
In this section we introduce the key elements of the blackfold approach. In particular, we are interested in describing stationary blackfolds carrying multiple (dipole) charges in asymptotic spacetimes with non-vanishing curvature and fluxes. To this end, we follow Refs.~\cite{Emparan:2009at,Emparan:2011br,Caldarelli:2010xz, Armas:2016mes,Armas:2012bk,Armas:2018ibg}, generalising certain aspects where appropriate. As the ideal order dynamics of blackfolds is characterised by higher-form hydrodynamics on dynamic submanifolds with an effective fluid stress tensor and various higher-form currents, we also discuss in some detail the dynamics of perfect fluids with q-form currents.

\subsection{Basic philosophy and (super)gravity actions}\label{Basicphilosophyandregimeofapplicability}
The blackfold approach is an effective theory that allows to perturbatively construct large classes of black hole solutions with a given asymptotic structure in higher-dimensional gravity whose near-horizon geometry is locally given by a (multi charged and/or spinning) boosted black p-brane solution. The existence of such blackfold regime is known to be ubiquitous in higher-dimensional gravity when the black hole horizon is characterised by two or more widely separated length scales.  

More concretely, if we denote by $\mathcal{R}$ and $\mathcal{L}$ the characteristic length scales associated with deformations of the black p-brane (e.g. curvature or temperature fluctuations across the worldvolume) and of the asymptotic background spacetime (e.g. background curvature or modulations of background electro-magnetic fields) respectively, while collectively denoting by $r_{c}$ the smallest characteristic length scale associated with the black brane (e.g. horizon radius or charge density radii), then for the blackfold approximation to be valid one requires\footnote{A more rigorous analysis of the length scales involved can be performed by requiring higher-derivative terms of the blackfold effective action to be subleading \cite{Armas:2015kra}.}
\begin{equation} \label{approx}
    r_{c} \ll \min(\mathcal{R,L})\ .
\end{equation} 
The existence of this scale separation can be used to set up a matched asymptotic expansion and perturbatively extract the black hole metric in the following way. Denoting by $r$ a suitably defined radial coordinate with $r=0$ the curvature singularity, then in the near-zone, $r\ll\mathcal{R}$, the solution is approximated by long-wavelength deformations of an exact black p-brane solution, where $r_{c}/\mathcal{R}$ is the small perturbative parameter. In cases where the exact solution is not known, we perform a parallel perturbative expansion in powers of $r_{c}/\mathcal{L}$ such that we can approximate the near-zone by a known flat black p-brane solution. In turn, in the far-zone, $r\gg r_{c}$, we solve the Einstein equations of the (super)gravity action in powers of $r_{c}/r$ around a given asymptotic background and for a given source (in this case a probe brane with the geometry of the singularity in the near-zone).  Finally, these two regions have to be identical in the \textit{overlap-zone} at $r_{c}\ll r \ll \mathcal{R}$, where the integration constants of the two zones are matched (see \cite{Camps:2012hw,DiDato:2015dia,Emparan:2007wm, Nguyen:2021srl} for specific examples).

A necessary requirement for the existence of such black hole solutions, arising from the matched asymptotic expansion, is to solve a set of constraint equations, namely that blackfold equations. Finding solutions to this latter set of equations is the focus of this paper. In section \ref{Higher-dimensionalgravity}, we obtain new charged and rotating asymptotically (A)dS blackfold solutions in Einstein-dilaton gravity with a $(q+1)$-form gauge field and a cosmological constant $\Lambda$ with action
\begin{equation} \label{actionads}
    I = \frac{1}{16\pi G}\int_{\mathcal{M}_{D}}\Big(\star (R-2\Lambda) -\frac{1}{2}d\phi\wedge\star d\phi -\frac{1}{2}e^{a_{q}\phi}F_{q+2}\wedge\star F_{q+2} \Big) \ ,
\end{equation} where $a_{q}$ is an arbitrary dilaton coupling. An exact $q$-charged black p-brane solution of the action above is not known and thus in this case we will employ the double perturbative regime mentioned above where the characteristic background length scale is $\mathcal{L}=|\Lambda|^{-1/2}$. In this context, the near-region is approximated at leading order by the flat $q$-charged black p-brane solution found in \cite{Caldarelli:2010xz}. On the other hand, the asymptotic background we are interested in is an (A)dS solution of the action \eqref{actionads} with vanishing dilaton and gauge field. Therefore, we require our solutions to asymptote to the global AdS metric in the form
\begin{equation}\label{globalads}
    ds_{AdS_{D}}^{2}= -f(r)dt^{2} + \frac{dr^{2}}{f(r)} + r^{2}d\Omega_{D-2}^{2}\ ,\ \ \ f(r)=1+\frac{r^{2}}{L^{2}}\ ,
\end{equation} where, $d\Omega_{D-2}^{2}$ is the metric of a unit sphere and $L$ is the AdS radius. We also consider the case of dS which can be obtained from \eqref{globalads} by Wick rotation, i.e. $L\to i L$.

In section \ref{TypeII/M-theory}, we focus on finding charged and rotating blackfold solutions of type IIB/M-theory. Using the democratic formulation \cite{becker_becker_schwarz_2006}, the bosonic sector of Type IIB is characterised by the following action \begin{equation}\label{actiontypeii}
    I = \frac{1}{16\pi G}\int_{\mathcal{M}_{10}}\Big(\star R -\frac{1}{2}d\phi\wedge\star d\phi -\frac{1}{2}e^{-\phi}H_{3}\wedge\star H_{3} -\frac{1}{4}\sum_{q}e^{a_{q}\phi}F_{q+2}\wedge\star F_{q+2} \Big) \ ,
\end{equation} where the dilaton coupling is now fixed such that $a_{q}=(3-q)/2$, while $H_{3}$ and $F_{q+2}$ are the NSNS and RR field strengths respectively and $q=-1,3,5,7$. In addition we must impose self-duality for the five-form flux, namely $F_5=\star F_5$. In turn, the bosonic part of the M-theory action is given by \cite{becker_becker_schwarz_2006}
\begin{equation}\label{actionmtheory}
    I = \frac{1}{16\pi G}\int_{\mathcal{M}_{11}}\Big(\star R -\frac{1}{2}F_{4}\wedge\star F_{4}\Big) -\frac{1}{6}\int A_{3}\wedge F_{4}\wedge\star F_{4} \ ,
\end{equation}  
where $F_{4}=dA_{3}$ is the field strength associated to the gauge field $A_{3}$. Due to the presence of multiple gauge fields, black branes can carry multiple charges, contrary to those in \eqref{actionads}. Within the realm of ten and eleven dimensional supergravity we focus on black hole solutions that asymptote to AdS$_{l}\times S^{m}$, where $(l,m)=\{(4,7),(5,5),(7,4)\}$ and with non-vanishing $F_{l}$ flux but trivial dilaton. In particular we parameterise the background metric as $ds^2=ds^{2}_{AdS_{l}}+ds^{2}_{S^{m}}$ with 
\begin{align} \label{eq:AdSS5metric}
     ds^{2}_{AdS_{l}}&=-f(r)dt^{2}+f(r)^{-1}dr^{2}+r^{2}d\Omega^{2}_{l-2} \ , \\
      ds^{2}_{S^{m}}&=L^{2}(d\zeta^{2}+\cos^{2}\zeta d\phi^{2}+\sin^{2}\zeta d\Omega^{2}_{m-2})\ ,
\end{align}
where $f(r)\equiv 1+\frac{r^{2}}{\tilde{L}^{2}}$ with $\tilde{L}=2L/(l-3)$, $d\Omega^{2}_{l-2}$ and $d\Omega^{2}_{m-2}$ are the line elements of the unit sphere, respectively parameterised by the angles $\theta_{i}$ and $\rho_{i}$. In these coordinates, the background gauge fields for $AdS_l$ and $S^m$ respectively, are given by 
\begin{align} \label{gaugefields}
    A_{l-1}&=-\frac{r^{l-1}}{\Tilde{L}}dt\wedge d\Omega_{l-2}\ ,  \\
    A_{m-1}&=\beta_{m}(L\sin\zeta)^{m-1}d\phi\wedge d\Omega_{m-2} \ ,
\end{align} 
where $\beta_{m} = (-1)^{D-m-1}$. This background is a solution of both type IIB and eleven dimensional supergravity. 

\subsection{Blackfold dynamics}\label{blackfolddynamics}
In the blackfold approximation, we consider a probe p-brane embedded in $D$-dimensional spacetime with a localised stress energy tensor of the form
\begin{equation}
    \tilde{T}^{\mu\nu}(x) = \int_{\mathcal{W}_{p+1}}d^{p+1}\sigma\tilde{\delta}(x)T^{\mu\nu}(\sigma) \ ,
\end{equation} 
where $\tilde{\delta}(x)=\frac{\sqrt{-\gamma}}{\sqrt{-g}}\delta^{(D)}(x-X(\sigma))$ is the reparametrisation invariant delta function. In the blackfold approach the stress energy tensor is given by an effective stress energy tensor describing the black brane solution and is defined as the quasi-local Brown and York stress tensor \cite{Brown:1992br}. At leading order in the blackfold expansion, the stress energy tensor is that of an ideal fluid in local thermodynamic equilibrium and whose equation of state is extracted from a given black p-brane solution. The probe branes we consider are also charged and hence carry at least one localised $(q+1)$-form current of the form   
\begin{equation} \label{eq:spacetimecurrent}
    \tilde{\mathcal{J}}_{q+1}(x)=\int_{\mathcal{W}_{p+1}}d^{p+1}\sigma\tilde{\delta}(x)\mathcal{J}_{q+1}(\sigma) \ ,
\end{equation} 
which couples to the field strength $F_{q+2}$. In general, the probe brane can also carry a dilatonic current but we will not consider this here.

The equations of motion for a charge probe brane can be obtained by coupling it to the gravity actions we introduced above. Given that the backgrounds we consider only have one non-vanishing flux turned on, the dynamics of the probe brane is given by 
 \begin{align}\label{geneom}
    \nabla_{\mu}\tilde{T}^{\mu\nu}&=\frac{1}{(q+1)!}F^{\nu\alpha_{1}...\alpha_{q+1}}\tilde{\mathcal{J}}_{\alpha_{1}...\alpha_{q+1}} \ , \\ \nabla_{\mu}\tilde{\mathcal{J}}^{\mu\alpha_{1}...\alpha_{q}}&=0\ .
\end{align} 
The first equation arises as the Ward identity associated with diffeormorphism invariance where on the right hand side is the Lorentz force experienced by the probe brane (see \cite{Armas:2016mes} for a detailed discussion of probe branes in supergravity backgrounds). The second equation arises as the Ward identity associated to gauge transformations of the higher-form gauge fields and implies the existence of a conserved charge $Q_q$ obtained by integrating the $(q+1)$-form current over a co-dimension $(D-q-1)$ surface. In the contexts we consider here, probe branes may carry several of such currents and conserved charges. Projecting eqs.\eqref{geneom} along the brane worldvolume leads to
\begin{equation}\label{intrinsiceqn}
    D_{a}T^{ab}=0 \ , \ \ \ D_{a}\mathcal{J}^{ab_{1}...b_{q}}=0\ ,
\end{equation} 
where we assumed that the Lorentz force vanishes along the worldvolume. The extrinsic equations are instead obtained by projecting along the perpendicular directions and read
\begin{equation} \label{cartereqn}
    T^{ab}K_{ab}^{\ \ \mu}=\perp_{\nu}^{\ \mu}F^{\nu a_{1}...a_{q+1}}\mathcal{J}_{a_{1}...a_{q+1}} \ ,
\end{equation} 
where $K_{ab}^{\ \ \mu}=\partial_{a}X^{\alpha}\partial_{b}X^{\beta}$ is the extrinsic curvature tensor given by 
\begin{equation} \label{projextrcurv}
    K_{ab}^{\ \ \mu}=\perp_{\nu}^{\ \mu}\Big(\partial_{a}\partial_{b}X^{\nu}+\Gamma_{\alpha\beta}^{\nu}\partial_{a}X^{\alpha}\partial_{b}X^{\beta}\Big)\ ,
\end{equation} 
and where we introduced the affine connection $\Gamma_{\alpha\beta}^{\nu}$ associated to the background metric $g_{\mu\nu}$. Eq.~\eqref{cartereqn} can be interpreted as a force-balance equation for the probe p-brane or as the dynamical equation for the transverse scalars $\perp^\mu_\nu X^\nu(\sigma)$. 

\subsection{Perfect fluids with (multiple) $q$-brane current(s)}\label{Perfect fluids with q-brane currents}
In order to understand the properties of the effective blackfold higher-form fluids, it is necessary to introduce the main concepts of a perfect fluid containing a (multiple) $q_{l}$-brane current(s), with $l=1,...,m$. The first thing to notice is that the introduction of $q_{l}$-form conserved currents breaks the symmetry group of the isotropic fluid. Specifically, the symmetry group is broken as $SO(p)\rightarrow SO(p-\sum_{l}^{m}q_{l})\times SO(q_{l})\times ... \times SO(q_{l})$, which clearly implies that the presence of only a 0-, $p$- or $(0,p)$-form currents will not disrupt the isotropy of the fluid. On the other hand, the presence of any other type of $q_{l}$ current(s) will render the fluid anisotropic. In this study we will be interested in the presence of either a $q$-form charge or a combination of $(q,p)$-form charges\footnote{As a consequence, the index $l$ in $q_{l}$ is rendered moot and thus it will be dropped hereafter.}. This means that at most the symmetry group will be broken to $SO(p-q)\times SO(q)$. 

Associated with each charge there is a $q+1$-brane current of the form \eqref{eq:spacetimecurrent}. We are interested in the particular case in which each of these currents can be parameterised as
\begin{equation}\label{currentvoldefn}
    \mathcal{J}_{q+1}=\mathcal{Q}_{q}\mathcal{V}_{q+1}\ ,
\end{equation} with each of the $q$-brane currents foliating the worldvolume into sub-worldvolumes $\mathcal{C}_{q+1}\subset\mathcal{W}_{p+1}$ with volume form $\mathcal{V}_{q+1}=u\wedge v^{(1)}\wedge...\wedge v^{(q)}$, where $v^{(i)}$ are orthonormal spacelike vectors along which the (dissolved) current lies and normal to the timelike fluid velocity $u^a$ satisfying $u^a u_
a=-1$. More general currents than \eqref{currentvoldefn} have been studied in \cite{Armas:2019asf, Armas:2023tyx}.  In the case of $q=p$, the current is a top form and we deduce that the current continuity eq. \eqref{intrinsiceqn} requires that \begin{equation}\label{chargeconsp}
    \partial_{a}\mathcal{Q}_{p}=0 \ ,
\end{equation}  implying that the charge density is a constant along the worldvolume. Said differently, this means that the charge has no local degrees of freedom and therefore is not a collective variable of the fluid. On the other hand, for the $q\neq p$ cases the charge will have a non-trivial time evolution. In fact, if we project the current continuity equation \eqref{intrinsiceqn} along the timelike and spacelike vectors, we obtain the charge continuity equations \begin{align}\label{chargeconsveqn}
    D_{a}(\mathcal{Q}_{q}u^{a}) + \mathcal{Q}_{q}u^{a}\dot{v}_{a} &=0 \ , \\
    D_{a}(\mathcal{Q}_{q}v^{a,i}) + \mathcal{Q}_{q}v^{a,i}\dot{v}_{a}-\mathcal{Q}_{q}v^{a,i}\dot{u}_{a} &=0 \ ,
\end{align} where $\dot{v}_{a}\equiv v^{b,j}D_{b}v_{a,j}$ and the Einstein summation convention over the indices $(i,j,...)$ is assumed. At this point we would like to remark some important aspects \cite{Caldarelli:2010xz,Emparan:2011hg}. First, while it is true that $\mathcal{Q}_{q}$ are local degrees of freedom, they are constant along $\mathcal{C}_{q+1}$ and can vary only in direction transverse to the current. For this reason they are known as ``quasi-local". Second, for $q'\leq q$, we do not require that $C_{q'+1}\subseteq C_{q+1}$. Third, the anisotropy induces a difference in pressures in directions parallel and transverse to the $q$-form currents which will be proportional to the charge density \begin{equation}\label{diffpressure}
    P_{\perp}-P_{\parallel}=\Phi_{q}\mathcal{Q}_{q} \ ,
\end{equation} where $\Phi_{q}$ is the string/brane chemical potential conjugate to the charge density. Finally, in a set-up where multiple $q\neq p$ charges are present, each of the equations in \eqref{chargeconsveqn} are valid for each charge. 

In order to continue our journey, we first introduce the projector onto $\mathcal{C}_{q+1}$ in terms of the fluid velocity and the spacelike vectors as \begin{equation}\label{hath}
    \hat{h}_{ab}^{(q)}=-u_{a}u_{b}+(1-\delta_{q,0})v_{a}^{i}v_{b,i}\ ,
\end{equation} such that $\hat{h}_{ab}^{(0)}=-u_{a}u_{b}$, $\hat{h}_{ab}^{(1)}=-u_{a}u_{b}+v_{a}v_{b}$ and $\hat{h}_{ab}^{(p)}=\gamma_{ab}$ reproduce the analysis done in \cite{Caldarelli:2010xz,Emparan:2011hg}. Note that by construction\footnote{\label{footnote1}As stated at the beginning of this chapter, the $q$-brane currents foliate the worldvolume into sub-worldvolumes $\mathcal{C}_{q+1}$. This can be shown by first defining a set of orthonormal one-forms orthogonal to the projector \eqref{hath} and contracting it with the current continuity equation. Then by using Frobenius' theorem, one can show that these one-forms are surface-forming, implying the existence of an integral over the vectors $u$ and $v^{i}$ on the $(q+1)$-dimensional submanifold. It is precisely on these submanifolds, with induced metric $\hat{h}^{(q)}_{ab}$ and volume form as defined in \eqref{currentvoldefn}, that the charge lies. Projecting the contracted equation along the vectors leads to \eqref{conditiononvec}. For further details we refer the reader to \cite{Armas:2018ibg}.}, $\hat{h}^{(q)}_{ab}$ is also the induced metric on $\mathcal{C}_{q+1}$ and by defining the perpendicular projector to the sub-worldvolume (as defined in section \ref{introduction}), we can exploit the current continuity equation \eqref{intrinsiceqn} to prove that the vectors obey the following relations\cref{footnote1} \begin{equation}\label{conditiononvec}
    \hat{\perp}^{a}_{\ b}(v^{c}_{[i}D_{c}v^{b}_{j]})=\hat{\perp}^{a}_{\ b}(v^{c}_{i}D_{c}u^{b}-u^{c}D_{c}v^{b}_{i})=0\ .
\end{equation} 

At this point, we use all of the information above to write down the stress energy tensor for a fluid with an $SO(p-q)\times SO(q)$ symmetry as  \begin{equation}\label{anistropytensor}
    T_{ab}=(\varepsilon+P_{\parallel})u_{a}u_{b}+(P_{\parallel}-P_{\perp})\hat{h}_{ab}^{(q)}+P_{\perp}\gamma_{ab} \ ,
\end{equation} where $\varepsilon$ is the energy density. In the case of $0$-, $p$- and $(0,p)$-form currents one can easily show that \eqref{anistropytensor} reduces to the isotropic fluid stress energy tensor. In addition, locally this anisotropic fluid obeys the following thermodynamic relations\footnote{Note that in general, a fluid with more than one $q\neq p$ obeys $d\varepsilon = \mathcal{T}ds + \sum_{q}^{p-1}\Phi_{q} d\mathcal{Q}_{q}$ but given that we are interested in fluids with at an $SO(p-q)\times SO(q)$ symmetry, the sum is redundant.}
\begin{equation}
    \varepsilon + P_{\perp}= \mathcal{T}s + \Phi_{q}\mathcal{Q}_{q}\ , 
\end{equation} and the Gibbs-Duhem relations 
\begin{equation}
    d\varepsilon = \mathcal{T}ds + \Phi_{q} d\mathcal{Q}_{q}\ , \ \ \ dP_{\perp}=sd\mathcal{T}+\mathcal{Q}_{q}d\Phi_{q}\ , \ \ \ dP_{\parallel}=sd\mathcal{T}-\mathcal{Q}_{q}d\Phi_{q}\ ,
\end{equation}  where, \eqref{diffpressure} is none other than the integral of the last two equations and the $p$-brane charge is excluded because it is constant on the worldvolume, thus not a collective variable of the fluid, as previously discussed. 

Finally, we can rewrite the conservation for stress energy tensor \eqref{intrinsiceqn}, by projecting the equation along the velocity and spacelike vectors. In doing so, we obtain the conservation of current entropy and Euler force equations \begin{align} \label{newintrinsiceqn}
    D_{a}(su^{a})&=0\ , \\
    (\hat{h}^{ab}+u^{a}u^{b})(\dot{u}_{b}+D_{b}\ln{\mathcal{T}})&=0 \ , \\
    \mathcal{T}s\hat{\perp}^{ab}(\dot{u}_{b}+D_{b}\ln{\mathcal{T}})-\Phi_{q}\mathcal{Q}_{q}(\hat{K}_{(q)}^{a}-\hat{\perp}^{ab}D_{b}\ln{\Phi_{q}})&=0\ ,
\end{align} where, $\hat{K}^{a}_{(q)}=\hat{h}^{cb}\hat{K}_{cb}^{\ \ a}$ is the mean extrinsic curvature of the worldvolume $\mathcal{C}_{q+1}$. Finding a solution to these equations is equivalent to solving the intrinsic equations. 

We would like to conclude this part by counting the number of the time dependent equations of the fluid and the independent degrees of freedom. The conservation of the stress tensor \eqref{intrinsiceqn} and the 
charge conservation equations \eqref{chargeconsveqn} and \eqref{conditiononvec} provide us with $(p+1)$ and $(p-q)q+1$ dynamical equations respectively. The independent degrees of freedom associated to the system are the $p$ independent components of $u$ (since $u^{a}u_{a}=-1$ removes a degree of freedom); $(p-q)q$ independent components of $v^{i}$ and two degrees of freedom associated to the temperature $\mathcal{T}$ and chemical potential $\Phi_{q}$. Hence, the degrees of freedom associated to the system match exactly the number of dynamical equations. 

\subsubsection*{Boundaries}\label{boundaries}
Before dwelling into technicalities, we would like to remark that the analysis done here is valid when the blackfold carries a (combination of) $q<p$-brane current(s). The $p$-current is a particular case in the sense that the conservation of $Q_{p}$ does not permit any boundaries. An escape route would be to consider brane intersections such that the charge required is carried on the other brane, but these setups will not be considered here. 

As was described in \cite{Emparan:2009at}, let $f(\sigma^{a})$ be a level-set function such that $f(\sigma^{a})>0$ in $\mathcal{W}_{p+1}$ and $f(\sigma^{a})=0$ when on the boundary $\partial\mathcal{W}_{p+1}$. Thus, let $-\partial_{a}f$ be the one-form normal to the boundary which points away from the fluid. In order for the fluid to remain within the boundary, we require that \begin{equation}
    u^{a}\partial_{a}f|_{\partial\mathcal{W}_{p+1}}=0\ ,
\end{equation} i.e., the fluid velocity is parallel to the boundary. 

Given that we are considering charged blackfolds we need to also obey the current conservation law. Note that in the case of $q=0$, the condition above is enough, while for $0<q<p$, we require that 
\begin{equation} \label{quvf}
\prod_{j=1}^{q}\mathcal{Q}_{q}u^{[a}v_{j}^{b_{j}]}\partial_{a}f|_{\partial\mathcal{W}_{p+1}}=0\ ,
\end{equation} 
which further implies that\footnote{To be more explicit we shall consider the case of $q=2$, such that the vectors under consideration are the timelike $u_{a}$ and the spacelike ones $v_{a}$ and $w_{a}$. Then we see that eq. \eqref{quvf} implies that $\mathcal{Q}_{2}u^{[a}v^{b}w^{c]}\partial_{a}f|_{\partial\mathcal{W}_{p+1}}=0$. Now, if we contract with $u_{b}$ and $w_{c}$ we obtain the current along the direction of $v_{a}$ and we see that this implies $\mathcal{Q}_{2}v^{a}\partial_{a}f|_{\partial\mathcal{W}_{p+1}}=0$. Repeating the same procedure we reach the same conclusion for the current along the direction of the vector $w_{a}$, as advertised in the main text.} 
\begin{equation}
    \mathcal{Q}_{q}v_{j}^{a}\partial_{a}f|_{\partial\mathcal{W}_{p+1}}=0\ ,
\end{equation} meaning that the current has to be parallel to the boundary. Note that if we pick $q=1$ we recover the work in \cite{Caldarelli:2010xz}. Likewise, the conservation of the stress energy tensor requires
\begin{equation}
    P_{\perp}|_{\partial\mathcal{W}_{p+1}}=0\ ,
\end{equation} 
implying that there is no surface tension on the boundary, in turn leading to \begin{equation}
    r_{0}|_{\partial\mathcal{W}_{p+1}}=0\ ,
\end{equation} 
i.e. the thickness of the horizon must approach zero at the boundary. The two relations above are also obeyed in the case of neutral blackfolds. The latter condition is a necessary condition for regularity of the black hole horizon and we refer the reader to \cite{Emparan:2009at} for further details.

\subsection{Stationarity and conserved charges}\label{stationarityandconservedcharges}
In this paper we restrict to stationary blackfold configurations as we aim to find new stationary black hole solutions. Stationarity requires the existence of a background timelike Killing vector field $k$. Letting $\xi$ and $\chi$ denote the timelike Killing vector generator of asymptotic time and the spacelike Killing vector generator of space translations respectively, then we parametrise $k$ according to
\begin{equation}\label{fluidvelocity}
    k = \xi + \Omega\chi\ ,
\end{equation}
with constant $\Omega$. Generically, there can be multiple angular velocities on different rotation planes. We assume that the worldvolume projection of the  Killing vector on the background, $k^{\mu}\partial_{\mu}$, given by the pullback $k_{a}=\partial_{a}X^{\mu}k_{\mu}$ is a Killing vector field of the worldvolume itself. It was proven in \cite{Caldarelli:2008mv} that for stationary configurations in the absence of dissipative effects, the intrinsic fluid equations require that the fluid velocity is proportional to a worldvolume Killing vector. To this end, we can define the fluid velocity as \begin{equation}\label{defnu}
    u^{a}\partial_{a}=\frac{k^{a}\partial_{a}}{|k|} \ , \ \ \ |k|=\sqrt{-\gamma_{ab}k^{a}k^{b}}\ .
\end{equation} 
In the case of $0<q<p$ charges, we have to also define the orthonormal spacelike vectors $v^{a}\partial_{a}$ along which the current lies, which need to obey \eqref{conditiononvec}. As shown in \cite{Armas:2018ibg}, we will assume the existence of $q$ mutually commuting spacelike Killing vectors $\psi_{i}^{\mu}\partial_{\mu}$ that commute with $k^{\mu}\partial_{\mu}$, such that we can define a set of orthonormal vectors\footnote{Generically, it is not required to have the existence of spacelike Killing vectors in equilibrium, see \cite{Armas:2018ibg, Armas:2018atq, Armas:2018zbe, Armas:2023tyx}.}, 
\begin{equation}\label{fluidspacevelocity}
    v_{i}^{a}\partial_{a}=\mathbb{M}_{i}^{\ j}\zeta_{j}^{a}\partial_{a}\ , \ \ \ \mathbb{M}_{i}^{\ k}\mathbb{M}_{j}^{\ l}\zeta_{k}^{a}\zeta_{a,l}=\delta_{ij}\ ,\ \ \ \zeta_{j}^{a}\partial_{a}\equiv\Big(\psi_{j}^{a}\partial_{a}-\frac{k_{b}\psi^{b}_{j}}{k^{2}}k^{a}\partial_{a}\Big) 
\end{equation} which satisfy $v_{i}^{a}u_{a}=0$ and $v_{i}^{a}v_{ja}=\delta_{ij}$, as required. The matrices $\mathbb{M}_{i}^{\ j}$ are determined by the second condition up to a residual SO$(q)$ symmetry group enjoyed by the spacelike vectors. Notice that in general the $v_{i}^{a}\partial_{a}$ vectors may not to be Killing vectors of the worldvolume, since $k_{b}\psi^{b}_{j}/{k^{2}}$ might vary along directions transverse to the Killing vectors.

Having defined our vectors, we solve the intrinsic eqs. \eqref{newintrinsiceqn} (or equivalently eqs. \eqref{intrinsiceqn}) for the local temperature $\mathcal{T}(\sigma)$ and the chemical potential $\Phi(\sigma)$. The stationarity condition requires that these two physical quantities cannot be time dependent. In order to achieve a solution we first see that the intrinisc equations can be rewritten as \begin{align}\label{intrinsicln}
    (\hat{h}^{ab}+u^{a}u^{b})D_{b}\ln(|k|{\mathcal{T}})&=0 \ , \\
    \mathcal{T}s\hat{\perp}^{ab}D_{b}\ln(|k|{\mathcal{T}})-\Phi_{q}\mathcal{Q}_{q}\hat{\perp}^{ab}(\dot{v}_{b}-D_{b}\ln{(|k|\Phi_{q}}))&=0\  ,
\end{align} where we have used the fact that $\dot{u}_{b}=D_{b}\ln{|k|}$. Now, in the case of $q=0$ and $q=p$, the equations simplify drastically and the solution is provided by \begin{equation}\label{solutiontemperature}
    \mathcal{T}(\sigma)=\frac{T}{|k|}\ ,
\end{equation} where $T$ is the global temperature. In the cases of $0<q<p$, the solution above has to be coupled with a solution for $\Phi_{q}$. In order to achieve this, we use the fact that $\dot{v}_{b}=D_{b}\ln{|\mathbb{M}|}$,\footnote{By exploiting the Killing nature of $k$ and $\psi$, note that $v^{ia}D_{(a}v_{b)i}=\dot{v}_{b}=\delta^{ij}v_{i}^{a}\zeta_{ka}D_{b}\mathbb{M}_{j}^{\ k}$ and from \eqref{fluidspacevelocity} we have that $v^{a}_{i}\zeta_{ka}=(\mathbb{M}^{-1})_{k}^{\ j}\delta_{ij}$. Combining the two results, we see that $\dot{v}_{b}=(\mathbb{M}^{-1})_{k}^{\ j}D_{b}\mathbb{M}_{j}^{\ k}=D_{b}\ln{|\mathbb{M}|}$.} where $|\mathbb{M}|$ is the determinant of the matrix $\mathbb{M}_{i}^{\ j}$. Then, the equations simplify considerably and the solution is simply \begin{equation} \label{solutionphiq}
    \Phi_{q}(\sigma)=\frac{\phi_{q}(\sigma)}{|\hat{h}^{(q)}|^{1/2}}\ ,
\end{equation} where, $\phi(\sigma)$ is a field which can only vary along $\mathcal{C}_{q}$ and $|\hat{h}^{(q)}|^{1/2}=|k||\mathbb{M}|$ is the volume element on $\mathcal{C}_{q+1}$. One can easily check that in the case of $q=1$, where $|\mathbb{M}|=|\zeta|$, the equation above reduces to the one studied in \cite{Caldarelli:2010xz}. We would like to point out that this solution is valid also for static cases. The pair of worldvolume functions $(\mathcal{T}(\sigma),\Phi_{q}(\sigma))$ (or alternatively  $(r_{0}(\sigma),\alpha(\sigma))$), together with the vector $u(\sigma)$, and when needed the vectors $v(\sigma)$, are enough to fully specify the solution for the stationary fluid, as already noted at the end of section \ref{Perfect fluids with q-brane currents}.

Independently of the fact that the fluid considered solves the intrinsic equations as described above, we can relate the chemical potential to the global potential via \cite{Caldarelli:2010xz,Emparan:2011hg}\begin{equation}\label{firstdefn}
    \Phi_{H}^{(q)}=\int_{\mathcal{C}_{q}}d^{q}\sigma|\hat{h}^{(q)}|^{1/2}\Phi_{q}(\sigma)\ .
\end{equation} In the case in which the solutions are provided by \eqref{solutiontemperature} and \eqref{solutionphiq}, the integral above is trivially solved. In order to achieve this, we exploit \eqref{conditiononvec} to show that \begin{equation}
    v^{a,i}\dot{v}_{a}=v^{a,i}\dot{u}_{a}=u^{a}\dot{v}_{a}=0\ ,
\end{equation} which combined with the fact that $D_{a}u^{a}=D_{a}v_{i}^{a}=0$ imply that \eqref{chargeconsveqn} reduces to $v^{a}_{i}D_{a}\mathcal{Q}=u^{a}D_{a}\mathcal{Q}=0$, i.e. the charge does not vary along the fluid vectors. Thus, together with \eqref{solutiontemperature}, one concludes that $\Phi_{q}$ cannot vary along $\mathcal{C}_{q}$, i.e. $\phi(\sigma)=\phi$ in \eqref{solutionphiq}. Therefore, the global chemical potential $\Phi_{H}^{(q)}$ is simply obtained by integrating over the worldvolume element.

In this paper we will define $\xi$ and $\chi$ such that they are canonically normalised with the latter describing angular rotations with closed orbits of periodicity $2\pi$. Thus, on $\mathcal{W}_{p+1}$
\begin{equation}
    -\xi^{2}|_{\mathcal{W}_{p+1}} = R_{0}^{2}(\sigma) \ , \ \ \ \chi^{2}|_{\mathcal{W}_{p+1}}= R^{2}(\sigma)\ ,
\end{equation} 
where $R_{0}$ is the redshift factor between infinity and the worldvolume while $R$ is the proper radius of the orbits on the worldvolume. Furthermore, we will assume that the normal to the hypersurfaces $\mathcal{B}_{p}\subset\mathcal{W}_{p+1}$ is constructed out of $\xi$, such that \begin{equation}
    n^{a}\partial_{a}=\frac{1}{R_{0}}\xi^{a}\partial_{a} \ .
\end{equation} 
Hence, we can define the rapidity $\eta$ of the fluid velocity with respect to the worldvolume time generated by $n$ as 
\begin{equation} \label{defnrapidity}
    n^{a}u_{a}=-\cosh{\eta} \ .
\end{equation} 
This quantity will be of pivotal importance in order to solve the extrinsic equations when considering blackfolds carrying more than one $q$-brane current.  

We now have enough information to study the conserved charges associated with the Killing vectors $\xi$ and $\chi$ for stationary blackfolds in backgrounds with fluxes. It can be proven  \cite{Armas:2012bk} that the conserved charged associated with a Killing vector field $k$ is given by 
\begin{equation}
    \mathcal{Q}_{k} = \int_{\mathcal{B}_{p}}dV_{p}(T^{\mu\nu}+\mathcal{V}^{\mu\nu})n_{\mu}k_{v}|_{x^{\mu}=X^{\mu}}\ ,\ \ \ 
\end{equation} 
where $dV_{p}$ is the spatial volume element and $n_{\mu}$ is the unit normal to the Cauchy slice of constant time and 
\begin{equation}
    \mathcal{V}^{\mu\nu} = \frac{1}{q!}A^{\nu}_{\ \alpha_{1}...\alpha_{q}}\mathcal{J}^{\mu\alpha_{1}...\alpha_{q}}\ .
\end{equation} 
Thus, we can define the rest mass and the angular momentum as \begin{equation}\label{fluxthermo}
    M = \int_{\mathcal{B}_{p}}dV_{p}(T^{\mu\nu}+\mathcal{V}^{\mu\nu})n_{\mu}\xi_{v}|_{x^{\mu}=X^{\mu}}\ , \ \ \  J = -\int_{\mathcal{B}_{p}}dV_{p}(T^{\mu\nu}+\mathcal{V}^{\mu\nu})n_{\mu}\chi_{v}|_{x^{\mu}=X^{\mu}}\ .
\end{equation} 
Just as for the equations of motion, we see that the background fields also play a role in the conserved charges. This is expected since the probe branes are electrically (or magnetically) coupled to the gauge fields. When the fluxes are not present, these expressions are simplified by setting $\mathcal{V}^{\mu\nu}=0$.  In the case of $q<p$, the charge density is not constant over all $\mathcal{W}_{p+1}$ and in order to define it, we need to  consider the $q$ spatial directions in $\mathcal{C}_{q+1}$ orthogonal to $n^{a}\partial_{a}$, which we will denote by $\mathcal{B}_{p-q}$. In doing so, we can define a unit $q$-form $\omega_{q}$ orthogonal to the normal vector \cite{Emparan:2011hg}
\begin{equation}
    \omega_{q}=\frac{-\mathcal{V}_{q+1}\cdot n}{\sqrt{-\hat{h}_{ab}^{(q)}n^{a}n^{b}}} \ ,
\end{equation} 
and define the total charge $Q_{q}$ as the integral of its charge density over these $\mathcal{B}_{p-q}$ directions, which are transverse to the current, such that
\begin{equation}\label{globalchargedfn}
    Q_{q}=-\int_{\mathcal{B}_{p-q}}dV_{p-q}\mathcal{J}_{q+1}\cdot
(n\wedge\omega_{q}) = \int_{\mathcal{B}_{p-q}}dV_{p-q}\sqrt{-\hat{h}_{ab}^{(q)}n^{a}n^{b}}\mathcal{Q}_{q}\ . 
\end{equation} 
We can also re-write the global potential \eqref{firstdefn} as 
\begin{equation} \label{globalphidfn}
    \Phi_{H}^{(q)}=\int_{\mathcal{C}_{q}}dV_{q}\frac{R_{0}}{\sqrt{-\hat{h}_{ab}^{(q)}n^{a}n^{b}}}\Phi_{q}\ .
\end{equation} 
The equations above are also valid for the $q=p$, which reduce to very trivial integrals. Furthermore, it is worth noticing that in our constructions, we will always consider the case of the velocity fluid being parallel to the string current. In such cases, one can easily prove that $\hat{h}_{ab}^{(q)}n^{a}n^{b}=-1$, implying that $\Phi_{H}^{(q)}$ undergoes a gravitational redshift $R_{0}$. In other cases, it does not simplify so easily and Lorentz-boost redshift factors have to be also considered (for further details see \cite{Emparan:2011hg}).

\subsection{Action and thermodynamics}\label{actionandthermodynamics}
In \cite{Emparan:2011hg} it was shown that in the case of stationary blackfolds in the absence of any external gauge field the extrinsic equations \eqref{cartereqn} with vanishing Lorentz force can be obtained from the action \begin{equation}
\label{actiongibbs}
    I = -\int_{\mathcal{W}_{p+1}}\star_{(p+1)}\mathcal{G}\ , 
\end{equation} 
with the local Gibbs free energy \begin{equation}\label{defngibbs}
    \mathcal{G}=\varepsilon-\mathcal{T}s-\sum^{p}_{q}\Phi_{q}\mathcal{Q}_{q}\ .
\end{equation} 
In this ensemble, the extrinsic equations are obtained by varying the action while keeping $(T,\ \Omega,\ \{\Phi_{H}^{(q)}\})$ fixed. For some cases of interest in this paper, it is convenient to introduce the action in another ensemble, where the $\{Q_{q}\}$ charges are kept constant. In order to do so, we start with \eqref{actiongibbs} and perform a Legendre transform. The change of ensemble is dictated by the requirement of gauge invariance and worldvolume diffeomorphism invariance of the action (see \cite{Armas:2016mes} for further details). In the case of a brane with a $Q_{p}$ charge in the presence of an $A_{p+1}$ gauge field in the background (which is the case of interest in this study), performing the Legendre transformation leads to a change in the global chemical potential. This contribution is captured by adding  the pullback of $A_{p+1}$ onto the worldvolume $\mathbb{P}[A_{p+1}]$ to \eqref{firstdefn} such that \cite{Armas:2016mes}
\begin{equation}
\bar{\Phi}_{H}^{(p)}=R_{0}\int_{\mathcal{B}_{p}}\star_{(p)}\Phi_{p}-\int_{\mathcal{B}_{p}}\mathbb{P}[A_{q+1}]\ .
\end{equation} 
Using all this information, we perform the Legendre transformation by  \begin{equation}\label{actionforbrane}
     \tilde{I} = -\int_{\mathcal{B}_{p}}\star_{(p)}R_{0}(\mathcal{G}+\sum_{q\neq p}\Phi_{q}\mathcal{Q}_{q}) -\bar{\Phi}_{H}^{(p)}\mathcal{Q}_{p}\ = \int_{\mathcal{B}_{p}}\star_{(p)}(R_{0}P_{\parallel} -\mathbb{P}[A_{q+1}])\ ,
\end{equation} 
where we have factored out the integral over time and used the thermodynamic relations described in section \ref{Perfect fluids with q-brane currents}. This leads to the required action
\begin{equation}\label{actionbrane}
    I = \int_{\mathcal{W}_{p+1}}(\star_{(p+1)}P_{\parallel}+Q_{p}\mathbb{P}[A_{p+1}])\ .
\end{equation} 
It is straightforward to show that the equations \eqref{cartereqn} are obtained by varying \eqref{actionbrane} while keeping the set of global parameters $(T,\Omega,\{Q_{q}\})$ constant. As we will explicitly argue in later sections, when only a $p$-form charge is present the action reduces to the DBI (M-brane) action in the extremal limit.\footnote{A more general treatment of these actions can be accomplished by exposing their higher-form or higher-group structures \cite{Armas:2023tyx, Armas:2024caa}.}

By Wick rotating and integrating the action above along the time circle, it can be shown that it is equivalent to the thermodynamic action 
\begin{equation} \label{thermopot}
    \tilde{I}_{E} = M-\Omega J -TS \ , 
\end{equation} 
where $\tilde{I}_{E}=I_{E}/\beta$ with $\beta$ being the inverse temperature and $S$ is the global entropy obtained by integrating the entropy current $su^{\mu}$ over $\mathcal{B}_{p}$ \begin{equation} \label{fluxentropy}
    S = -\int_{\mathcal{B}_{p}}dV_{p}su^{a}n_{a}\ .
\end{equation} 
Hence, by noticing that $T$ and $\Omega$ are just integration constants, we conclude that the extrema of the action obey the first law of thermodynamics $dM=TdS+\Omega dJ$ at fixed $\{Q_{q}\}$. Finally, one can show that once the explicit effective thermodynamic quantities have been determined, a Smarr relation can be obtained. This will be shown in later sections together with specific microscopic realisations for the thermodynamic quantities.

\section{Einstein-dilaton gravity with higher-form gauge fields}\label{Higher-dimensionalgravity}
In this section we obtain new perturbative charged and rotating black hole solutions that are asymptotically (A)dS in Einstein-dilaton gravity with a $(q+1)$-form gauge field and a cosmological constant \eqref{actionads}. These solutions will carry a single electric charge ($q=0$), string dipole charge ($q=1$) or $p$-brane charge and have different horizon topologies. In asymptotically dS space, these solutions can also be static. In addition, we discuss the possible extremal limits that these solutions can attain.

\subsection{Single charged blackfold effective fluids}
\label{Blackfoldeffectivefluids}
The near-horizon geometries of the black hole solutions of interest are given by the $q$-charged $p$-brane solutions found in \cite{Caldarelli:2010xz,Emparan:2011hg}. The thermodynamic properties of the effective fluids dual to these near-horizon geometries can be characterised by their horizon size $r_{0}$ and charge parameter $\alpha$. The thermodynamic quantities introduced in the previous section are parameterised in terms of $r_{0}$ and $\alpha$ according to
\begin{gather}\label{localthermo}
    \varepsilon = \frac{\Omega_{n+1}}{16\pi G}r_{0}^{n}(1+n+nN\sinh^{2}{\alpha})\ ,  \\ \mathcal{T}=\frac{n}{4\pi r_{0}\cosh^{N}{\alpha}}\ , \ \  \ s= \frac{\Omega_{n+1}}{4G}r_{0}^{n+1}\cosh^{N}{\alpha}\ , \\ \mathcal{Q}_{q}=\frac{\Omega_{n+1}}{16\pi G}n\sqrt{N}r_{0}^{n}\sinh{\alpha}\cosh{\alpha}\ , \ \  \ \Phi_{q}=\sqrt{N}\tanh{\alpha}\ .
\end{gather} While the above are valid for any $q$, the pressure(s) are case specific. In the case of $0<q<p$ we have  \begin{gather}\label{localpressure}  
   P_{\parallel}=-\frac{\Omega_{n+1}}{16\pi G}r_{0}^{n}(1+nN\sinh^{2}{\alpha}) \ , \ \ \ P_{\perp}= -\frac{\Omega_{n+1}}{16\pi G}r_{0}^{n}\ ,
\end{gather} while for the $q=0$ case, we have that $P_{\parallel}=0$ and $P_{\perp}\equiv P$ and for the $q=p$ case, we have that $P_{\perp}=0$ and $P_{\parallel}\equiv P$. This is in accordance with the fact that the presence of these charges do not disrupt the isotropy of the fluid. The parameter $N$ is related to the dilaton coupling and since we require $a_{q}\geq 0$, it is bounded from above according to\footnote{In order to connect with section \ref{sec:supergravity}, we note that for D-branes in $D=10$ type IIB string theory and M-branes in $D=11$ M-theory, we have that $N=1$, $p=-1,1,3,5,7$ and $p=2,5$, respectively. In this section, we keep $N$ as a general parameter as to find solutions to the general action \eqref{actionads}.}
\begin{equation}
    a_{q}^{2}=\frac{4}{N}-\frac{2(p+1)n}{D-2} \implies N \leq 2\Big(\frac{n+p+1}{n(p+1)}\Big) \ .
\end{equation}
The Gibbs free energy can be written as 
\begin{equation}
    \mathcal{G}= \frac{\Omega_{n+1}}{16\pi G}r_{0}^{n}\ ,
\end{equation} 
where from \eqref{localthermo} and \eqref{localpressure} we notice that $\mathcal{G}=-P_{\perp}$ and $\mathcal{G}=-(P+\Phi_{p}\mathcal{Q}_{p})$  in the case of $0\leq q<p$ and $q=p$ respectively. It is possible to exploit the local thermodynamic quantities to obtain some simple and useful relations, in particular
\begin{align}
    \varepsilon &= -(n+1)P_{\perp}+\Phi_{q}\mathcal{Q}_{q}\ , \\
    \varepsilon &= -(n+1)P-n\Phi_{p}\mathcal{Q}_{p}\ , 
\end{align} 
which implies that the charge provides an additional tension to the worldvolume, but interestingly in opposite ways for the different cases. Note that using the equations above together with \eqref{diffpressure} permits us to write the effective stress-energy tensor as \begin{equation}\label{stressimportant}
    T_{ab}=\mathcal{T}s\Big(u_{a}u_{b}-\frac{1}{n}\gamma_{ab}\Big)-\Phi_{q}\mathcal{Q}_{q}\hat{h}_{ab}^{(q)}\ ,
\end{equation} 
valid identically for all $q$. 

Focusing in the case of equilibrium stationary configurations, we can make use  of \eqref{solutiontemperature}, \eqref{solutionphiq} and the local thermodynamics described above to obtain
\begin{equation}\label{intrinsicsln}
    r_{0}= \frac{n|k|}{4\pi T}\Big(1-\frac{\Phi_{H}^{2}}{c^{2}N|\hat{h}^{(q)}|}\Big)^{N/2}\ , \ \ \ \tanh{\alpha}=\frac{\Phi_{H}}{\sqrt{N}c|\hat{h}^{(q)}|^{1/2}}\ ,
\end{equation} 
where $c$ is a constant such that
\begin{equation} \label{casesh}
    c|\hat{h}^{(q)}|^{1/2}=\begin{cases}
        |k| \ \ \ \text{for} \ \ \ q=0\ , \\
        \Omega_{q}|k||\mathbb{M}| \ \ \ \text{for} \ \ \ 0<q<p, \\
        \Omega_{p}|\gamma| \ \ \ \text{for} \ \ \ q=p\ , 
    \end{cases}
\end{equation} where $\Omega_{q}$ denotes the integration over the angular directions. 
Eqs.~\eqref{intrinsicsln} and \eqref{casesh} provide the equilibrium solutions to the intrinsic equations when the background does not contain any non-trivial fluxes or dilaton fields. Given an equilibriun configuration, we can obtain global thermodynamic quantites as specified in the previous section as well as the Smarr relation
\begin{equation}\label{smarrrelation}
    (D-3)M-(D-2)(TS+\Omega J)-(D-q-3)\Phi_{H}^{(q)}Q_{q}=\mathcal{T}_{tot}\ ,
\end{equation} 
where $\mathcal{T}_{tot}$ is the total tension (or binding energy) given by 
\begin{equation}\label{defntott}
    \mathcal{T}_{tot} = -\int_{\mathcal{B}_{p}}dV_{p}R_{0}(\gamma^{ab}+n^{a}n^{b})T_{ab}\ .
\end{equation} 
This quantity is zero in the case of asymptotically flat backgrounds, as already shown in \cite{Emparan:2011hg,Caldarelli:2010xz} but not necessarily the case for (A)dS \cite{Armas:2010hz}. Indeed, this quantity can also be obtained by direct variation of the global free energy $F=M-TS-\Omega J-\Phi_H^{(q)}Q_q$ with respect to the background scale $L$ of (A)dS (see \cite{Armas:2015qsv}) and is related to black hole volume and spacetime pressure \cite{Kubiznak:2014zwa, Karch:2015rpa, Armas:2015qsv}.

\subsection{Extremal limits}\label{extremallimits1}
Extremal limits can be obtained by simultaneously considering $r_{0}\rightarrow0$ and $\alpha\rightarrow\infty$, while keeping $Q_{q}$ constant. In this context, the effective fluids described above admit different types of extremal limits depending on whether the fluid velocity remains timelike in the limit or becomes null. When considering such limits it is important to distinguish between blackfold solutions carrying $q=p$-brane charge and $q<p$. In the former case, such solutions do not admit boundaries and thus the extremal limit is uniform over all $\mathcal{B}_{p}$. On the other hand, for $q<p$ extremality can be reached locally at the boundaries, while keeping the rest of $\mathcal{B}_{p}$ non-extremal.   

\subsubsection*{Timelike fluid velocity} 
In this case, the extremal limit is taken such that $\mathcal{T}s\rightarrow0$ and $\Phi_{q}\rightarrow \sqrt{N}$ while $u^\mu$ remains timelike rendering the stress tensor \eqref{stressimportant} in the form 
\begin{equation}
    T_{ab}=-\sqrt{N}\mathcal{Q}_{q}\hat{h}^{(q)}_{ab}\ .
\end{equation} 
A close inspection of \eqref{localthermo} and \eqref{localpressure} reveals that in the case $q=p$ we have that $-\sqrt{N}\mathcal{Q}_{p}\hat{h}^{(p)}_{ab}=P\gamma_{ab}$ in which case, since the charge is constant on the worldvolume, the stress tensor describes a brane with uniform tension whose action \eqref{actionbrane} is proportional to the volume and with vanishing gauge field\footnote{In the case of string theory where $N=1$ these correspond to D-branes and the action reduces to the Dirac-Born-Infeld action with vanishing gauge fields.}. 

In this limit, we can make use of the Smarr relation and the fact that the action \eqref{actiongibbs} vanishes to obtain the following relations \begin{equation}\label{relationsp}
      M =(q+1)\Phi_{H}^{(q)}Q_{q}-\mathcal{T}_{tot}\ ,\ \ \ \Omega J = q\Phi_{H}^{(q)}Q_{q}-\mathcal{T}_{tot}\ .
\end{equation} In a static case, the charge potential energy will be proportional to the total tension of the black hole, which further implies that the mass is proportional to the charge. Furthermore, we notice that for $q=0$ the angular momentum is non-zero, thus we can have a rotating extremal black hole with an electric charge. This was not possible in the case of asymptotically flat case as seen in \cite{Caldarelli:2010xz}, implying that the tensional energy associated to the length scale of the background is enough to keep the black hole in a stable state. Finally, we can make use of these two relations to obtain \begin{equation}\label{mqpj}
    M = \Phi_{H}^{(q)}Q_{q} + \Omega J,
\end{equation} meaning that in this limit the total energy can be written independently of the total tension and as a function of the kinetic and charge potential energies. We would like to remark that the expression above is similar to typical BPS relations.

\paragraph{Extremality at the boundary.} \label{extremalboundary} This case occurs when $q\neq p$ and was first discussed in \cite{Caldarelli:2010xz}. We noted that the extremal limit is reached by $r_{0}\rightarrow0$ and $\alpha\rightarrow\infty$, while keeping $Q_{q}$ constant. We also showed in section \ref{boundaries} that at the boundary we require that $r_{0}|_{\partial\mathcal{W}_{p+1}}=0$, which for charged blackfolds can occur either by approaching a null velocity, i.e. $|k|=0$, or at some specific value of $\Phi_{H}$ as one can see in \eqref{intrinsicsln}. The latter case occurs when $\alpha$ diverges and thus we reach the extremal limit at the boundary while leaving the velocity timelike: the blackfold is locally extremal. Particular examples of this phenomenon will be presented in section \ref{diskandannulus}.

\subsubsection*{Null fluid velocity} \label{nullwavebrane}
In this case we require that $r_{0}^{n/2}u^{a}$ remains finite while taking the extremal limit. This can be accomplished introducing the momentum density $\mathcal{K}$ and a vector $l$ that remains finite in this limit 
\begin{equation}\label{rescaling}
    \sqrt{\mathcal{T}s}u^{a} = \mathcal{K}^{1/2}l^{a},
\end{equation} such that when $r_{0}\rightarrow 0$, $l^{a}l_{a}=0$ implying that $l$ is a lightlike vector, i.e. the fluid on the brane is moving at the speed of light. In this case the stress tensor becomes 
\begin{equation}\label{nullstressenergy}
    T_{ab} = \mathcal{K}l_{a}l_{b}-\sqrt{N}\mathcal{Q}_{q}\hat{h}_{ab}^{(q)}\ ,
\end{equation} 
meaning that the brane supports a lightlike momentum wave in addition to the usual ground state. A consequence of this limit is that we obtain $\varepsilon+P_{\parallel}=\mathcal{T}s=0$. For the limit to be well defined in the $q=0$ case, we note that $\hat{h}^{(0)}_{ab}=-u_{a}u_{b}$ and the induced metric diverges. Thus we also need to impose that $\Phi_{0}\mathcal{Q}_{0}\rightarrow0$, implying that the null-wave limit can be obtained only if the charge goes to zero. In the case of $0<q<p$, where $\hat{h}^{(q)}_{ab}(u,v)$, we require the string currents to be parallel to the directions of the wave such that the induced metric remains finite in this limit. In this case the charge survives the limit. Finally, for $q=p$, where $\hat{h}^{(q)}_{ab}(u,v)=\gamma_{ab}$, the charge also survives in the limit. For further details, we refer the reader to \cite{Emparan:2011hg}.  

By letting $n^{a}l_{a}=-1$\footnote{Note that given the definition of $\mathcal{K}$ in \eqref{rescaling}, this normalisation implies that $\mathcal{K}=\mathcal{K}(R_{0},r_{0})$. One can remove the dependence from $R_{0}$ by letting $n^{a}l_{a}=-R_{0}$.}, it is easy to see that \eqref{fluxthermo} and \eqref{defntott} in the absence of the background gauge field, reduce to  \begin{align} \label{integralsnullwave}
    \mathcal{T}_{tot}&= -\int_{\mathcal{B}_{p}}dV_{p}R_{0}(\mathcal{K}-q\sqrt{N}\mathcal{Q}_{q})\ , \\ 
    M &= \int_{\mathcal{B}_{p}}dV_{p}R_{0}(\mathcal{K}+\sqrt{N}\mathcal{Q}_{q})\ , \\
    J &= \int_{\mathcal{B}_{p}}dV_{p}\mathcal{RK}\ , \ \ \ \mathcal{R}\equiv l_{a}\chi^{a}\ ,
\end{align} where, the latter can be viewed as the lever arm radius for the momentum. In the case when the integrands are constant over the worldvolume,  we obtain \eqref{relationsp} for the null-brane,\begin{align}\label{relationsnull}
     M &= (q+1)\Phi_{H}^{(q)}Q_{q} - \mathcal{T}_{tot}\ = \frac{1}{q}(V_{p}R_{0}(q+1)\mathcal{K} + \mathcal{T}_{tot})\ , \\
    \frac{J}{\mathcal{R}} &= \frac{1}{R_{0}}(q\Phi_{H}^{(q)}Q_{q} - \mathcal{T}_{tot})= V_{p}\mathcal{K} \ .
\end{align} where, one can see that the total tension contributes to the virialisation of total energy of the brane at equilibrium between kinetic and potential energy. Combining the two \begin{equation}\label{mqpjnull}
    M=V_{p}R_{0}\sqrt{N}Q_{p}+R_{0}\frac{J}{\mathcal{R}}\ ,
\end{equation} which permits us to write the total energy in terms of the charge and angular momentum. It is worth noticing that these relations receive corrections at next orders in the blackfold expansion as shown in \cite{Emparan:2008qn,Blanco-Pillado:2007eit}. However, since this work is restricted to ideal order, the equations above remain valid.

\subsection{Odd-sphere solutions}\label{oddspheres}
Here we will construct perturbative black hole solutions carrying either an electric charge, $q<p$ dipole charge or a $q=p$-brane charge with a $\mathbb{S}^p\times \mathbb{S}^{n+1}$ horizon topology with $p$ an odd number that asymptote to global AdS with metric \eqref{globalads}. We will embed the odd sphere $S^{2k+1}$ with $p=2k+1$\footnote{Note that the particular case of $k=0\implies p=1$ is the case of a black string and can show that in the neutral case we reproduce the results in \cite{Caldarelli:2008pz}.} in a $(p+1)$-dimensional subspace of AdS$_{D}$ with constant radius and parametrise the sphere using $k+1$ Cartan angles $\phi_{i}$ and $k$ independent direction cosines $\mu_{i}$, such that 
\begin{equation} \label{eq:embeddingodd}
    \sigma^{0}=t = \tau\ , \ \ \ r=R\ , \ \ \ \sigma_{i}=\phi_{i}\ ,\ \ \ \sigma_{j}=\mu_{i}\ , 
\end{equation} 
with $i=(1,...,k+1)$ and $j=(k+2,...,p+1)$. This leads to the induced  metric \begin{equation}\label{embeddsingle}
    \gamma_{ab}d\sigma^{a}d\sigma^{b} = -f(R)d\tau^{2} + R^{2}\sum_{i=1}^{k+1}(d\mu_{i}^{2}+\mu_{i}^{2}d\phi_{i}^{2})\ , \ \ \ \sum_{i=1}^{k+1}\mu_{i}^{2} = 1 \ .
\end{equation} 
We consider configurations that are rotating along all Cartan angles with the same angular velocity $\Omega$, such that 
\begin{equation}\label{ksingle}
    k^{a}\partial_{a} = \partial_{\tau} + \Omega\partial_{\Phi}\ , \ \ \ \partial_{\Phi}\equiv\sum_{i=1}^{k+1}\partial_{\phi_{i}}\ ,
\end{equation} 
from which the fluid velocity can be extracted from \eqref{fluidvelocity}. For $q=1$ we need to also introduce the spacelike vector $v^{a}\partial_{a}$, which can be defined by using \eqref{fluidspacevelocity}. In order to do so we start with $\psi^{\mu}\partial_{\mu}=\Omega\partial_{\Phi}$, which commutes with $k^{\mu}\partial_{\mu}$, such that \begin{equation}\label{zetasingle}
    \zeta = \frac{1}{|k|^{2}} \Big(\Omega^{2}R^{2}\partial_{\tau}+f(R)\Omega\partial_{\Phi}\Big)\ ,
\end{equation} 
concluding our search. For other values of $q$, namely $1<q<p$ we discuss the construction for static geometries in appendix \ref{staticsolutions}.

We now have all the main ingredients in order to write the down the action describing this class of blackfold configurations. We will consider the ensemble where the global potential $\Phi_{H}^{(p)}$ is kept fixed and thus consider the action \eqref{actiongibbs}. Hence, by using the above, \eqref{intrinsicsln}, \eqref{fluxentropy} and the local thermodynamics in section \ref{Blackfoldeffectivefluids}, the equilibrium action is 
\begin{equation}
    \tilde{I}_{E}= \frac{\Omega_{n+1}V_{p}}{16\pi G}\sqrt{f(R)}\Big(\frac{n}{4\pi T}\Big)^{n}(f(R)-\Omega^{2}R^{2})^{n/2}\Big(1-\frac{\Phi_{H}^{2}}{c^{2}N|\hat{h}^{(q)}|}\Big)^{nN/2}\ ,
\end{equation} 
where $V_{p}\equiv\Omega_{p}R^{p}$ is the volume of $\mathcal{B}_{p}$ and $c^{2}|\hat{h}^{(q)}|$ is given in \eqref{casesh}. Finally, by extremising the action with respect to $R$, we solve the extrinsic equations. We will present the result in terms of the known local thermodynamic quantities \eqref{localthermo} in order to treat all $q$ cases simultaneously
\begin{equation}\label{omega01p}
    \Omega^{2} =\frac{f(R)}{R^{2}} \frac{p\mathcal{T}s+qn\Phi_{q}\mathcal{Q}_{q}+\frac{R^{2}}{L^{2}}\Xi_{q}}{(n+p)\mathcal{T}s+(\delta_{q,0}+q)n\Phi_{q}\mathcal{Q}_{q}+\frac{R^{2}}{L^{2}}\Xi_{q}}\ , 
\end{equation} 
where $\Xi_{q}=\mathcal{T}s(1+n+p)+n(1+q)\Phi_{q}\mathcal{Q}_{q}$. We have explicitly checked that \eqref{omega01p} holds for $q=0,1,p$ but we expect it to be satisfied for all $q$. Indeed, as shown in appendix \ref{staticsolutions} condition \eqref{omega01p} holds for arbitrary $q$ in dS space when $\Omega=0$. The global thermodynamics, found by using \eqref{fluxthermo}, \eqref{fluxentropy}, \eqref{globalchargedfn} and \eqref{globalphidfn}, can be written in a very neat way in terms of the known local thermodynamics \eqref{localthermo} and by exploiting the rapidity \eqref{defnrapidity}, in particular
\begin{align} \label{globalthermoq=p}
     M &= V_{p}R_{0}\Big(\cosh^{2}{\eta}(\mathcal{T}s+\delta_{q,0}\Phi_{q}\mathcal{Q}_{q}) + \frac{\mathcal{T}s}{n}+(1-\delta_{q,0})\Phi_{q}Q_{q}\Big)\ , \\ 
    J&= V_{p}R(\mathcal{T}s+\delta_{q,0}\Phi_{q}\mathcal{Q}_{q})\cosh{\eta}\sinh{\eta}\ , \\
    S &= V_{p}s\cosh{\eta}\ , \ \ \ T = \mathcal{T}R_{0}\sech{\eta}\ ,
\end{align} 
while the charge and chemical potentials take particular values for different values of $q$, namely
\begin{align}\label{globaltermophiq}
   \Phi_{H}^{(0)} &= R_{0}\sech{\eta}\Phi_{0}\ , \ \ \ Q_{0} = V_{p}\cosh{\eta}\mathcal{Q}_{0}\ ,\\ \Phi_{H}^{(1)} &= 2\pi R R_{0}\Phi_{1}\ ,\ \ \ Q_{1} = \frac{V_{p}}{2\pi R}\mathcal{Q}_{1}\ , \\ \Phi_{H}^{(p)} &= V_{p}R_{0}\Phi_{p}\ ,  \ \ \ Q_{p} = \mathcal{Q}_{p}\ .
\end{align} 
Finally, it is instructive to explicitly compute the total tension as a function of the local thermodynamic quantities (and not of the rapidity). Using \eqref{defntott} we have 
\begin{equation}\label{ttotp}
    \mathcal{T}_{tot}= -\frac{V_{p}}{n}\sqrt{f(R)}\frac{R^{2}}{L^{2}}\Xi_{q}\ ,
\end{equation} 
where as seen already in other cases \cite{Armas:2010hz,Armas:2016mes} does not vanish in AdS. Most importantly, we can see that $\mathcal{T}_{tot}\propto R^{2}/L^{2}$, implying that in the flat limit the total tension goes to zero as required. Having obtained the global thermodynamics, one can easily see that the Smarr relation \eqref{smarrrelation} is satisfied.

\paragraph{de Sitter space.} The results for the dS case are easily obtainable from the AdS analysis by Wick rotating $L\rightarrow iL$. In particular, we see that the equilibrium condition \eqref{omega01p} is given by 
\begin{equation}\label{omegads1}
    \Omega^{2} =\frac{f'(R)}{R^{2}} \frac{p\mathcal{T}s+qn\Phi_{q}\mathcal{Q}_{q}-\frac{R^{2}}{L^{2}}\Xi_{q}}{(n+p)\mathcal{T}s+(\delta_{q,0}+q)n\Phi_{q}\mathcal{Q}_{q}-\frac{R^{2}}{L^{2}}\Xi_{q}}\ ,  \ \ \ f'(R)=1-\frac{R^{2}}{L^{2}}\ .
\end{equation} 
Because of the various subtractions appearing in \eqref{omegads1} it is not guaranteed that $\Omega$ takes real values. For this reason, these solutions can only exist if $R$ is constrained according to 
\begin{equation}\label{possibleR}
    \Big(\frac{(n+p)\mathcal{T}s+(\delta_{q,0}+q)n\Phi_{q}\mathcal{Q}_{q}}{\Xi_{q}}\Big)^{1/2} < \frac{R}{L} \leq 1 \ \ \ \lor \ \ \  \frac{R}{L} \leq \Big(\frac{p\mathcal{T}s+qn\Phi_{q}\mathcal{Q}_{q}}{\Xi_{q}}\Big)^{1/2}\ ,
\end{equation} 
where we have excluded the lower bound of the first solution as it would imply that $\Omega\rightarrow\infty$, which is a regime where the validity of this analysis breaks down. The upper bound of the first solution is a special one as it would imply that $r_{0}=0$, which can only occur in the extremal cases or on the boundary. In fact, by sending $\alpha\rightarrow\infty$ in \eqref{omega01p}, one can indeed show that this bound is saturated. The dependence of the radius on the charge parameter can easily be obtained by rewriting the above relations using \eqref{localthermo}.  

Similarly, the global thermodynamic quantities are can be derived from \eqref{globalthermoq=p} by Wick rotation. In particular, notice that the total tension is proportional to $L^{2}$ and for this reason the resultant tension will have the opposite sign 
\begin{equation}
    \mathcal{T}_{tot}= \frac{V_{p}}{n}\sqrt{f(R)}\frac{R^{2}}{L^{2}}\Xi_{q}\ ,
\end{equation} 
which is an expected result given the different background geometries considered. When relating this to black hole chemistry \cite{Kubiznak:2014zwa, Karch:2015rpa, Armas:2015qsv}, the sign flip arises due to the spacetime pressure having opposite sign.

Finally, note that differently from AdS, in dS space we can also have a static solution, i.e. $\Omega = 0$, which occurs when the radius saturates the second bound in \eqref{possibleR}. In this case, $\tanh{\eta}=0\implies\eta=0$, which simplifies the thermodynamics considerably.

\paragraph{Extremal limit.}\label{oddadsextr}
In the extremal limit $\mathcal{T}s\rightarrow0$ the equilibrium conditions \eqref{omega01p} and \eqref{omegads1} simplify. Starting with the $q=0$ case, the equilibrium condition reduces to
\begin{equation} \label{eq:extremaleq}
    \Omega^{2} =\pm\frac{1}{L^{2}}\ .
\end{equation} 
In the case of AdS (+ sign in \eqref{eq:extremaleq}), one can easily check that this does not lead to $|k|\rightarrow 0$, meaning that we are in the timelike extremal limit. One can therefore use the explicit thermodynamic relations and show that \eqref{mqpj} is obeyed and given by 
\begin{equation}\label{nonbps}
    M = \frac{J}{L}+\sqrt{Nf(R)}Q_{0}\ ,
\end{equation} 
where, the proportionality factor $\sqrt{N}$ reminds us that we are not in a supersymmetric theory and this is thus a BPS-like relation. On the other hand, in the case of dS (- sign in \eqref{eq:extremaleq}), we see that the extremal limit does not exist, which is in accordance with the possible values of $R$ shown in \eqref{possibleR}. We would like to remark, that $R\rightarrow0$ is also another possible solution to the equilibrium condition but this solution is excluded since it is beyond the regime of validity of the blackfold approximation. Further details on the regime of applicability are given in section \ref{regimeofapplicability}.

When considering other values of $q$, we see that $\Omega^{2}R^{2} = f(R)$ in the extremal limit, which in the case of the AdS imposes a lower bound on the value of the angular velocity $\Omega>1/L$. This simply tells us that even in the extremal limit there cannot be a static black hole in AdS, as expected. Next, notice that in this limit $\tanh\eta = 1 \implies \eta \rightarrow\infty$, meaning that the local boost becomes lightlike. Thus in this case the limit is achieved when the fluid is moving at the speed of light. In fact, by identifying the momentum density as $\mathcal{K}=\mathcal{T}s\cosh{\eta}\sinh{\eta}$, we see that the global thermodynamic quantities \eqref{globalthermoq=p} are the integrated versions of \eqref{integralsnullwave}. In order to continue the analysis, we first define the lightlike vector as 
\begin{equation} \label{nullvectorodd}
    l = \frac{1}{\sqrt{f(R)}}\partial_{\tau} + \frac{1}{R}\partial_{\Phi}\ ,
\end{equation}  
such that $\chi_{a}l^{a}=R$. We can combine the relations \eqref{relationsnull} to write the radius in terms of the global thermodynamic quantities \begin{equation}\label{rjmt}
    R = \frac{L(q+1)J}{\sqrt{L^{2}(qM-\mathcal{T}_{tot})^{2}-(q+1)^{2}J^{2}}}\ ,
\end{equation} 
which sets a lower bound on the angular momentum $J$. This is a consequence of the bound $\Omega>1/L$ previously found on the angular velocity. This leads to a simple relation between all the global thermodynamic quantities 
\begin{equation} \label{qmtj}
    Q_{q}=\frac{(M+\mathcal{T}_{tot})\big(L^{2}(qM-\mathcal{T}_{tot})^{2}-(q+1)^{2}J^{2}\big)^{\frac{2q+1}{2}}}{\sqrt{N}\Omega_{q}\big(L(q+1)\big)^{q+1}J^{q}(qM-\mathcal{T}_{tot})}\ .
\end{equation} 
By fixing two out of the five parameters, one can study how the thermodynamic quantities impose bounds on each other. Such bounds will be saturated by the equations \eqref{relationsnull}. Finally, we can explicitly check that the relation \eqref{mqpjnull} is obeyed and given by 
\begin{equation}
     M=V_{q}\sqrt{Nf(R)}Q_{q}+\sqrt{f(R)}\frac{J}{R}\ ,
\end{equation} 
where we can see a further departure from a BPS-like relation due to the fact that now we are considering dipole charges. 

The analysis for the dS case proceeds in a similar manner but with some important differences. Interestingly enough, the equilibrium radius is now given by
\begin{equation}\label{rjmtds}
    R = \frac{L(q+1)J}{\sqrt{L^{2}(qM-\mathcal{T}_{tot})^{2}+(q+1)^{2}J^{2}}}\ ,
\end{equation} 
which does not require a lower bound on the angular momenta. This is possible since in dS there exists a static limit as we demonstrated above. Note that as $J\rightarrow 0$ we obviously depart from the null extremal limit and a different analysis is required. To do so, we first take the static limit and then apply the extremal one yielding\footnote{Note that the order is non-commutative. This is an artefact of the fact that the timelike extremal limit does not exist.} 
\begin{equation}
    \lim_{\alpha\rightarrow\infty}(\lim_{\Omega\rightarrow0}\Omega^{2})\implies \frac{R^{2}}{L^{2}} = \frac{p}{1+p} \ \ \ \lor \ \ \ \frac{R^{2}}{L^{2}} =1\ .
\end{equation} 
The latter solution means that the radius of the blackfold coincides with the one of the dS spacetime, i.e. the black hole horizon approaches the cosmological horizon. It is therefore not a surprise that in this case $r_{0} = 0$, since we are hitting a limiting surface/boundary in spacetime \cite{Armas:2015kra}. For this particular choice of radius all thermodynamic quantities vanish and therefore this solution must be discarded for dS space. In contrast, the null extremal limit is valid.

\subsection{Product of odd-spheres solutions}\label{productofoddspheres}
We will now generalise the previous solution to a product of odd-spheres, $\mathcal{B}_{p}=\prod^{b}_{a=1}S^{pa}$ with $p_{a}=2k_{a}+1$ and $\sum_{a=1}^{b}p_{a}=p$. These solutions have horizon topology $\prod_a \mathbb{S}^{p_a}\times \mathbb{S}^{n+1}$. We will embed these spheres in a $p+b$ subspace of AdS$_{D}$ with metric \eqref{globalads} and therefore the number of possible odd-spheres is limited by the dimensions of the transverse space, $b\leq n+2$. Thus, following the embedding procedure of the previous case, we embed the product of spheres according to
\begin{equation}\label{embeddprod}
    \gamma_{ab}d\sigma^{a}d\sigma^{b} = -f(R)d\tau + \sum_{a=1}^{b}\sum_{i=1}^{k+1}R_{a}^{2}(d\mu_{a,i}^{2}+\mu_{a,i}^{2}d\phi_{a,i}^{2})\ ,\ \ \ R^{2}\equiv\sum_{a=1}^{b}R^{2}_{a}\ , \ \ \ \sum_{i=1}^{k_{a}+1}\mu_{i}^{2} = 1\ .
\end{equation} 
Setting each sphere to rotate on all angles $\phi_{i}$ with equal velocity, i.e. $\Omega_{a,i}=\Omega_{a}$, the timelike Killing vector is\begin{equation}\label{kprod}
    k = \partial_{\tau} + \sum_{a=1}^{b}\Omega_{a}\partial_{\Phi_{a}}\ , \ \ \ \partial_{\Phi_{a}}\equiv\sum_{i=1}^{k+1} \partial_{\phi_{a,i}}\ .
\end{equation} For the $q=1$ case, we further consider the Killing vector $\psi^{\mu}\partial_{\mu}=\sum_{a}^{m}\Omega_{a}\partial_{\Phi_{a}}$, leading to the spacelike vector 
\begin{equation}\label{zetaprod}
     \zeta = \frac{1}{|k|^{2}}\sum_{a=1}^{b}\Big(\Omega_{a}^{2}R_{a}^2\partial_{\tau}+f(R)\Omega_{a}\partial_{\Phi_{a}}\Big)\ . 
\end{equation} 
Hence, following the steps of the previous section, we write \eqref{actiongibbs} as \begin{equation} 
    \tilde{I}_{E}= \frac{\Omega_{n+1}V_{p}}{16\pi G}\Big(\frac{n}{4\pi T}\Big)^{n}(f(R)-\sum_{a=1}^{b}\Omega_{a}^{2}R_{a}^{2})^{n/2}\Big(1-\frac{\Phi_{H}^{2}}{c^{2}N|\hat{h}^{(q)}|}\Big)^{nN/2}\ ,
\end{equation} 
with $V_{p}=\prod_{a=1}^{m}V_{p_{a}}$ and $c^{2}|\hat{h}^{(q)}|$ having the same structure as \eqref{casesh}. When varying with respect to $R_{a}$, we obtain a set of $b$ equations which are solved for a collection of equilibrium conditions. In what follows we will group the $q=0,\ p$ solutions together, while the $q=1$ has a different form 
\begin{align}\label{omegaproduct}
    \Omega_{a}^{2} &=\frac{f(R)}{R_{a}^{2}} \frac{p_{a}\mathcal{T}s+q_{a}n\Phi_{q}\mathcal{Q}_{q}+\frac{R_{a}^{2}}{L^{2}}\Xi_{q}}{(n+p)\mathcal{T}s+(\delta_{q,0}+q)n\Phi_{q}\mathcal{Q}_{q}+\frac{R^{2}}{L^{2}}\Xi_{q}}\ , \ \ \ q_{a}=(0,p_{a}) \ , \\
    \Omega_{a}^{2} &=\frac{f(R)}{R_{a}^{2}} \frac{\Big( p_{a}\mathcal{T}s+\frac{R_{a}^{2}}{L^{2}}\Xi_{1}\Big)\Big(p\mathcal{T}s+n\Phi_{1}\mathcal{Q}_{1}+\frac{R^{2}}{L^{2}}\Xi_{1}\Big)}{\Big(p\mathcal{T}s+\frac{R^{2}}{L^{2}}\Xi_{1}\Big)\Big((n+p)\mathcal{T}s+n\Phi_{1}\mathcal{Q}_{1}+\frac{R^{2}}{L^{2}}\Xi_{1}\Big)}\ , \ \ \ q=1\ .
\end{align} 
It is easy to see that if we had to set $b=1$, we recover exactly the result for a single odd-sphere. Once more, we will exploit the rapidity \eqref{defnrapidity} to write the physical quantities in a neat way, which is now given by 
\begin{equation}
    \tanh{\eta}=\frac{\sum_{a=1}^{b}\Omega_{a}R_{a}}{\sqrt{f(R)}}\ .
\end{equation} 
The thermodynamic quantities that follow from here coincide (up to the new definition of $V_{p}$) with the ones obtained for the single odd-sphere case in \eqref{globalthermoq=p}, \eqref{globaltermophiq} and \eqref{ttotp} except for the angular momentum, which, for each sphere is given by \begin{equation}
    J_{a}= V_{p}\frac{\Omega_{a}R_{a}^{2}}{|k|}\mathcal{T}s\cosh{\eta}\ ,
\end{equation} which in the case of $b=1$ yields the angular momentum obtained in \eqref{globalthermoq=p}. These charges obey the Smarr relation \eqref{smarrrelation}.

\paragraph{de Sitter space.} As in the previous case, the results in dS are obtained just by the Wick rotating $L\rightarrow iL$. The extra conditions on $R$ in order to require $\Omega^{2}\geq0$ are given by 
\begin{itemize}
    \item for $q_{a}=(0,p_{a})$ \begin{gather}\label{possibleRa0pa}
     \frac{R_{a}}{L} \leq \Big(\frac{p_{a}\mathcal{T}s+q_{a}n\Phi_{q}\mathcal{Q}_{q}}{\Xi_{q}}\Big)^{1/2}\ \ \ \lor\ \\ \frac{R_{a}}{L} \geq \Big(\frac{p_{a}\mathcal{T}s+q_{a}n\Phi_{q}\mathcal{Q}_{q}}{\Xi_{q}}\Big)^{1/2}\ \ \ \land \ \ \ \Big(\frac{(n+p)\mathcal{T}s+(\delta_{q,0}+q)n\Phi_{q}\mathcal{Q}_{q}}{\Xi_{q}}\Big)^{1/2} < \frac{R}{L} < 1 \ ;  
    \end{gather}
    \item for $q=1$ \begin{align}\label{possibleRa1}
     \frac{R_{a}}{L} \leq \Big(\frac{p_{a}\mathcal{T}s}{\Xi_{1}}\Big)^{1/2}\ \ \ \ &\lor\ \ \ \frac{R_{a}}{L} \geq \Big(\frac{p_{a}\mathcal{T}s}{\Xi_{1}}\Big)^{1/2}\ \ \ \land \\ \Bigg(\frac{R}{L} \leq  \Big(\frac{p\mathcal{T}s+n\Phi_{1}\mathcal{Q}_{1}}{\Xi_{1}}\Big)^{1/2} \ \ \ &\lor \ \ \ 
     \Big(\frac{(n+p)\mathcal{T}s}{\Xi_{1}}\Big)^{1/2} < \frac{R}{L} < 1 \Bigg) \ ;  
    \end{align}
\end{itemize} 
where, as for the single odd-sphere case we excluded the lower bounds that do not satisfy the validity of this analysis. Also, the saturated bound $R/L=1$ is valid only for extremal cases. The static solution $\Omega_a=0$ occurs at a fixed radius corresponding to the saturated bounds of the first inequalities in the above solutions. It is possible to show that the thermodynamic quantities are non-trivial in the static case. In fact, when setting $b=1$ such that $p_{a}=p$ we recover exactly the single odd-sphere thermodynamic quantities.

\paragraph{Extremal limit.}
Following the analysis done for the single odd-sphere, we see that for the $q=0$ case we have that $\Omega_{a}^{2}=\pm1/L^{2}$. In the case of AdS, this implies that in the timelike extremal limit we can explicitly prove that the thermodynamic quantities obey \begin{align} \label{prodmqpj}
    M &= \sum_{a}^{b}\Omega_{a}J_{a} + \Phi_{H}^{(q)}Q_{q} \ , \\ \implies M &= \sqrt{f(R)}\frac{J}{L}+\sqrt{Nf(R)}Q_{0} \ ,
\end{align} 
where the first line is simply a generalisation to \eqref{mqpj} for the product of odd-spheres and in the second line we used $J\equiv\sum_{a}^{b}J_{a}$. On the other hand, for dS, we see that this limit is not a valid one due to \eqref{possibleRa0pa}. In turn, for $q>0$, we have that 
\begin{equation}
    \Omega_{a}^{2} =  \frac{f(R)}{R_{a}^{2}}\frac{p_{a}+(1+p)\frac{R_{a}^{2}}{L^{2}}}{p+(1+p)\frac{R^{2}}{L^{2}}} \equiv \frac{f(R)}{R_{a}^{2}}\rho_{a} \ ,
\end{equation} 
with $\sum_{a=1}^{b}\Omega_{a}^{2}R_{a}^{2}=f(R)$, implying that the fluid velocity is lightlike. We define the null vector as 
\begin{equation}
    l = \frac{1}{\sqrt{f(R)}}\partial_{t} + \sum_{a=1}^{b}\Big(\frac{\rho_{a}}{R_{a}^{2}}\Big)^{1/2}\partial_{\Phi_{a}}\ ,
\end{equation} 
such that $l_{a}\chi^{a}=\sum_{a=1}^{b}R_{a}\sqrt{\rho_{a}}$ while the angular momentum is given by 
\begin{equation}
    J_{a}=V_{p}\mathcal{K}\sum_{a=1}^{b}R_{a}\sqrt{\rho_{a}}\ .
\end{equation} 
Hence, we can relate the mass to the angular momentum according to
\begin{equation}\label{finalnullprod}
    M = \frac{q+1}{q}\Big(\sum_{a=1}^{b}\frac{\sqrt{f(R)}J_{a}}{R_{a}\sqrt{\rho_{a}}}+\frac{\mathcal{T}_{tot}}{q+1}\Big)\ .
\end{equation} 
The two equations above are the generalisations of the results for a single odd-sphere which can be recovered by setting $b=1$. Finally, the above can be repeated for the dS case, which in the case of a static black hole leads to 
\begin{align}
    \lim_{\alpha\rightarrow\infty}(\lim_{\Omega\rightarrow0}\Omega^{2})&\implies \frac{R_{a}^{2}}{L^{2}} = \frac{p_{a}}{1+p} \ \ \ \lor \ \ \ \frac{R^{2}}{L^{2}} =1~.
\end{align} 
An analysis to the AdS case reveals similar results as for a single odd-sphere in dS.

\subsection{Disk and annulus solutions}\label{diskandannulus}
In this section we consider embedding a disk geometry in the background \eqref{globalads} with $q=0,1$. The case $q=0$ was already obtained in \cite{Armas:2015qsv} and we merely reviewed it here. The flat space limit of these configurations reduces to the solutions found in \cite{Caldarelli:2010xz}. Embedding the disk leads to the following worldvolume metric
\begin{equation}
    \gamma_{ab}d\sigma^{a}d\sigma^{b}=-f(r)d\tau^2+f(r)^{-1} dr^2+r^2d\phi^2\ .
\end{equation} 
This embedding corresponds to a minimal surface and for this reason it solves trivially the extrinsic equations \cite{Armas:2015kra}. Setting the blackfold to rotate with a constant velocity $\Omega$ along the angle $\phi$, we have the timelike Killing vector 
\begin{equation}
    k^{a}\partial_{a} = \partial_{\tau} + \Omega \partial_{\phi}\ .
\end{equation} 
Given the Killing vector and the embedding it is easy to see that $|k|$ will vanish at particular values of $r$. These introduce natural boundaries in the geometry in which the fluid is moving at the speed of light. For the string current, we also define the spacelike vector along which the current lies by considering $\psi^{\mu}\partial_{\mu}=\partial_{\phi}$, such that \begin{equation}
    \zeta^{a}\partial_{a} = \frac{1}{|k|^{2}}(\Omega r^{2}\partial_{\tau}+f(r)\partial_{\phi}) \ , 
\end{equation} 
completing the requirements to solve the intrinsic equations. We now focus the two cases separately. 

\paragraph{Electric charge ($q=0$).} In this case, the horizon radius and chemical potentials \eqref{intrinsicsln} are given by
\begin{align}
    r_{0}(r) &= \frac{n}{4\pi T}\left(f(r)-\Omega^2 r^2\right)^\frac{1-N}{2} \left(f(r)-\Omega^2 r^2-\frac{\Phi_H^2}{N}\right)^\frac{N}{2}\ ,\\
\tanh \alpha(r)&= \frac{\Phi_{H}}{\sqrt{N}\sqrt{f(r)-\Omega^2 r^2}}\ . 
\end{align} 
As mentioned above these geometries have boundaries determined by the locations where $r_0=0$ as discussed in section \ref{boundaries}. In the $q=0$ this happens at a single location, namely
\begin{equation}
    r = r_{max}=\sqrt{\frac{1-\frac{\Phi_H^2}{N}}{\Omega^2-\frac{1}{L^2}}}\ ,
\end{equation} 
which sets the maximum radius of the disk. In fact, one can check that the requirement that $r_{0}$ is real and positive implies that $0\leq r\leq r_{max}$. Hence in this case, the topology is one that of a sphere $\mathbb{S}^{n+1}$ fibbered over the disk $D^{2}$ leading to a $\mathbb{S}^{n+3}$ horizon topology. Having defined the boundaries of our geometry, we can write the action of the blackfold as \begin{equation} \begin{aligned}\label{actiondisk}
    \tilde{I}_{E}&=\frac{ \Omega_{(n+1)}}{8 G}\left(\frac{n}{4\pi T}\right)^{n}\int_{0}^{r_{max}} dr  r\left(f(r)-\Omega^2 r^2\right)^\frac{n-n N}{2} \left(f(r)-\Omega^2 r^2-\frac{\Phi_H^2}{N}\right)^\frac{n N}{2} \\
    & =\frac{\Omega_{(n+1)}}{8 G}\left(\frac{n}{4\pi T}\right)^{n}\frac{x^{2+nN}}{\left(2+nN\right)\left(\Omega^2-\frac{1}{L^2}\right)}{}_2 F_{1}\left(1, \frac{1}{2}(N-1) n ; \frac{N n}{2}+2 ;x^{2}\right) \ ,
\end{aligned} \end{equation} 
where, in the first line we have already performed the integral over $\phi\in[0,2\pi]$ and $x\equiv\left(1-\frac{\Phi_H^2}{N}\right)^{1/2}$. The global thermodynamics are found by using \eqref{fluxthermo}, \eqref{fluxentropy}, \eqref{globalchargedfn}, \eqref{globalphidfn} and \eqref{defntott}. Although they can be obtained analytically, the dependence of the integrand on the radius ensures that the results are rather cumbersome. For this reason, we have decided to present the physical parameters in appendix \ref{physicaldiskannulus}. 

\paragraph{String charge ($q=1$).} In the case of $q=1$ the horizon thickness and chemical potential are given by
\begin{align}
    r_{0}(r) &= \frac{n}{4\pi T}\left(f(r)-\Omega^2 r^2\right)^\frac{1}{2} \left(1-\frac{\Phi_H^2}{Nf(r)(2\pi r)^{2}}\right)^\frac{N}{2}\ ,\\
    \tanh \alpha(r)&= \frac{\Phi_{H}}{2\pi\sqrt{N}\sqrt{f(r)}r}\ . 
    \end{align} 
When checking for the boundaries, we see that the radius is bounded both from above and below according to 
\begin{equation}
    r_{min}=\frac{L}{\sqrt{2}}\sqrt{\sqrt{1+\frac{\Phi_H^2}{NL^2 \pi^2}}-1}\leq r \leq r_{max}=\sqrt{\frac{1}{\Omega^2-\frac{1}{L^2}}}\ ,
\end{equation} 
implying that instead of the geometry of a disc we are now dealing with the geometry of an annulus. In this case the transverse $\mathbb{S}^{n+1}$ is fibbered over the annulus leading to the horizon topology of a black ring $\mathbb{S}\times \mathbb{S}^{n+2}$. The validty requirement that $r_{max}-r_{min}\gg r_{c}$ implies that $s^{n+2}$ is a prolate sphere elongated in the directions transverse to the strings. The action for this configuration is 
\begin{equation}\label{actionannulus}
    \tilde{I}_{E}=\frac{ \Omega_{(n+1)}}{8  G} \left(\frac{n}{4\pi T}\right)^n\int_{r_{min}}^{r_{max}} dr r  \left(f(r)-\Omega^2 r^2\right)^\frac{n}{2} \left(1-\frac{\Phi_H^2}{Nf(r)\left(2\pi r\right)^2}\right)^\frac{n N}{2}\ ,
\end{equation} 
where, we have integrated over $\phi\in[0,2\pi]$. This integral as well as similar ones for the global thermodynamic properties in appendix \ref{physicaldiskannulus} cannot be performed analytically but can be calculated numerically. 

Finally, we would like to remark that the generalisation of these embeddings, i.e. even-ball and hollow-ball solutions respectively, which contain multiple angular velocities, can be obtained in (A)dS. Nonetheless, since the process is relatively straightforward and the solutions are rather cumbersome, we omit the details. The dS solutions are obtained by $L\rightarrow iL$ leading to slightly different bounds but with the same horizon topology.

\paragraph{Extremal limit in AdS.} In section \ref{extremalboundary}, we argued that these two geometries are examples of blackfolds that reach extremality locally. Let us start with the disk geometry and consider $r=r_{max}$. Then one can show that the radius $r$ at which $|k|= 0$, is larger than $r_{max}$, implying that the brane hits the boundary before the velocity becomes lightlike. Furthermore, we have $\Phi{(r_{max})}=\sqrt{N}$ at the boundary, implying that the charge parameter is diverging while the velocity remains finite. Thus, we conclude that the blackfold is locally extremal on the boundary. For the string charge case, the whole procedure can be repeated with $r=r_{min}$, where the blackfold is locally extremal. On the other hand, the $r=r_{max}$ boundary is a case in which the fluid becomes lightlike and therefore this phenomenon is not present on this boundary. 

If we want to study how the extremal limit is reached over the whole worldvolume then we see that this implies $\Phi_{H}\rightarrow\sqrt{N}$ as $\alpha\rightarrow\infty$. In the case of the disk we have that, $r_{max}\sqrt{\Omega^{2}-1/L^{2}}\rightarrow0$ leaving us with two options. The first option is $r_{max}\rightarrow0$, which implies that the disk disappears. More precisely, this is an artefact of the fact that this limit is not within the validity of blackfold approach and hence this solution is excluded. The second option is to keep the radius fixed and $\Omega^{2}\rightarrow1/L^{2}$, which on increasing the AdS radius, slows down the rotation of the blackfold. In the flat limit $L\rightarrow\infty\implies\Omega\rightarrow0$, which very neatly leads to the analysis done in \cite{Caldarelli:2010xz}. In the case of the annulus, we have already argued that $r_{max}-r_{min}\gg r_{c}$, which translates into \begin{equation}\label{phiannulus}
    \Phi_{H}\leq \frac{2\pi\Omega\sqrt{N}}{\Omega^{2}-1/L^{2}}\ ,
\end{equation} 
where, the bound is saturated in the extremal limit. When the bound is saturated, this implies that $r_{min}\rightarrow r_{max}$, i.e. the annulus becomes infinitely thin. As a result, the null fluid velocity regime is attained. Furthermore, we notice that the thermodynamic quantities go smoothly to zero if we keep $T$ finite, but if we simultaneously also require that $T\rightarrow 0$, then the thermodynamic quantities remain finite $\forall\  n\geq 1 \land N\geq 1$, except for the entropy which diverges for $N>1$.

\paragraph{Extremal limit in dS.} We now consider extremal blackfold solutions in dS spacetimes. In the case of the disk solution wee see that in the extremal limit $r_{max}\sqrt{\Omega^{2}+1/L^{2}}\rightarrow0$, which can only occur if $r_{max}\rightarrow0$, which is outside the blackfold regime. Hence there is no such extremal solution in dS. In the case of the annulus, one can show that the bound imposed is the Wick rotated version of \eqref{phiannulus}. It is important to note that when finding this bound we exclude a new upper bound since this bound is larger than the upper bound already set by $r_{max}$. The analysis done in this case is identical to the one done for the AdS one.

\paragraph{Static solutions in dS.} Finally, we consider the cases in which the solutions are static ($\Omega=0$). In the disk solution, we see that $r_{max}=L\sqrt{1-\Phi_{H}^{2}/N}$ and, differently from the uncharged case \cite{Armas:2010hz}, the horizon radius does not reach the cosmological one at $r_{max}=L$. In the case of the annulus solution, the new upper bound which was previously excluded is now considered, since it is smaller than $r_{max}(\Omega=0)$. Thus, we have \begin{equation}
    r_{min}=\frac{L}{\sqrt{2}}\sqrt{1-\sqrt{1-\frac{\Phi_H^2}{NL^2 \pi^2}}}\leq r \leq r'_{max}=\frac{L}{\sqrt{2}}\sqrt{1+\sqrt{1-\frac{\Phi_H^2}{NL^2 \pi^2}}}\  ,
\end{equation} where one can see that the maximum value of $r'_{max}$ is recovered for the uncharged case where $r_{max}=r'_{max}=L$. Hence, we conclude that just like for the disk geometry, the horizon radius of the annulus solution does not reach the cosmological horizon.  

\section{Type IIB supergravity and M-theory}\label{TypeII/M-theory} \label{sec:supergravity}
In this section we are interested in obtaining new black hole solutions in type IIB supergravity and M-theory that asymptote to AdS$_{l}\times S^{m}$ with $(l,m)=\{(4,7),(5,5),(7,4)\}$\footnote{Note that the dimensions of the background geometry is related to the dimensions of the blackfold via $p=l-2$ and $n=m-1$, such that $D=p+n+3=l+m$.}, with nontrivial $F_{l}$ flux and a constant dilaton as in \eqref{eq:AdSS5metric}-\eqref{gaugefields}. Given this background, we focus in particular in obtaining solutions with $q=1,\ p,\ (1,p)$-brane charges and with odd-sphere and product of odd-spheres horizon topologies. These solutions have interesting extremal limits that we discuss. At the end we also consider solutions carrying multiple higher-form charges via long-wavelength deformations of black brane bound states. We discuss the thermodynamic properties, stress tensor, and extremal limits for such bound states below.

\subsection{Multi-charged blackfold effective fluids}\label{Blackfoldswithmultipleq-branecurrents}
Here we give the necessary details for describing multi-charged fluids. This is a generalisation of the analysis done in section \ref{Blackfoldeffectivefluids} and based on \cite{Emparan:2011hg}. It can be shown that for certain classes of type II/M-theory the stress energy tensor of a black brane with multiple currents corresponds, at leading order in perturbation theory, to a perfect fluid with multiple currents. This can be written as
\begin{equation}
    T_{ab}=\mathcal{T}su_{a}u_{b}-\mathcal{G}\gamma_{ab}-\sum_{q}^{p}\Phi_{q}\mathcal{Q}_{q}\hat{h}_{ab}^{(q)} \ ,
\end{equation} 
where we used \eqref{defngibbs}. The single-charged blackfold effective fluids describing these black branes are given in \eqref{localthermo} with $N=1$ for D$p$ and M$p$ branes, while for D$0$-D$p$ and F1-D$p$, we refer the reader to \cite{Emparan:2011hg}. It is possible to verify that, apart from the local thermodynamic relations, the energy density and Gibbs free energy density obey the following relations
\begin{equation}\label{stressenergymult}
    \varepsilon = \frac{n+1}{n}\mathcal{T}s + \sum_{q}\Phi_{q}\mathcal{Q}_{q}\ \implies\ \mathcal{G}=\frac{1}{n}\mathcal{T}s\ .
\end{equation} 
By exploiting this identity we can re-write the stress-energy tensor as 
\begin{equation} \label{stressimportantq}
    T_{ab}=\mathcal{T}s\Big(u_{a}u_{b}+\frac{1}{n}\gamma_{ab}\Big) - \sum_{q}\Phi_{q}\mathcal{Q}_{q}\hat{h}_{ab}^{(q)}\ ,
\end{equation} 
which is a generalisation of \eqref{stressimportant}. We can further use these relations to obtain the Smarr relation 
\begin{equation}\label{smarrrelation2}
    (D-3)M-(D-2)(TS+\Omega J)-\sum_{q}^{p}(D-q-3)\Phi_{H}^{(q)}Q_{q}=\mathcal{T}_{tot}\ ,
\end{equation} 
where $\mathcal{T}_{tot}$ is the total tension, or binding energy, given by \begin{equation}\label{defntott2}
    \mathcal{T}_{tot} = -\int_{\mathcal{B}_{p}}dV_{p}\Big(R_{0}T+ (D-2)\mathcal{L}_{p}+(T^{\mu\nu}+\mathcal{V}^{\mu\nu})n_{\mu}\xi_{\nu}|_{x^{\mu}=X^{\mu}} \Big) \  ,
\end{equation} 
where $T\equiv\gamma_{ab}T^{ab}$ and $\mathcal{L}_{p}\equiv\mathcal{Q}_{p}\mathbb{P}[A_{l-1}]$ times potential redshift factors. Note that, as expected, the presence of the gauge field has modified the total tension. Finally, we can use \eqref{stressimportantq} to re-write the projected extrinsic equation \eqref{cartereqn} as \begin{equation}\label{extrinsicmoreq}
    \mathcal{T}s\Big(\Dot{u}^{\mu}-\frac{1}{n}K^{\mu}\Big)-\sum_{q}\Phi_{q}\mathcal{Q}_{q}\hat{K}^{\mu}_{(q)}= \mathcal{F}^{\mu} \ ,
\end{equation} 
where $\mathcal{F}^{\mu}=\perp_{\nu}^{\ \mu}F^{\nu a_{1}...a_{q+1}}\mathcal{J}_{a_{1}...a_{q+1}}/(q+1)!$ is the projected force term. Note that the mean extrinsic curvature has the following properties \begin{equation}\label{propertiesK}
    \hat{K}_{(0)}^{\mu}=-\perp^{\mu}_{\ \nu}\dot{u}^{\nu}\ , \ \ \ \hat{K}_{(q)}^{\mu}=-\perp^{\mu}_{\ \nu}(\dot{u}^{\nu}-\dot{v}^{\nu})\ , \ \ \ \hat{K}^{\mu}_{(p)}=K^{\mu}\ ,
\end{equation} which will be useful to solve the extrinsic equation. Equation \eqref{extrinsicmoreq} is valid for geometries in the presence of fluxes. If the latter vanishes then the equations simplify by setting the force term to zero. 

\subsection{Extremal limits}
In this section we consider the extremal limits of blackfolds carrying multiple currents, one of which is necessarily the one carrying the top charge $\mathcal{Q}_{p}$. As seen in eq.~\eqref{actionandthermodynamics}, the gauge field plays an active role in describing the action of the non-extremal black brane and we expect this to be true also in the extremal case. In fact, in the case of string theory one can show that the non-extremal action reduces to the DBI action, where the gauge field is present (as already shown in various occasions \cite{Armas:2016mes, Armas:2012bk,Armas:2013ota}). These extremal limits are a direct generalisation of section \ref{extremallimits1} and are obtained by setting $\alpha\rightarrow\infty$, $r_{0}\rightarrow0$ and keeping $Q_{q}$ fixed.

\subsubsection*{Timelike fluid velocity}\label{subluminalvelflux}
We start with the simple case of a blackfold containing just $q=p$. In the extremal limit the action \eqref{actionforbrane} reduces to 
\begin{equation}
    I = -\int_{\mathcal{W}_{p+1}}\mathcal{Q}_{p}(\star_{(p+1)}\textbf{1}-\mathbb{P}[A_{l-1}])\ ,
\end{equation} 
where we used the fact that in this limit $P=-\mathcal{Q}_{p}$ as can be seen from \eqref{localthermo}. Note that since $\mathcal{Q}_{p}$ is a constant, this is indeed the DBI action in the presence of a gauge field if we identify the charge as the tension of the brane. If we now specialise to the case where the quantities are constant along the worldvolume and the geometry has a constant radius $R$, then we can integrate the action and together with \eqref{thermopot}, we find that \begin{equation}\label{moja}
    M = \Omega J + \Phi_{H}^{(p)}Q_{p}(1-\Tilde{A})\ , \ \ \ \Tilde{A}\equiv\frac{A_{\sigma^{0}...\sigma^{p}}}{R_{0}R^{p}}\ .
\end{equation} 
Here we have used the fact that in the extremal limit $R_{0}V_{p}Q_{p}=\Phi^{(p)}_{H}Q_{p}$. Furthermore, using the Smarr relation, we split the equation above into two to relate the mass and the angular momentum to the total tension as \begin{align}\label{mqpjback}
    M &= \Phi_{H}Q_{p}((n+p+1)\Tilde{A}+p+1)-\mathcal{T}_{tot}, \\
    \Omega J &= \Phi_{H}Q_{p}((p+n)\Tilde{A}+p)-\mathcal{T}_{tot}. 
\end{align} The relations above are the generalisation of \eqref{relationsp} and \eqref{mqpj} for blackfolds in the presence of background fields. 

These results can be extended to the case in which the brane contains more than one current smeared on it. Nonetheless, we are still in the regime where the asymptotic background contains only the $F_{l}$ flux. Thus, these ``extra" brane currents will not appear as pullbacks in the action. For this reason, the generalisation is simply given by \begin{equation}\label{mpqojqcurrentback}
    M = \Omega J + \sum_{q}^{p}\Phi_{H}^{(q)}Q_{q}- \Phi_{H}^{(p)}Q_{p}\Tilde{A}\ .
\end{equation} 
Similarly to above, it is possible to split the above equation into two relating the thermodynamic quantities to the total tension. As we will see later, these formulas will be explicitly reproduced once the global thermodynamics have been found both for the cases with a flux being present or not. In fact, for the latter case one has to simply set $\tilde{A}=0$ and obtain the generalisation of the analysis done in section \ref{extremallimits1}.

Finally, we would like to remark that while these equations are very similar in structure to the BPS bounds, the charges are dipole in nature and therefore are not the conserved charges entering the BPS relations. These equations reduce to the BPS relations in the case of only a $q=0$ charge being present in the absence of the gauge field. 

\subsubsection*{Null fluid velocity}\label{nullwavebraneflux}
In the case in which the fluid velocity approaches the speed of light, we re-scale an appropriate combination of the velocity field and thermodynamic quantities as to obtain finite results. 
The external gauge field is not affected by this re-scaling, implying that the results from section \ref{extremallimits1} follow in this case as well. In particular, adopting the same re-scaling \eqref{rescaling} and given the stress-energy tensor \eqref{stressenergymult}, we have that 
\begin{equation}
    T_{ab}=\mathcal{K}l_{a}l_{b}-\sum_{q}\Phi_{q}\mathcal{Q}_{q}\hat{h}_{ab}^{(q)}\ .
\end{equation} 
Following the same footsteps as in section \ref{nullwavebrane}, but this time keeping tack of $\mathcal{V}^{\mu\nu}$ and assuming that it is constant on the worldvolume we can see that  \begin{align} \label{intthermonullflux}
     \mathcal{T}_{tot} &= -\int_{\mathcal{B}_{p}}dV_{p}R_{0}(\mathcal{K}-\sum_{q}q\Phi_{q}\mathcal{Q}_{q}+\tilde{\mathcal{V}}_{\mathcal{T}})\ , \\
      M &= \int_{\mathcal{B}_{p}}dV_{p}R_{0}(\mathcal{K}+\sum_{q}\Phi_{q}\mathcal{Q}_{q}+\tilde{\mathcal{V}}_{M})\ , \\
      J &= \int_{\mathcal{B}_{p}}dV_{p}(\mathcal{RK}+\tilde{\mathcal{V}}_{J})\ ,
\end{align} 
where $\tilde{\mathcal{V}}_{M}=\mathcal{V}^{\mu\nu}n_{\mu}\xi_{\nu}/R_{0}$ and similarly for $\tilde{\mathcal{V}}_{J}$ but $\chi_{\nu}$ replaces $\xi_{\nu}$ and $\tilde{\mathcal{V}}_{\mathcal{T}}=\tilde{\mathcal{V}}_{M}+(D-2)\mathcal{L}_{p}$. Assuming the case in which the integrands are constants, we are led to
\begin{equation}\label{mjqnullflux}
    M = V_{p}R_{0}\Big(\tilde{\mathcal{V}}_{M}-\frac{\tilde{\mathcal{V}}_{J}}{\mathcal{R}}+\sum_{q}^{p}\Phi_{q}\mathcal{Q}_{q}\Big)+R_{0}\frac{J}{\mathcal{R}} \ .
\end{equation}  The equations shown here are the generalised version of the equations presented in section \ref{nullwavebrane} to include the case of multiple charges and a background gauge field. 

\subsection{Odd-sphere solutions}\label{oddspheresflux}
We now look for odd-sphere solutions as in section~\eqref{oddspheres} but now in Type II/M-theory that asymptote to the spacetime \eqref{eq:AdSS5metric}. There are two distinct ways to embed odd-sphere solutions of probe black $p$-branes: the first is to embedded the sphere in the AdS$_{l}$ part of the spacetime and the second is to embedded it in the $S^{m}$ part. The solutions presented are derived using the general form of the stress tensor and currents given in section \ref{Blackfoldswithmultipleq-branecurrents} for fluids carrying multiple charges. In section \ref{examplesmtypeii} we will be studying in further detail specific examples of such fluids arising in supergravity with their associated local thermodynamic quantities.

\paragraph{Embedding in AdS$_{l}$.} Starting with the embedding in AdS$_{l}$, we have that $p=l-2=2k+1$. We can write the unit sphere in terms of $k$ direction cosines $\mu_{i}$ and $k+1$ Cartan angles $\theta_{i}$. Then the embedding used is 
\begin{equation}
    \sigma^{0}=t = \tau\ , \ \ \ r=R\ , \ \ \ \sigma_{i}=\theta_{i}\ ,\ \ \ \sigma_{j}=\mu_{i}\ , \ \ \ \zeta =\rho_{i}=0\ ,
\end{equation} 
with $i=(1,...,k+1)$ and $j=(k+2,...,p+1)$, implying that the metric on the black brane is identical to \eqref{embeddsingle}. Setting the blackfold to rotate along all $\theta_{i}$ angles with the same velocity $\Omega$, we see that the vectors $k^{a}\partial_{a}$ and $\zeta^{a}\partial_{a}$ (for $q=1$) are given by \eqref{ksingle} and \eqref{zetasingle}, respectively. Hence, after finding the fluid velocity and the spacelike vector along which the current lies, we solve the extrinsic equations \eqref{extrinsicmoreq}. By using \eqref{propertiesK}, we see that the following non-zero components 
\begin{align}\label{dotukf}
    \dot{u}^{r} &= \cosh^{2}{\eta}\Big(\frac{R}{\Tilde{L}^{2}}-\Omega^{2}R\Big)\ , \ \ \ \dot{v}^{r}=\frac{R}{\tilde{L}^{2}}\sinh^{2}{\eta}-\frac{f(R)}{R}\cosh^{2}{\eta} \ , \\ K^{r}&=-\Big(\frac{R}{\Tilde{L}^{2}}+\frac{p}{R}f(R)\Big)\ , \ \ \ \mathcal{F}_{p+2}^{r}= \frac{l-1}{\Tilde{L}}\sqrt{f(R)}\mathcal{Q}_{p}\ ,
\end{align} 
are the required elements to solve the extrinsic equations. We can group the equilibrium condition for the $1,\ p,\ (1,p)$-cases into one equation 
\begin{equation} \label{omegaflux}
   \Omega^{2} =\frac{f(R)}{R^{2}} \frac{p(\mathcal{T}s+n\Phi_{p}\mathcal{Q}_{p})+n\Phi_{1}\mathcal{Q}_{1}+\frac{R^{2}}{\Tilde{L}^{2}}\Xi_{1,p}-\frac{Rn(l-1)}{\Tilde{L}}\sqrt{f(R)}\mathcal{Q}_{p}}{(n+p)\mathcal{T}s+pn\Phi_{p}\mathcal{Q}_{p}+n\Phi_{1}\mathcal{Q}_{1}+\frac{R^{2}}{\Tilde{L}^{2}}\Xi_{1,p}-\frac{Rn(l-1)}{\Tilde{L}}\sqrt{f(R)}\mathcal{Q}_{p}}\ , 
\end{equation} 
where $\Xi_{1,p}\equiv\mathcal{T}s(1+n+p)+n(1+p)\Phi_{p}\mathcal{Q}_{p}+2n\Phi_{1}\mathcal{Q}_{1}$. The equilibrium solution \eqref{omegaflux} is valid for the cases of branes carrying $q=1,\ p$ charge as well as for the bound states carrying $(1,p)$-brane charge.\footnote{In the case $q=p$ these solutions extend the $\Omega=0$ solutions of \cite{Armas:2013ota} beyond the near-extremal limit.} By comparing the result with the previous sections, we see that the background flux have a non-trivial effect on the equilibrium condition, as expected. Given this solution we need to impose the requirement that $\Omega^{2}\geq0$. In order to do so, one has to consider the different black brane systems separately with their respective local thermodynamic quantities, which we detail in section~\ref{examplesmtypeii}. The result is that $\Omega>0\ \forall\ R>0$, and there exists no $R>0$ such that $\Omega=0$.   

Having confirmed that the solution is a valid one, we can now study the global thermodynamic quantities. By taking look at  \eqref{fluxthermo}, \eqref{fluxentropy}, \eqref{globalchargedfn}, \eqref{globalphidfn} and \eqref{defntott2}, we see that only $M$ and $\mathcal{T}_{tot}$ change in the presence of the background flux, while the rest of the thermodynamic quantities will have the same structure as in \eqref{globalthermoq=p} and \eqref{globaltermophiq}. In particular, since $\mathcal{V}^{\mu\nu}n_{\mu}\chi_{\nu}|_{x^{\mu}=X^{\mu}}=0$, the angular momentum $J$ has the same structure as in \eqref{globalthermoq=p}. On the other hand \begin{align}\label{mjsback}
    M &= V_{p}\Big(R_{0}(\cosh^{2}{\eta}(\mathcal{T}s+ \delta_{\bar{q},0}\Phi_{\bar{q}}\mathcal{Q}_{\bar{q}})+\frac{\mathcal{T}s}{n} + \Phi_{p}Q_{p}+(1-\delta_{\bar{q},0})\Phi_{\bar{q}}\mathcal{Q}_{\bar{q}})-\frac{R\mathcal{Q}_{p}}{\Tilde{L}}\Big)\ , \\ 
    \mathcal{T}_{tot}&= - \frac{V_{p}}{n}\Big(R_{0}\frac{R^{2}}{L^{2}}\Xi_{\bar{q},p}+n\mathcal{Q}_{p}\frac{R}{L}(R_{0}^{2}(l-1)+1) \Big)\ .
\end{align} Once more, we see that the total tension is different than zero and proportional to $1/L$ as in the previous cases. Furthermore, the global thermodynamic quantities obey the Smarr relation \eqref{smarrrelation2}.

\paragraph{Embedding in $S^{m}$.}
Let us now embed the odd-sphere in the $S^{m}$ part of the spacetime. For $p=m-2=2k+1$, we can write the unit sphere in terms of $k$ direction cosines $\mu_{i}$ and $k+1$ Cartan angles $\phi_{i}$. The embedding is given by 
\begin{equation}
    \sigma^{0}=t = \tau\ , \ \ \ r=\theta_{i}=0\ , \ \ \ \sigma_{i}=\rho_{i}\ ,\ \ \ \sigma_{j}=\mu_{i}\ ,\ \ \  \zeta=\zeta_{0}\ , 
\end{equation}
with $i=(1,...,k+1),\ j=(k+2,...,p+1)$ and $\zeta_{0}$ being a constant, implying that metric on the black brane is 
\begin{equation}
    \gamma_{ab}d\sigma^{a}d\sigma^{b} = -d\tau^{2} + R^{2}d\Omega_{m-2}~~,
\end{equation} 
where $R^{2}=L^{2}\sin\zeta_{0}$. We immediately notice that given our embedding, the pullback of the flux form $F^{\mu}_{\ a_{1}...a_{m}}=0$, implying that the force term is zero. This in turn implies that the extrinsic \eqref{extrinsicmoreq} has a vanishing right-hand-side. Furthermore, we notice that 
\begin{align}\label{zeroforce}
    \mathcal{V}^{\mu\nu}n_{\mu}\xi_{\nu}&=0 \ \ \ \text{since } A^{t}\cdot \xi_{t}=0\ , \\
    \mathcal{V}^{\mu\nu}n_{\mu}\chi_{\nu}&=0 \ \ \ \text{since } A^{\chi_{i}}\cdot J^{t}=0\ ,
\end{align} 
implying that the thermodynamics are not affected by the background flux, which should not come as a surprise since the force term does not interfere with the dynamics of the blackfold. Altogether this implies that the solutions are remarkably simple and are identical to the ones obtained in the flat spacetime case in \cite{Emparan:2011hg} since, apart from the flux not playing a role, the gravitational redshift is trivial, that is $R_{0}\equiv\sqrt{-\xi^{2}}|_{\mathcal{W}_{p+1}}=1$. In fact, one can show (setting the blackfold to rotate over all angles with the same angular velocity) that the equilibrium condition is given by 
\begin{equation}\label{omegaoddone}
    \Omega^{2} =\frac{1}{R^{2}} \frac{p(\mathcal{T}s+n\Phi_{p}\mathcal{Q}_{p})+n\Phi_{1}\mathcal{Q}_{1}}{(n+p)\mathcal{T}s+pn\Phi_{p}\mathcal{Q}_{p}+n\Phi_{1}\mathcal{Q}_{1}}\ , 
\end{equation} 
and we will restrain from showing the physical quantities as they can be read off from the flat case.\footnote{In the case $q=p$ these solutions coincide with the $\Omega=0$ solutions of \cite{Armas:2013ota}.} In fact, from the point of view of the effective fluid, the embedding at the origin of AdS ($r=0$) is interpreted as an one in Minkowski in spherical coordinates, with ``no communication" with the AdS$_{l}$ part of the spacetime. Furthermore, the fluid obeys ``free" dynamics, in the sense that the force term does not play a role and therefore both the equilibrium condition and the thermodynamics do not change compared with the flat space case, in contrast with the thermal and spinning giant graviton solutions of \cite{Armas:2012bk,Armas:2013ota}. 

\paragraph{Extremal limits.} In the case of blackfolds embedded in the $S^{m}$ part of the spacetime, we refer the reader to the flat cases \cite{Caldarelli:2010xz,Emparan:2011hg}. Moving on to blackfolds embedded in the AdS$_l$ part of the spacetime, we have that $\eta\rightarrow\infty$, implying an extremal limit with null fluid velocity. In this limit, we can define $\mathcal{K}=\mathcal{T}s\sinh{\eta}\cosh{\eta}$ and explicitly show that the thermodynamic quantities are given by the integrated versions of \eqref{intthermonullflux}. Furthermore, defining the null vector as \eqref{nullvectorodd} such that $l^{a}\chi_{a}=R$,  one can show that \eqref{mjqnullflux} is indeed obeyed. Unfortunately, in the case of multiple charges being present, there is no simple equation relating all thermodynamic quantities and fluxes, even in the case of a branes with just $p$-brane charges. 

\subsection{Product of odd-spheres}\label{prododdsphereflux}
In this section we look for solutions whose horizons are products of odd-sphere geometries and asymptote to \eqref{eq:AdSS5metric}. The product of odd-spheres considered here is slightly different than the ones considered in section~\ref{productofoddspheres}. In particular, in section~\ref{productofoddspheres} we embedded the product of odd-spheres in a subspace of AdS$_{D}$, while here we embed it partly in AdS$_{l}$ and partly in $S^{m}$. As a result, one of the main differences is the fact that we cannot set $b=1$ to recover the solutions of a single odd-sphere. Furthermore, when dealing with the product of odd-spheres, we require that $p=\sum_{a}^{b}p_{a}$ with $p_{a}=2k_{a}+1$ and given that $(l,m)$ are fixed, we have a limited number of possibilities to consider. Here we will consider $\prod_{a}^{3}\mathbb{S}^{p_{a}}$ and $\prod_{a}^{2}\mathbb{S}^{p_{a}}$ horizon topologies, leaving higher products to future work.

Interestingly, the topology of these configurations is such that they will not couple to the background fluxes. This is due to one of the following two reasons. First, the embedding is such that the pullback of the flux form is zero and likewise for $\mathcal{V}^{\mu\nu}$, as already observed in \eqref{zeroforce}. Second, recall that the current carried by the blackfolds is a $(p+1)$-form, while the background contains an $F_{l}$ field strength. It so happens that for most of these embeddings $(p+1)\geq l$ implying that $\mathcal{F}^{\mu}=F^{\mu}\cdot \mathcal{J}=0$. The same reasoning can be used to show that $\mathcal{V}^{\mu\nu}=0$. This implies that the thermodynamic quantities are simple and we give explicit formulae at the end of this section.

\subsubsection*{$\prod_{a}^{3}\mathbb{S}^{p_{a}}$ horizon topologies} Considering the background \eqref{eq:AdSS5metric}, we are restricted to $\mathbb{S}^{1}\times \mathbb{S}^{1}\times \mathbb{S}^{3}$ and $\mathbb{S}^{1}\times \mathbb{S}^{1}\times \mathbb{S}^{1}$. There exists different scenarios that can lead to geometries with such topologies. The first scenario is to embed the $\mathbb{S}^{1}\times \mathbb{S}^{1}$ pair in the $S^m$ part of the spacetime and the other sphere in the AdS$_l$ spacetime. The second scenario is to do the opposite, i.e. $\mathbb{S}^{1}\times \mathbb{S}^{1}$ in the  AdS$_l$ part of the spacetime and the other sphere in $S^{m}$. The third and fourth scenarios are case specific to $\mathbb{S}^{1}\times \mathbb{S}^{1}\times \mathbb{S}^{3}$, where for the former case we embed the $\mathbb{S}^{1}\times \mathbb{S}^{3}$ pair in the $S^m$ part of the spacetime and the $\mathbb{S}^{1}$ in the AdS$_l$ part of the spacetime; whilst for the latter we do the opposite. In order to keep the notation between all cases similar and easily comparable, it is useful (but not strictly necessary for all cases) to parametrise the metric of the unit sphere $d\Omega_{m-2}^{2}\subset ds^{2}_{S^{m}}$ in \eqref{eq:AdSS5metric} as \begin{equation}\label{parasphere}
    d\Omega_{m-2}^{2}=d\delta^{2}_{1}+\cos^{2}\delta_{1} d\delta_{2}^{2}+\sin^{2}\delta_{1} d\Omega_{m-4}^{2}\ .
\end{equation} 
\paragraph{$\mathbb{S}^{1}\times \mathbb{S}^{1}\times \mathbb{S}^{p_a}$ with $p_a=1,3$ and $\mathbb{S}^{1}\times \mathbb{S}^{1}$ embedded into $S^m$.}
We will start with the first scenario and consider the type IIB background, where $d\Omega_{m-4}=d\rho_{1}$. In the case of $\mathbb{S}^{1}\times \mathbb{S}^{1}\times \mathbb{S}^{1}$ we use the embedding \begin{equation}
    \sigma^{0}=t = \tau\ , \ r=R_{1}\ , \ \sigma_{1}=\theta_{1}\ ,\ \theta_{2,3}=\pi/2\ , \ \zeta = \zeta_{0}\ ,  \ \sigma_{2}=\phi\ , \ \delta_{1}=\delta_{0}, \ \sigma_{3}=\delta_{2}\ ,\ \rho_{1}=0\ ,
\end{equation} where, the coordinate $\theta_{1}$ has the Killing vector $\partial_{\theta_{1}}$ associated to it and $\delta_{0}$ is a constant; while for  $\mathbb{S}^{3}\times \mathbb{S}^{1}\times \mathbb{S}^{1}$, we use \begin{equation}\label{embeds3s1s1}
   \sigma^{0}=t = \tau\ ,\ r=R_{1}\ , \ \sigma_{i}=\theta_{i}\ ,\ \sigma_{j}=\mu_{i}\ , \ \zeta = \zeta_{0}\ , \ \sigma_{5}=\phi\ ,\ \delta_{1}=\delta_{0}, \   \sigma_{6}=\delta_{2}\ ,\ \rho_{1}=0\ ,
\end{equation} with $i=(1,2)$, $j=(3,4)$. In the case of M-theory, we will consider the $\mathbb{S}^{3}\times \mathbb{S}^{1}\times \mathbb{S}^{1}$ horizon topology in AdS$_{7}\times S^{4}$, where $d\Omega_{m-4}=0$. Furthermore, we parametrise the metric of the unit sphere $d\Omega_{5}^{2}\subset ds^{2}_{AdS_{7}}$ in \eqref{eq:AdSS5metric} as \begin{equation}\label{parasphereads}
    d\Omega_{5}^{2}=d\omega_{1}^{2}+\cos^{2}\omega_{1} d\omega_{2}^{2}+\sin^{2}\omega_{1} d\Omega_{3}^{2}\ .
\end{equation} The embedding in this case is provided by \eqref{embeds3s1s1} with the addition of $\omega_{1}=\pi/2$. By letting $p_{a}=\{1,3\}$, the induced metric can be written for all three cases as \begin{equation}
     \gamma_{ab}d\sigma^{a}d\sigma^{b} = -f(R_{1})d\tau^{2}+R_{1}^{2}d\Omega_{p_{a}}^{2} + R_{2}^{2}d\phi^{2}+R_{3}^{2}d\delta_{2}^{2}\ ,
\end{equation} where $R_{2}=L\cos\zeta_{0}$ and $R_{3}=L\sin\zeta_{0}\cos\delta_{0}$. Setting the blackfold to rotate along all angles with equal angular velocity $\Omega_1$ in the $\mathbb{S}^{p_{a}}$ directions and two different ones $(\Omega_2,\Omega_3)$ along the $\mathbb{S}^{1}$ directions, we obtain \begin{equation}\label{comps3s1s1}
    k^{a}\partial_{a}=\partial_{\tau}+\Omega_{1}\partial_{\Theta}+\Omega_{2}\partial_{\phi}+\Omega_{3}\partial_{\delta_{2}}\ , \ \ \zeta^{a}\partial_{a}=\frac{1}{|k|^{2}}\Big(\sum_{a=1}^{3}\Omega^{2}_{a}R_{a}^{2}\partial_{\tau}+f(R_{1})(\Omega_{1}\partial_{\Theta}+\Omega_{2}\partial_{\phi}+\Omega_{3}\partial_{\delta_{2}})\Big) \ ,
\end{equation}  where $\partial_{\Theta}=\sum_{i}^{p_{a}}\partial_{\theta_{i}}$. The extrinsic equations \eqref{extrinsicmoreq} provides us with a set of 3 equations, for $\mu=\{r,\zeta,\delta_{1}\}$. The non-zero components are \begin{align}\label{3components}
    \dot{u}^{r} &= \frac{f(R_{1})}{|k|^{2}}\Big(\frac{R_{1}}{L^{2}}-\Omega_{1}^{2}R_{1}\Big)\ , \ \ \ \dot{u}^{\zeta}=\frac{R_{2}}{L^{2}|k|^{2}}\Big(\Omega_{2}^{2}c_{\zeta}+\frac{\Omega_{3}^{2}R^{2}_{3}}{c_{\zeta}}\Big)\ ,\ \ \ \dot{u}^{\delta_{1}}=\frac{\Omega_{3}^{2}R_{3}c_{\delta_{1}}}{|k|^{2}}\ , \\ \dot{v}^{r}&=\frac{1}{|k|^{2}}\Big(\frac{R_{1}}{L^{2}}\sum_{a=1}^{2}\Omega^{2}_{a}R_{a}^{2}-\frac{\Omega_{1}^{2}R_{1}f(R_{1})}{\sum_{a=1}^{2}\Omega^{2}_{a}R_{a}^{2}}\Big)\ , \ \ \ \dot{v}^{\zeta}=\frac{f(R_{1})}{\sum_{a=1}^{3}\Omega^{2}_{a}R_{a}^{2}}\dot{u}^{\zeta}\ ,\ \ \ \dot{v}^{\delta_{1}}=\frac{f(R_{1})\dot{u}^{\delta_{1}}}{\sum_{a=1}^{3}\Omega^{2}_{a}R_{a}^{2}}\ , \\ K^{r}&=-\Big(\frac{R_{1}}{L^{2}}+\frac{p_{a}}{R_{1}}f(R_{1})\Big)\ , \ \ \ K^{\zeta}=\frac{c_{\zeta}}{R_{2}}+\frac{R^{2}_{3}}{R_{2}L^{2}c_{\zeta}}\ \ , \ \ \ K^{\delta_{1}}=\frac{c_{\delta_{1}}}{R_{3}}\ ,
\end{align} where $c_{\delta_{1}}=\sqrt{L^{2}-R_{2}^{2}-R_{3}^{2}}/c^{2}_{\zeta}$ and $c_{\zeta}=\sqrt{L^{2}-R_{2}^{2}}$. Solving the set of equations for the three equilibrium conditions for blackfolds carrying a $1,\ p$ or $(1,p)$-brane charges, we have  
\begin{align}\label{s3s1s1omega}
     \Omega_{1}^{2} &=\frac{f(R_{1})}{R_{1}^{2}} \frac{\Big(p_{a}(\mathcal{T}s+n\Phi_{p}\mathcal{Q}_{p})+\frac{R_{1}^{2}}{\tilde{L}^{2}}\Xi_{1,p}\Big)\Big(p(\mathcal{T}s+n\Phi_{p}\mathcal{Q}_{p})+n\Phi_{1}\mathcal{Q}_{1}+\frac{R_{1}^{2}}{\tilde{L}^{2}}\Xi_{1,p}\Big)}{\Big(p(\mathcal{T}s+n\Phi_{p}\mathcal{Q}_{p})+\frac{R_{1}^{2}}{\tilde{L}^{2}}\Xi_{1,p}\Big)\Big((n+p)\mathcal{T}s+n(p\Phi_{p}\mathcal{Q}_{p}+\Phi_{1}\mathcal{Q}_{1})+\frac{R_{1}^{2}}{\tilde{L}^{2}}\Xi_{1,p}\Big)} \ , \\
    \Omega_{a}^{2} &=\frac{f(R_{1})}{R_{a}^{2}} \frac{\Big(\mathcal{T}s+n\Phi_{p}\mathcal{Q}_{p}\Big)\Big(p(\mathcal{T}s+n\Phi_{p}\mathcal{Q}_{p})+n\Phi_{1}\mathcal{Q}_{1}+\frac{R_{1}^{2}}{\tilde{L}^{2}}\Xi_{1,p}\Big)}{\Big(p(\mathcal{T}s+\Phi_{p}\mathcal{Q}_{p})+\frac{R_{1}^{2}}{\tilde{L}^{2}}\Xi_{1,p}\Big)\Big((n+p)\mathcal{T}s+p\Phi_{p}\mathcal{Q}_{p}+\Phi_{1}\mathcal{Q}_{1}+\frac{R_{1}^{2}}{\tilde{L}^{2}}\Xi_{1,p}\Big)} \ ,
\end{align} 
for $a=2,3$. Note that the expressions for $\Omega^{2}_{2}$ and $\Omega^{2}_{3}$ differ due to the radius $R_a$ appearing in the denominator. 

\paragraph{$\mathbb{S}^{1}\times \mathbb{S}^{1}\times \mathbb{S}^{p_a}$ with $p_a=1,3$ and $\mathbb{S}^{p_a}$ embedded into $S^m$.} In the second scenario, we start by parameterising the metric of the unit sphere $d\Omega_{l-2}^{2}\subset ds^{2}_{AdS_{l}}$ in \eqref{eq:AdSS5metric}  as in \eqref{parasphereads}.  In the case of M-theory, there exists no embedding such that the $\mathbb{S}^{1}\times \mathbb{S}^{1}\times \mathbb{S}^{3}$ horizon topology with this geometry is attainable for the stationary case. Thus, considering type IIB, where $d\Omega_{l-4}=d\theta$, the embedding for $\mathbb{S}^{1}\times \mathbb{S}^{1}\times \mathbb{S}^{1}$ is given by \begin{equation}
   \sigma^{0}=t = \tau\ , \ r=R\ ,\ \omega_{1}=\omega_{0}\ , \ \sigma_{1}=\omega_{2}\ , \ \sigma_{2}=\theta\ , \  \zeta = \zeta_{0}\ , \ \phi=\phi_{0}\ ,\  \sigma_{3}=\rho_{1}\ , \ \rho_{2,3}=\pi/2\ ,
\end{equation} where, the coordinate $\rho_{1}$ has the Killing vector $\partial_{\rho_{1}}$ associated to it and $\omega_{0}$ is a constant; while for the $\mathbb{S}^{1}\times \mathbb{S}^{1}\times \mathbb{S}^{3}$ case \begin{equation}\label{embeds1s1s3}
    \sigma^{0}=t = \tau\ , \  r=R\ ,\  \omega_{1}=\omega_{0}\ , \ \sigma_{1}=\omega_{2}\ , \ \sigma_{2}=\theta\ , \ \zeta = \zeta_{0}\ , \ \phi=\phi_{0}\ ,\ \sigma_{i}=\rho_{i}\ ,\ \sigma_{j}=\mu_{i}\ ,
\end{equation} with $i=(3,4)$, $j=(5,6)$. As usual, by letting $p_{a}=\{1,3\}$, we have that
\begin{equation}\label{embaddings1s1s3}
     \gamma_{ab}d\sigma^{a}d\sigma^{b} = -f(R)d\tau^{2}+ R_{1}^{2}d\theta^{2}+R_{2}^{2}d\omega^{2}+R_{3}^{2}d\Omega_{p_{a}}^{2} \ ,
\end{equation} where $R_{1}=R\cos\delta_{0}$, $R_{2}=R\sin\delta_{0}$ and $R_{3}=L\sin\zeta_{0}$. Note that by construction, the radii are linearly dependent according to  $R^{2}=R_{1}^{2}+R_{2}^{2}$, thus removing a degree of freedom. Setting the blackfold to rotate along all angles with equal angular velocities $\Omega_3$ in the $\mathbb{S}^{p_{a}}$ directions and two different ones ($\Omega_1,\Omega_2$) along the $\mathbb{S}^{1}$ directions, we obtain \begin{equation}
    k^{a}\partial_{a}=\partial_{\tau}+\Omega_{1}\partial_{\theta_{1}}+\Omega_{2}\partial_{\omega_{2}}+\Omega_{3}\partial_{\mathcal{P}}\ , \ \ \zeta^{a}\partial_{a}=\frac{1}{|k|^{2}}\Big(\sum_{a=1}^{3}\Omega^{2}_{a}R_{a}^{2}\partial_{\tau}+f(R)(\Omega_{1}\partial_{\omega_{2}}+\Omega_{2}\partial_{\phi}+\Omega_{3}\partial_{\mathcal{P}})\Big) \ ,
\end{equation} where $\partial_\mathcal{P}=\sum_{i}^{p_{a}}\partial_{\rho}$. Once more, the extrinsic equations provide us with three equations for $\mu=\{r,\omega_{1},\zeta\}$. The non-zero components are \begin{align}\label{comps1s1s3}
    \dot{u}^{r} &= \frac{f(R)}{|k|^{2}}\Big(\frac{R}{L^{2}}-\frac{\sum_{a}^{2}\Omega_{a}^{2}R_{a}^{2}}{R}\Big)\ , \ \ \ 
    \dot{u}^{\omega_{1}}=\frac{R_{1}R_{2}}{R^{2}|k|^{2}}(\Omega_{1}^{2}-\Omega_{2}^{2})\ , \ \ \ \dot{u}^{\zeta}= -\frac{\Omega_{3}^{2}R_{3}\sqrt{L^{2}-R_{3}^{2}}}{L^{2}|k|^{2}} \\  
    \dot{v}^{r}&=\frac{1}{|k|^{2}}\Big(\frac{R}{L^{2}}\sum_{a}^{3}\Omega^{2}_{a}R_{a}^{2}-\frac{\sum_{a}^{2}\Omega^{2}_{a}R_{a}f(R)}{\sum_{a}^{3}\Omega^{2}_{a}R_{a}^{2}}\Big)\ , \ \ \ \dot{v}^{\omega_{1}}=\frac{f(R)\dot{u}^{\omega_{1}}}{\sum_{a}^{2}\Omega^{2}_{a}R_{a}^{2}} \ ,\ \ \ \dot{v}^{\zeta}=\frac{f(R)\dot{u}^{\zeta}}{\sum_{a}^{2}\Omega^{2}_{a}R_{a}^{2}} \\
    K^{r}&=-\Big(\frac{R_{1}}{L^{2}}+2\frac{f(R)}{R_{1}}\Big)\ , \ \ \ K^{\omega_{1}}=\frac{1}{R^{2}}\Big(\frac{R_{2}}{R_{1}}-\frac{R_{1}}{R_{2}} \Big)\ , \ \ \ K^{\zeta}=\frac{p_{a}\dot{u}^{\zeta}}{\Omega_{3}^{2}R_{3}^{2}} .
\end{align} 
Solving the set of equations for the three equilibrium conditions for blackfolds carrying a $1,\ p$ or $(1,p)$-brane charges, we have \begin{align}\label{s1s1s3omega}
     \Omega_{a}^{2} &=\frac{f(R)}{R_{a}^{2}} \frac{\Big(\mathcal{T}s+n\Phi_{p}\mathcal{Q}_{p}+\frac{R_{a}^{2}}{\tilde{L}^{2}}\Xi_{1,p}\Big)\Big(p(\mathcal{T}s+n\Phi_{p}\mathcal{Q}_{p})+n\Phi_{1}\mathcal{Q}_{1}+\frac{R^{2}}{\tilde{L}^{2}}\Xi_{1,p}\Big)}{\Big(p(\mathcal{T}s+n\Phi_{p}\mathcal{Q}_{p})+\frac{R^{2}}{\tilde{L}^{2}}\Xi_{1,p}\Big)\Big((n+p)\mathcal{T}s+n(p\Phi_{p}\mathcal{Q}_{p}+\Phi_{1}\mathcal{Q}_{1})+\frac{R^{2}}{\tilde{L}^{2}}\Xi_{1,p}\Big)} \ , \\
    \Omega_{3}^{2} &=\frac{f(R)}{R_{3}^{2}} \frac{\Big(p_{a}(\mathcal{T}s+n\Phi_{p}\mathcal{Q}_{p})\Big)\Big(p(\mathcal{T}s+n\Phi_{p}\mathcal{Q}_{p})+n\Phi_{1}\mathcal{Q}_{1}+\frac{R^{2}}{\tilde{L}^{2}}\Xi_{1,p}\Big)}{\Big(p(\mathcal{T}s+\Phi_{p}\mathcal{Q}_{p})+\frac{R^{2}}{\tilde{L}^{2}}\Xi_{1,p}\Big)\Big((n+p)\mathcal{T}s+p\Phi_{p}\mathcal{Q}_{p}+\Phi_{1}\mathcal{Q}_{1}+\frac{R^{2}}{\tilde{L}^{2}}\Xi_{1,p}\Big)} \ ,
\end{align} 
where $a=1,2$ and we have used the constraint on the radii to simplify the solutions.

\paragraph{$\mathbb{S}^{1}\times \mathbb{S}^{1}\times \mathbb{S}^{3}$ with $\mathbb{S}^{1}\times \mathbb{S}^{3}$ embedded into $S^m$.} In this scenario, $\mathbb{S}^{1}\times \mathbb{S}^{3}$ is embedded in $S^m$ and $\mathbb{S}^{1}$ in AdS$_{l}$. In the case of M-theory we consider AdS$_{4}\times S^{7}$, with $d\Omega^{2}_{5}\subset ds^{2}_{S^{7}}$ parametrised as in \eqref{parasphere}. The embedding is provided by \begin{equation}\label{embeds1s1s3m}
    \sigma^{0}=t = \tau\ , \  r=R_{1}\ , \ \sigma_{1}=\theta_{1}\ , \ \theta_{2}=\pi/2\ , \ \zeta = \zeta_{0}\ , \ \sigma_{2}=\phi\ ,\ \delta_{1}=\delta_{0}\ ,\ \sigma_{i}=\rho_{i}\ ,\ \sigma_{j}=\mu_{i}\ ,
\end{equation} with $i=(3,4)$, $j=(5,6)$,  which leads to the induced metric \begin{equation}
     \gamma_{ab}d\sigma^{a}d\sigma^{b} = -f(R_{1})d\tau^{2}+ R_{1}^{2}d\theta_{1}^{2}+R_{2}^{2}d\phi^{2}+R_{3}^{2}d\Omega_{3}^{2} \ ,
\end{equation} where $R_{2}=L\cos\zeta_{0}$ and $R_{3}=L\sin\zeta_{0}\sin{\delta_{1}}$, Setting the blackfold to rotate along all angles with equal angular velocities $\Omega_3$ in the $\mathbb{S}^{p_{a}}$ ones and two different ones ($\Omega_1,\Omega_2)$ along the $\mathbb{S}^{1}$ directions, we obtain 
\begin{equation}    k^{a}\partial_{a}=\partial_{\tau}+\Omega_{1}\partial_{\theta_{1}}+\Omega_{2}\partial_{\phi}+\Omega_{3}\partial_{\mathcal{P}}\ , \ \ \zeta^{a}\partial_{a}=\frac{1}{|k|^{2}}\Big(\sum_{a=1}^{3}\Omega^{2}_{a}R_{a}^{2}\partial_{\tau}+f(R)(\Omega_{1}\partial_{\theta_{1}}+\Omega_{2}\partial_{\phi}+\Omega_{3}\partial_{\mathcal{P}})\Big) \ .
\end{equation} 
The extrinsic equations provide us with three equations for $\mu=\{r,\zeta,\delta_{1}\}$. The non-zero components are the ones in \eqref{comps3s1s1} with slight modifications: $p_{a}=1$ in $K^{r}$, $K^{\delta_{1}}$ is multiplied by 3 and \begin{equation}
    K^{\zeta}=\frac{c_{\zeta}}{R_{2}}+\frac{3 R_{3}^{2}}{R_{2}L^{2}c_{\zeta}}\ .
\end{equation} Solving the set of equations for the three equilibrium conditions for blackfolds carrying a $1,\ p$ or $(1,p)$-brane charges, we have \begin{align}\label{s1(s3s1)}
     \Omega_{1}^{2} &=\frac{f(R_{1})}{R_{1}^{2}} \frac{\Big(\mathcal{T}s+n\Phi_{p}\mathcal{Q}_{p}+\frac{R_{1}^{2}}{\tilde{L}^{2}}\Xi_{1,p}\Big)\Big(p(\mathcal{T}s+n\Phi_{p}\mathcal{Q}_{p})+n\Phi_{1}\mathcal{Q}_{1}+\frac{R^{2}}{\tilde{L}^{2}}\Xi_{1,p}\Big)}{\Big(p(\mathcal{T}s+n\Phi_{p}\mathcal{Q}_{p})+\frac{R_{1}^{2}}{\tilde{L}^{2}}\Xi_{1,p}\Big)\Big((n+p)\mathcal{T}s+n(p\Phi_{p}\mathcal{Q}_{p}+\Phi_{1}\mathcal{Q}_{1})+\frac{R_{1}^{2}}{\tilde{L}^{2}}\Xi_{1,p}\Big)} \ , \\
     \Omega_{2}^{2} &=\frac{f(R_{1})}{R_{2}^{2}} \frac{\Big((\mathcal{T}s+n\Phi_{p}\mathcal{Q}_{p})\Big)\Big(p(\mathcal{T}s+n\Phi_{p}\mathcal{Q}_{p})+n\Phi_{1}\mathcal{Q}_{1}+\frac{R_{1}^{2}}{\tilde{L}^{2}}\Xi_{1,p}\Big)}{\Big(p(\mathcal{T}s+n\Phi_{p}\mathcal{Q}_{p})+\frac{R_{1}^{2}}{\tilde{L}^{2}}\Xi_{1,p}\Big)\Big((n+p)\mathcal{T}s+n(p\Phi_{p}\mathcal{Q}_{p}+\Phi_{1}\mathcal{Q}_{1})+\frac{R_{1}^{2}}{\tilde{L}^{2}}\Xi_{1,p}\Big)} \ , \\
    \Omega_{3}^{2} &=\frac{f(R_{1})}{R_{3}^{2}} \frac{\Big(3(\mathcal{T}s+n\Phi_{p}\mathcal{Q}_{p})\Big)\Big(p(\mathcal{T}s+n\Phi_{p}\mathcal{Q}_{p})+n\Phi_{1}\mathcal{Q}_{1}+\frac{R_{1}^{2}}{\tilde{L}^{2}}\Xi_{1,p}\Big)}{\Big(p(\mathcal{T}s+\Phi_{p}\mathcal{Q}_{p})+\frac{R_{1}^{2}}{\tilde{L}^{2}}\Xi_{1,p}\Big)\Big((n+p)\mathcal{T}s+p\Phi_{p}\mathcal{Q}_{p}+\Phi_{1}\mathcal{Q}_{1}+\frac{R^{2}}{\tilde{L}^{2}}\Xi_{1,p}\Big)} \ ,
\end{align} which concludes the third scenario. 

\paragraph{$\mathbb{S}^{1}\times \mathbb{S}^{1}\times \mathbb{S}^{3}$ with $\mathbb{S}^{1}$ embedded into $S^m$.}In this fourth scenario, it is easy to see that the only background where such an embedding is attainable is in AdS$_{7}\times S^{4}$, where $d\Omega_{5}^{2}\subset ds^{2}_{AdS_{7}}$ is parametrised as \eqref{parasphereads}. The embedding is provided by \begin{equation}
\begin{split}
    \label{embeds3s1s1m}
   &\sigma^{0}=t = \tau\ , \  r=R_{1}\ ,\ \omega_{1}=\omega_{0}\ ,\ \sigma_{1}=\omega_{2} \ ,\ \sigma_{2,3}=\theta_{2,3}\ ,\ \sigma_{4,5}=\mu_{2,3}\ , \ \zeta = \zeta_{0}\ , \ \sigma_{6}=\phi\ , \\  &\delta_{1}=\pi/2\ ,
   \end{split}
\end{equation} 
which leads to the induced metric \begin{equation}
     \gamma_{ab}d\sigma^{a}d\sigma^{b} = -f(R)d\tau^{2}+R_{1}^{2}d\omega_{2}^{2}+ R_{2}^{2}d\Omega_{3}^{2}+R_{3}^{2}d\phi^{2} \ ,
\end{equation} where $R_{1}=R\cos\omega_{0}$, $R_{2}=R\sin\omega_{0}$ and $R_{3}=L\sin\zeta_{0}$. The radii obey the equation already noted before in the case of \eqref{embeds1s1s3}. Setting the blackfold to rotate along all angles with equal angular velocities $\Omega_2$ in the $\mathbb{S}^{p_{a}}$ ones and two different ones $(\Omega_1,\Omega_3)$ along the $\mathbb{S}^{1}$ directions, we obtain \begin{equation}
    k^{a}\partial_{a}=\partial_{\tau}+\Omega_{1}\partial_{\omega_{2}}+\Omega_{2}\partial_{\Theta}+\Omega_{3}\partial_{\phi}\ , \ \ \zeta^{a}\partial_{a}=\frac{1}{|k|^{2}}\Big(\sum_{a=1}^{3}\Omega^{2}_{a}R_{a}^{2}\partial_{\tau}+f(R)(\Omega_{1}\partial_{\omega_{2}}+\Omega_{2}\partial_{\Theta}+\Omega_{3}\partial_{\phi})\Big) \ ,
\end{equation} Once more, the extrinsic equations provide us with three equations for $\mu=\{r,\omega_{1},\zeta\}$. The non-zero components are the ones in \eqref{comps1s1s3} with slight modifications: the 2 in $K^{r}$ replaced by 4, $p_{a}=1$ in $K^{\zeta}$ and \begin{equation}
    K^{\omega_{1}}=\frac{1}{R^{2}}\Big(\frac{R_{2}}{R_{1}}-3\frac{R_{1}}{R_{2}} \Big)\ .
\end{equation} The 3 equations are solved for the three equilibrium conditions for blackfolds containing $1,\ p,$ or $ (1,p)$-brane charges, such that \begin{align}\label{(s3s1)s1}
     \Omega_{1}^{2} &=\frac{f(R)}{R_{1}^{2}} \frac{\Big(\mathcal{T}s+n\Phi_{p}\mathcal{Q}_{p}+\frac{R_{1}^{2}}{\tilde{L}^{2}}\Xi_{1,p}\Big)\Big(p(\mathcal{T}s+n\Phi_{p}\mathcal{Q}_{p})+n\Phi_{1}\mathcal{Q}_{1}+\frac{R^{2}}{\tilde{L}^{2}}\Xi_{1,p}\Big)}{\Big(p(\mathcal{T}s+n\Phi_{p}\mathcal{Q}_{p})+\frac{R^{2}}{\tilde{L}^{2}}\Xi_{1,p}\Big)\Big((n+p)\mathcal{T}s+n(p\Phi_{p}\mathcal{Q}_{p}+\Phi_{1}\mathcal{Q}_{1})+\frac{R^{2}}{\tilde{L}^{2}}\Xi_{1,p}\Big)} \ , \\
     \Omega_{2}^{2} &=\frac{f(R)}{R_{2}^{2}} \frac{\Big(3(\mathcal{T}s+n\Phi_{p}\mathcal{Q}_{p})+\frac{R_{2}^{2}}{\tilde{L}^{2}}\Xi_{1,p}\Big)\Big(p(\mathcal{T}s+n\Phi_{p}\mathcal{Q}_{p})+n\Phi_{1}\mathcal{Q}_{1}+\frac{R^{2}}{\tilde{L}^{2}}\Xi_{1,p}\Big)}{\Big(p(\mathcal{T}s+n\Phi_{p}\mathcal{Q}_{p})+\frac{R^{2}}{\tilde{L}^{2}}\Xi_{1,p}\Big)\Big((n+p)\mathcal{T}s+n(p\Phi_{p}\mathcal{Q}_{p}+\Phi_{1}\mathcal{Q}_{1})+\frac{R^{2}}{\tilde{L}^{2}}\Xi_{1,p}\Big)} \ , \\
    \Omega_{3}^{2} &=\frac{f(R)}{R_{3}^{2}} \frac{\Big((\mathcal{T}s+n\Phi_{p}\mathcal{Q}_{p})\Big)\Big(p(\mathcal{T}s+n\Phi_{p}\mathcal{Q}_{p})+n\Phi_{1}\mathcal{Q}_{1}+\frac{R^{2}}{\tilde{L}^{2}}\Xi_{1,p}\Big)}{\Big(p(\mathcal{T}s+\Phi_{p}\mathcal{Q}_{p})+\frac{R^{2}}{\tilde{L}^{2}}\Xi_{1,p}\Big)\Big((n+p)\mathcal{T}s+p\Phi_{p}\mathcal{Q}_{p}+\Phi_{1}\mathcal{Q}_{1}+\frac{R^{2}}{\tilde{L}^{2}}\Xi_{1,p}\Big)} \ ,
\end{align} where, we have used the constraint on the radii to simplify the equations. This concludes the analysis for the fourth and all of the possible scenarios.  

We would like to conclude this section by noticing that the above embeddings exhaust the possible $\prod_{a}^{3}\mathbb{S}^{p_{a}}$ horizon topologies and geometries, when consider stationary blackfolds with background \eqref{eq:AdSS5metric}. 

\subsubsection*{$\prod_{a}^{2}\mathbb{S}^{p_{a}}$ horizon topologies} The $\prod_{a}^{2}\mathbb{S}^{p_{a}}$ horizon topologies we are interested in are a subset of the $\prod_{a}^{3}\mathbb{S}^{p_{a}}$ case, obtainable by simply setting some of the embedding maps to zero. For example, in the case of $\mathbb{S}^{1}\times \mathbb{S}^{1}$ being embedded entirely in $S^{m}$, one uses the embedding map \eqref{embeds3s1s1} and simply sets $\sigma_{i}=\sigma_{j}=0$, such that the induced metric is \begin{equation}\label{m2brane}
     \gamma_{ab}d\sigma^{a}d\sigma^{b} = -f(R_{1})d\tau^{2}+ R_{2}^{2}d\phi^{2}+R_{3}^{2}d\delta_{2}^{2}\ ,
\end{equation} where $R_{2}=L\cos\zeta_{0}$. The equilibrium conditions can be simply read off from \eqref{s3s1s1omega} with $\Omega_{1}=0$.

\paragraph{Thermodynamics.} We now provide the thermodynamic quantities for all the classes of solutions found in this section. These are obtained using the formulae \eqref{fluxthermo}, \eqref{fluxentropy}, \eqref{globalchargedfn}, \eqref{globalphidfn} and \eqref{defntott2}. While the $S,\ T,\ Q_{q},\ \Phi_{q}$ can easily be read off from  \eqref{globalthermoq=p} and \eqref{globaltermophiq}, the other thermodynamic quantities require additional care. The $1$-, $p$-, and $(1,p)$-cases can be summarised in
\begin{align}\label{general1pthermo}
    M &= V_{p}R_{0}\Big(\mathcal{T}s\cosh^{2}{\eta} + \frac{\mathcal{T}s}{n}+\Phi_{1}Q_{1}+\Phi_{p}Q_{p}\Big)\ , \\ 
    J_{a}&= V_{p}\frac{\Omega_{a}R_{a}^{2}}{|k|}\mathcal{T}s\cosh{\eta}\ , \\
    \mathcal{T}_{tot}&=- \frac{V_{p}R_{0}}{n}\frac{R^{2}}{L^{2}}\Xi_{1,p}\ .
\end{align} 
A few remarks are in order. The thermodynamic properties above satisfy the Smarr relation \eqref{smarrrelation2}. Furthermore, the case of branes carrying only a $p$-brane charge is obtained by setting $\mathcal{Q}_{1}=0$ and one can easily check that the thermodynamic quantities obtained in this section do not reduce to the ones in section \ref{productofoddspheres}. The same is true when setting $\mathcal{Q}_{p}=0$, thus obtaining solutions for blackfolds carrying only $\mathcal{Q}_{1}$ charges. The difference in the thermodynamic quantities is due to the different geometries in which the blackfolds are embedded, as explained in more detail at the beginning of this section.

\paragraph{Extremal limits.} When considering the extremal limit of these solutions, we find that $\sinh{\eta}\rightarrow\infty\implies\tanh{\eta}=1$, and thus the fluid moves at the speed of light. Focusing on the case of a single odd-sphere by keeping $\mathcal{K}=\mathcal{T}s\cosh{\eta}\sinh{\eta}$ fixed, we have that the thermodynamic quantities in \eqref{general1pthermo} reduce to the integrated versions of \eqref{intthermonullflux} with $\mathcal{V}=0$. Furthermore, as in section \ref{productofoddspheres}, one can define the null vector and find an equivalent relation to \eqref{finalnullprod}.

\subsection{Specific configurations of odd-spheres}\label{examplesmtypeii}
In section \ref{oddspheresflux} and \ref{prododdsphereflux} we constructed solutions made of (products of) odd-sphere using the general currents and thermodynamic properties of section \ref{Blackfoldswithmultipleq-branecurrents}. We also mentioned in \ref{oddspheresflux} that further inspection was necessary in order to determine the validity of odd-sphere solutions in type II/M-theory. In this section we offer details on this analysis and their properties. In order to do so, we need the local thermodynamics of the effective fluids describing the black branes carrying the various charges. 

The thermodynamics of D$p$ and M$p$ branes are obtained from \eqref{localthermo} by setting $N=1$. On the other hand, the thermodynamics of D0-D$p$ and F1-D$p$ bound states can be found in \cite{Emparan:2011hg}. In section \ref{oddspheresflux}, we have found solutions for blackfolds with an $\mathbb{S}^{p}$ horizon topology with $p=2k+1$. This implies that by construction we require the $p$-form charge to be odd. Hence, these solutions can be used to study blackfold configurations that are based on long-wavelength deformations of F1, D1, D3 and F1-D3 branes/bound states in type IIB string theory and M5 branes in M-theory. On the other hand, in section \ref{prododdsphereflux}, we constructed blackfolds with an $\prod_{a}^{b}\mathbb{S}^{p_{a}}$ horizon topology with $p=\sum^{b}_{a} p_{a}=\sum^{b}_{a} (2k_{a}+1)$ and $b=(2,3)$, opening up to the possibility of an even $p$-form charge, such as M2 branes in M-theory. Likewise, these solutions can be used for D3, D5, F1-D3, F1-D5 branes/bound states in type IIB. In table \ref{branestypeiim} we summary the possible configurations, indicate their respective horizon topology and their physical parameters. 

\begin{table}[h!] 
\begin{center}\begin{tabular}{*4l}    \toprule
\textit{Theory} & \textit{Brane/Bound state} & \textit{Horizon topology} & \textit{Physical parameters}   \\\midrule
\multirow{4}{5em}{Type IIB}    & F1, D1 & $\mathbb{S}^{1}\times\mathbb{S}^{7}_{\perp}$& \multirow{2}{5em}{\ref{dpf1dpm5}}  \\ 
 & \multirow{2}{5em}{F1-D3, D3} & $\mathbb{S}^{3}\times\mathbb{S}^{5}_{\perp}$ &  \\
 &   & $\mathbb{S}^{1}\times \mathbb{S}^{1}\times\mathbb{S}^{1}\times\mathbb{S}^{5}_{\perp}$& \multirow{2}{8em}{\ref{d5f1d5}}\\ 
 & F1-D5, D5 &   $\mathbb{S}^{1}\times \mathbb{S}^{3}\times\mathbb{S}^{1}\times\mathbb{S}^{3}_{\perp}$& \\ 
 \multirow{3}{5em}{M-theory} & M2 & $\mathbb{S}^{1}\times \mathbb{S}^{1}\times\mathbb{S}^{7}_{\perp}$ & \ref{m2}\\
 & \multirow{2}{5em}{M5} & $\mathbb{S}^{5}\times\mathbb{S}^{4}_{\perp}$&\ref{dpf1dpm5} \\
 & & $\mathbb{S}^{1}\times \mathbb{S}^{3}\times\mathbb{S}^{1}\times\mathbb{S}^{4}_{\perp}$& \ref{d5f1d5} \\
\bottomrule
\hline \end{tabular} \end{center}
\caption{\textit{All possible blackfolds based on D$p$, M$p$ and F1-D$p$ black branes that give rise to solutions provided in sections \ref{oddspheresflux} and \ref{prododdsphereflux}. The label $\perp$ denotes the horizon topology transverse to the worldvolume.
Note that the $\prod_{a}^{3}\mathbb{S}^{p_{a}}$ and $\prod_{a}^{2}\mathbb{S}^{p_{a}}$ topologies include all the possible embeddings of section \ref{prododdsphereflux}.}}
\label{branestypeiim}
\end{table}
As discussed on general grounds in section \ref{oddspheresflux}, in the case of the odd-sphere solutions with a non-zero force term\footnote{In fact, the solutions in section \ref{prododdsphereflux} do not require any further analysis. In these cases $\Omega\geq0\ \forall\ R>0$ independently on the sign of $\alpha$.} we have to impose that $\Omega^{2}\geq0$ in \eqref{omegaflux}. This is done separately for the different black branes as it is more suitable to re-write \eqref{omegaflux} in terms of the charge parameters. In particular: 
\begin{itemize}
    \item D$p$- and M$p$-branes. Inspecting \eqref{localthermo} we see that \eqref{omegaflux} can be rewritten in terms of the charge parameter $\alpha$. Denoting $sh_{\alpha}\equiv\sinh{\alpha}$ and $ch_{\alpha}\equiv\cosh{\alpha}$, we obtain \begin{equation}\label{omegadpm}
        \Omega^{2} =\frac{f(R)}{R^{2}} \frac{p(1+nsh^{2}_{\alpha})+\frac{R^{2}}{\Tilde{L}^{2}}(1+n+p+n(1+p)sh^{2}_{\alpha})-\frac{Rn(l-1)}{\Tilde{L}}\sqrt{f(R)}sh_{\alpha}ch_{\alpha}}{n+p(1+nsh^{2}_{\alpha})+\frac{R^{2}}{\Tilde{L}^{2}}(1+n+p+n(1+p)sh^{2}_{\alpha})-\frac{Rn(l-1)}{\Tilde{L}}\sqrt{f(R)}sh_{\alpha}ch_{\alpha}}\ . 
    \end{equation} Assuming $\alpha>0$ we get that $\Omega>0\  \forall\ R>0$, but there exists no $R>0$ such that $\Omega=0$. For $\alpha<0$, only the terms proportional to $sh_{\alpha}ch_{\alpha}$ will change sign. This new solution is positive $\forall\ R>0$ and has no static case. 
    \item Taking the F1-D$p$ bound state and repeating the procedure just outlined, we obtain the equilibrium condition \eqref{thermof1dp}. We first notice that the solution is invariant under $\theta\rightarrow-\theta$ and close inspection reveals that for $\alpha>0$, $\Omega>0\  \forall\ R>0$, but no $R>0$ such that $\Omega=0$. For $\alpha\rightarrow-\alpha$, only the terms proportional to $sh_{\alpha}ch_{\alpha}\cos{\theta}$ change sign and the validity of the solution is identical.  
\end{itemize}
From the above analysis we have concluded that odd-sphere solutions are valid in both type II string theory and M-theory independently on the sign of $\alpha$ and furthermore that they do not admit a static limit. 

Based on the global thermodynamics of these blackfold solutions, which can be found in appendix \ref{physicalparameters}, one can easily show that the Smarr relation \eqref{smarrrelation2} is obeyed for all cases. Furthermore, one can study their extremal limits. In the case of D$p$, F1-D$p$ and M$p$ solutions, one can explicitly see that in this limit, $\tanh{\eta}=1$, hence the possibility of a null-wave brane construction and that the equations in section \ref{nullwavebraneflux} are obeyed.

\section{Discussion and outlook}\label{Discussion and Outlook}
In this paper we have used the long-wavelength effective theory, known as the blackfold approach, to construct new asymptotically (A)dS (multi-)charged and (non-)rotating non-extremal black holes in Einstein-dilaton gravity with higher-form gauge fields and in type II/M-theory and study their extremal limits. In table \ref{table2} we summarise the different horizon topologies of the perturbative black hole solutions that we found in this paper. One of these solutions corresponds to a family of static black hole solutions in dS (see appendix \ref{staticsolutions}) with arbitrary $q$-brane charge. In the case of type II/M-theory these solutions asymptote to AdS$_{l}\times S^{m}$ with $(l,m)=\{(4,7),(5,5),(7,4)\}$ and $F_{l}$ flux form.  In the case of $(l,m)=(5,5)$ these black hole solutions correspond to new thermal states of $\mathcal{N}=4$ Super Yang-Mills theory. We note that these solutions are perturbatively constructed directly in 10 or 11 dimensions and hence without the need of dimensional reduction or truncation usually employed to find exact solutions in this context. They also correspond to black hole solutions with a non-trivial horizon topology, which are usually typically hard to obtain when directly solving the Einstein equations.

\begin{table}[h!] 
\begin{center}\begin{tabular}{*5l}    \toprule
\textit{Action} & \textit{Background} & \textit{$q$-charge} & \textit{Horizon topology}& \textit{Solution}   \\\midrule
\multirow{5}{3em}{ \eqref{actionads}}    & \multirow{4}{4em}{(A)dS$_{D}$}  &\multirow{4}{4em}{$0,\ 1,\ p$}   & $\mathbb{S}^{p}\times \mathbb{S}_{\perp}^{n+1}$ & \eqref{omega01p} \\ & & & $\prod_{a}^{b}\mathbb{S}^{p_{a}}\times \mathbb{S}_{\perp}^{n+1}$ & \eqref{omegaproduct} \\ & & &  $\mathbb{S}^{n+3}$ & \eqref{actiondisk}\\  & & &  $\mathbb{S}\times \mathbb{S}^{n+2}$ & \eqref{actionannulus} \\ 
 & dS$_D$ & $q$ & $\prod_{a}^{b}\mathbb{S}^{p_{a}}\times \mathbb{S}_{\perp}^{n+1}$ & \eqref{qprod} \\ \midrule
 \multirow{5}{3em}{ \eqref{actiontypeii}}    & \multirow{5}{5em}{AdS$_{5}\times S^{5}$} & \multirow{5}{9em}{$1,\ p,\ (1,p)$} & $\mathbb{S}^{1}\times \mathbb{S}_{\perp}^{7}$ & \multirow{2}{6em}{\eqref{omegaflux}, \eqref{omegaoddone}} \\ & & & $\mathbb{S}^{3}\times \mathbb{S}_{\perp}^{5}$ &  \\ & & & $\mathbb{S}^{1}\times \mathbb{S}^{1}\times \mathbb{S}^{1}\times \mathbb{S}_{\perp}^{5}$ & \eqref{s3s1s1omega}, \eqref{s1s1s3omega} \\ & & & \multirow{2}{8em}{$\mathbb{S}^{1}\times\mathbb{S}^{1}\times \mathbb{S}^{3}\times \mathbb{S}_{\perp}^{3}$} & \eqref{s3s1s1omega}, \eqref{s1s1s3omega}, \\ & & & & \eqref{s1(s3s1)}, \eqref{(s3s1)s1} \\ \midrule
 \multirow{5}{3em}{ \eqref{actionmtheory}}    & \multirow{3}{5em}{AdS$_{4}\times S^{7}$} & \multirow{5}{4em}{$p$} & $\mathbb{S}^{1}\times\mathbb{S}^{1}\times \mathbb{S}_{\perp}^{7}$ & \eqref{m2brane} \\ & & & $\mathbb{S}^{5}\times \mathbb{S}_{\perp}^{4}$ & \eqref{omegaflux}, \eqref{omegaoddone} \\ & & & $\mathbb{S}^{3}\times\mathbb{S}^{1}\times\mathbb{S}^{1}\times \mathbb{S}_{\perp}^{4}$ & \eqref{s3s1s1omega}, \eqref{(s3s1)s1} \\ & \multirow{2}{5em}{AdS$_{7}\times S^{4}$} & & $\mathbb{S}^{1}\times \mathbb{S}^{1}\times \mathbb{S}_{\perp}^{7}$ & \eqref{m2brane} \\ & & & $\mathbb{S}^{3}\times\mathbb{S}^{1}\times\mathbb{S}^{1}\times \mathbb{S}_{\perp}^{4}$ & \eqref{s1(s3s1)} \\
 \bottomrule
 \hline \end{tabular} \end{center}
\caption{\textit{A summary of the actions solved with the corresponding asymptotic background solution, the charge carried by the blackfold, the horizon topology of the latter and the corresponding solutions. We note that in general $p_{a}=2k_{a}+1$ is odd and $b\leq n+2$. The label $\perp$ denotes the horizon topology transverse to the worldvolume. }}
\label{table2}
\end{table}

\paragraph{Regime of applicability.}\label{regimeofapplicability}
The blackfold approach is valid when we can locally approximate near the horizon the black hole by a black brane, which requires the horizon to be characterised by two or more widely separated length scales. For the approximation to be valid, we require that the smallest length scale associated to the horizon, in particular $r_c$ introduced in \eqref{approx}, is much smaller than all the curvature scales, intrinsic or extrinsic, associated to the worldvolume or to the background, namely
\begin{equation}\label{valcond}
    r_{c}\ll r_{int}\ , \ \ \ r_{c}\ll r_{ext}\ , \ \ \ r_{c}\ll\mathcal{L} \, 
\end{equation} 
where $r_{int}=|\mathcal{R}_{\gamma}|^{-1/2}$ with $\mathcal{R}_{\gamma}$ being the Ricci scalar of the worldvolume, $r_{ext}=|K^{\mu}K_\mu|^{-1/2}$. Note here that comparing with \eqref{approx} we have split $\mathcal{R}=\{r_{int},r_{ext}\}$. 

As shown in \cite{Caldarelli:2010xz}, when dealing with charged blackfolds in Einstein-dilaton gravity with higher-form gauge fields, the small characteristic length scales to consider are $r_{\varepsilon}$ and $r_{\mathcal{Q}}$ which are the energy density and charge density radii of the black brane, respectively. Nonetheless, one can show that $r_{\mathcal{Q}}$ is parametrically smaller, thus we consider $r_{c}=r_{\varepsilon}\sim r_{0}(\cosh{\alpha})^{2/n}$. Next, we notice that for the embeddings \eqref{embeddsingle} and \eqref{embeddprod} $r_{int}=R/\sqrt{6}$ and that $r_{ext}$ is parametrically larger than $r_{int}$. Thus, the requirements in \eqref{valcond} reduce to the first and last one, with $\mathcal{L}=|\Lambda|^{-1/2}\sim L$. Therefore, we require that \begin{equation}\label{highergrval}
    \frac{r_{0}}{R}\ll (\cosh{\alpha})^{-2/n}\ \ \ \text{and} \ \ \  \frac{r_{0}}{L}\ll (\cosh{\alpha})^{-2/n}\ .
\end{equation}
From the above, we see that $R\rightarrow0$ is not
within the regime of validity of the approximation. This is exactly why the $R=0$ solutions in the extremal limits of odd-spheres and product of odd-spheres in (A)dS, argued in sections \ref{oddspheres} and \ref{productofoddspheres} were excluded. Finally, in the cases considered in section \ref{diskandannulus}, we notice that $r_{int}=L/\sqrt{6}$, while $r_{ext}=0$ since the embeddings are minimal surfaces. This implies that the second condition in \eqref{highergrval} is the only condition present at leading order. 

In the case of type II/M-theory, apart from the validity of the probe approximation, we also have to require the validity of the supergravity description of the black probe branes. In order to deal with the former, we take a look at the functions $f,\ H,\ D$ defining the brane solutions in \cite{Caldarelli:2010xz,Emparan:2011hg} and by exploiting the fact that $0\leq\{\sin^{2}\theta,\cos^{2}\theta\}\leq1$, we conclude that the largest characteristic scale is $r_{c}^{n}=r_{0}^{n}\sinh^{2}{\alpha}$. Once again, when calculating the intrinsic and extrinsic characteristic scales of the embeddings we notice that the latter are parametrically larger than the former. Moreover, in all of the embeddings considered in section \ref{TypeII/M-theory}, once can show that $r_{int}=R/\sqrt{6}$ or $r_{int}= R_{1}/\sqrt{6}$ depending whether the embeddings are odd-spheres or product of odd-spheres respectively. In products of odd-spheres one can show that $R_2$ is of the same order of magnitude as $R_1$ so it is sufficient to consider the condition $r_{int}= R_{1}/\sqrt{6}$. In these cases, the blackfold approximations require that $r_{0}/R_{(1)}\ll(\sinh{\alpha})^{-2/n}$, implying that $R\rightarrow0$ (or alternatively $R_{1}\rightarrow0$) is not a valid limit. In the case of $R\sim L$, we have that $r_{0}/L\ll(\sinh{\alpha})^{-2/n}$, which is exactly the last condition in \eqref{valcond}.

It is useful to have a better understanding of the validity of these configurations in the near-extremal and extremal limits in which $r_{c}^{n}=r_{0}^{n}\sinh^{2}{\alpha}\sim r_{0}^{n}\sinh{\alpha}\cosh{\alpha}$ for sufficiently large $\alpha$. We will use the D$p$ and M$p$ branes embedded in the AdS$_{l}$ part of the spacetime in section \ref{TypeII/M-theory} as an example and treat the three cases of D$p$, M2 and M5 simultaneously. Recall that in supergravity the AdS$_{l}\times S^{m}$ geometry arises from having $N\gg1$ coincident branes placed at the origin of the spacetime, with $N$ being related to the brane tension $T_{l-2}$ and AdS radius $L$. At the same time, given a number of probe branes $N_{l-2}$, the charge is quantised as $\mathcal{Q}_{l-2}=N_{l-2}T_{l-2}$. Note that we require $N_{l-2}\ll N$ since otherwise the backreaction of the former would be greater than the latter one. Hence, with the aid of \eqref{localthermo}, the fact that $G\sim l_{p}^{l+m-2}$ and $T_{l-2}\sim1/l_{p}^{l-1}$, with $l_{p}$ being the Planck length, we have that 
\begin{equation}
    r_{c}\sim N_{l-2}^{\frac{1}{l-1}}T_{l-2}^{-\frac{1}{m-1}}\ .
\end{equation} 
Thus, after excluding the $R\rightarrow 0$ case and using all of the information above, we see that in the regime of $R\sim L$, we can summarise the validity of the approach as 
\begin{equation}
    1\ll N_{D_{p}}\ll N\ ,\ \ \ 1\ll N_{2}\ll N^{2}\ , \ \ \ 1\ll N_{5}^{2}\ll N\ .
\end{equation} 
Note that the analysis automatically encapsulates the requirement $N_{l-2}\ll N$. This same procedure can be repeated for other types of branes and embeddings.  

\paragraph{Future directions.} The work presented here suggests several directions to pursue. In particular, the new black hole solutions found are valid at ideal order but it would be interesting to study higher-order corrections and obtain explicit metrics at first order as in \cite{Camps:2008hb,Emparan:2009vd,Armas:2011uf,Emparan:2007wm,Armas:2014bia,Armas:2014rva,Armas_2013, Armas:2013aka, Armas:2012jg,Armas:2012ac,Nguyen:2021srl}. It would also be interesting to study thermal and dynamical stability under both fluid and elastic perturbations of the worldvolume as in \cite{Caldarelli:2012hy, Gath:2013qya, DiDato:2015dia, Armas:2019iqs, Nguyen:2019syc}. This perturbative analysis could serve as a starting point for finding more non-trivial stationary solutions, in particular in AdS$_5\times S^5$, in which the angular velocity is different in the various angular directions. Additional non-trivial solutions could be obtained by using different bound states or black branes with intrinsic spins \cite{Cvetic:1999ne, Harmark:1999xt} as a starting point, and by considering the possibility of multiple disconnected horizons - the simplest example being the uplift of AdS black saturn solutions \cite{Caldarelli:2008pz, Armas:2015qsv} to 10D supergravity. Together with exact and numerical results (e.g. \cite{Dias:2022eyq}) this endeavour could shine light on the various phases of $\mathcal{N}=4$ Super Yang-Mills theory. Other interesting directions to pursue include: (1) finding other non-trivial geometries due to the presence of $q<p$-brane charges such as the prolate rings and spheres obtained in \cite{Caldarelli:2010xz} for $q=1$ or helicoidal geometries as in \cite{Armas:2015nea}; and (2) considering other non-trivial supergravity asymptotic backgrounds with non-vanishing NS-NS fluxes or non-constant dilaton profiles as in \cite{Armas:2018rsy, Armas:2019asf,Armas:2022bkh}.

\acknowledgments
We thank Maria Peletidou for collaboration in the initial stages of this project, in particular in section 3 initiated during her MSc thesis at U. Amsterdam. JA is partly supported by the
Dutch Institute for Emergent Phenomena (DIEP) cluster at the University of Amsterdam
via the DIEP programme Foundations and Applications of Emergence (FAEME). GPN is partly supported by the Tertiary Education Scholarship Scheme (TESS). GPN would like to thank the PhD council, in particular A. van Spaendonck, and the IoP manager, Dr. J. van Mameren, for their understanding and willingness to lend a helping hand. Your actions are an example to follow.

\appendix
\section{Static solutions in dS$_D$ with arbitrary $q$-charge} \label{staticsolutions}
In this appendix we focus on static solutions of Einstein-dilaton gravity with higher-form gauge fields adopting a spherical geometry for the worldvolume according to \eqref{eq:embeddingodd}. Our goal is to show that there is a general class of non-rotating solutions ($\Omega=0$) with arbitrary $q$-charge with $q\le p$ and horizon topology $\prod_{a}^{b}\mathbb{S}^{p_{a}}\times \mathbb{S}^{n+1}$, where $p_a=2k_a+1$ is odd, that asymptote to dS space in Einstein-dilaton gravity with higher-form gauge fields. These solutions are static versions of those in sections \ref{oddspheres} and \ref{productofoddspheres} carrying arbitrary $q$-charge.

We consider the more general case of product of odd-spheres of section \ref{productofoddspheres} and embed the blackfold as in \eqref{embeddprod}. For general $q$ we can define the following vectors  \begin{equation}\label{vectorsqprod}
    u^{b}\partial_{b}=\frac{1}{\sqrt{f(R)}}\partial_{\tau}\ ,\ \ \ v_{a}^{b}\partial_{b}=\frac{1}{R_{a}}\partial_{\Phi_{a}}\ ,
\end{equation} 
such that we will have a set of $q\leq b$ (with $b\leq n+2$) spacelike vectors $v$, which can be shown to obey \eqref{conditiononvec}. The solutions to the intrinsic equations are described in section \ref{Perfect fluids with q-brane currents}. Following the procedure in section \ref{oddspheres}, we can write down the integrated action as 
\begin{equation}
    \tilde{I}_{E}= \frac{\Omega_{n+1}V_{p}}{16\pi G}\Big(\frac{n}{4\pi T}\Big)^{n}f'(R)^{\frac{n+1}{2}}\Big(1-\frac{\Phi_{H}^{2}}{4\pi^{4}NV_{q}^{2}}\Big)^{nN/2}\ ,
\end{equation} 
where we have used the fact that $c|\hat{h}^{(q)}|^{1/2}=V_{q}=(2\pi)^{q} \prod_{a}^{q}R_{a}\sqrt{f'(R)}$ and we note that $f'(R)$ was defined in \eqref{omegads1}. Varying the action with respect to each $R_a$ we find a set of $q$ equations of motion, which when summed yield \begin{equation}\label{qprod}
    \frac{R}{L} = \Big(\frac{p+qnN\sinh^{2}{\alpha}}{1+n+p+(q+1)nN\sinh^{2}{\alpha}}\Big)^{1/2}\ ,
\end{equation} 
where $R^{2}\equiv\sum_{a=1}^{b}R^{2}_{a}$. Equation \eqref{qprod} is exactly the upper bound of equation \eqref{possibleRa1} when $q=1$ (or alternatively by taking \eqref{possibleRa0pa} and summing over all $R_{a}$). The thermodynamic properties of these configurations can be obtained from \eqref{globalthermoq=p} by setting $\Omega=0$. This provides circumstantial evidence that the general solution of \eqref{omega01p} may be valid for all $q$ also in the rotating case. 

\section{Thermodynamic properties}\label{physicalparameterstot}
In this appendix we provide explicit details of the global thermodynamic properties of many of the solutions found throughout the bulk of paper, in particular the mass, angular momenta, entropy, charges, temperature, chemical potentials and total tension.
\subsection{Disk and annulus solutions}\label{physicaldiskannulus}
Here we give the global thermodynamics of the disk and annulus solutions of section \ref{diskandannulus}.

\paragraph{Disk solution.}
\begin{align}
    M&=\frac{\Omega_{(n+1)}x^{n N}}{8 \pi G \left(\Omega^2-\frac{1}{L^2}\right)^2}\left(\frac{n}{4\pi T}\right)^n \left[\left(\Omega^2-\frac{1}{L^2}\right)\frac{\Phi_H^2}{N}{}_2 F_1\left(1,1+\frac{n}{2}\left(N-1\right);1+\frac{n N}{2};x^{2}\right)\right. \nonumber \\
    &+x^{2}\left(\frac{1}{L^2}+\frac{\frac{-1-2n}{L^2}+\left(1+n\right)\Omega^2-\frac{n\left(N-1\right)}{L^2}\frac{\Phi_H^2}{N}}{2+nN}{}_2 F_1\left(1,1+\frac{n}{2}\left(N-1\right);2+\frac{n N}{2},x^{2}\right)\right) \nonumber \\
    &+\left.\frac{2\left(\frac{n+1}{L^2}-\Omega^2\right)x^{4}}{\left(2+n N\right)\left(4+n N\right)}{}_2 F_1\left(2,1+\frac{n}{2};3+\frac{n N}{2};x^{2}\right)\right]\ ,\\
    J&=\frac{ \Omega_{(n+1)}\Omega}{4 G \left(\Omega^2-\frac{1}{L^2}\right)^2}\left(\frac{n}{4\pi T}\right)^{n}\left[\frac{\Phi_H^2 x^{n N+2} }{N\left(n N+2\right)}{}_2 F_1\left(2,1+\frac{n}{2}\left(N-1\right);\frac{n N}{2}+2;x^{2}\right)\right. \nonumber \\ 
    &+\left. \frac{n  x^{n N+4} }{\left(n N+2\right)\left(n N+4\right)}{}_2 F_1\left(2,1+\frac{n}{2}\left(N-1\right);\frac{n N}{2}+3;x^{2}\right)\right]\ , \\
    S&=\frac{n \Omega_{(n+1)}}{8 G T}\left(\frac{n}{4\pi T}\right)^{n}\frac{x^{2+nN}}{\left(2+nN\right)\left(\Omega^2-\frac{1}{L^2}\right)}{}_2 F_{1}\left(1, \frac{1}{2}(N-1) n ; \frac{N n}{2}+2 ;x^{2}\right)\ , \\
    Q&=\frac{ \Phi_H \Omega_{(n+1)}}{8 G N}\left(\frac{n}{4\pi T}\right)^{n}\frac{x^{nN}}{\left(\Omega^2-\frac{1}{L^2}\right)}{ }_{2} F_{1}\left(1, \frac{1}{2}(N-1) n ; \frac{N n}{2}+1 ; x^{2}\right)\ , \\
    \mathcal{T}_{tot}&=\frac{\Omega_{(n+1)}x^{n N+2}}{8 G\left(\Omega^2-\frac{1}{L^2}\right)^2}\left[\left(\Omega^2-\frac{1}{L^2}\right)\frac{\Phi_H^2}{N}{}_2 F_1\left(1,1+\frac{n}{2}\left(N-1\right);1+\frac{n N}{2};x^{2}\right)\right. \nonumber \\
    &+\frac{\frac{4+n}{L^2}-\left(2+n\right)\Omega^2}{\left(2+n\right)}+\frac{n}{\left(2+n\right)\left(2+nN\right)}\Big(\left(2+n\right)\left(\Omega^2-\frac{1}{L^2}\right) \\ \nonumber
    &+\left(\frac{4+n-2N}{L^2}-\Omega^2 \left(2+n\right)\right)\frac{\Phi_H^2}{N}\Big)\times \left.{}_2 F_1\left(1,1+\frac{n}{2}\left(N-1\right);2+\frac{n N}{2};x^{2}\right)\right]\ ,
\end{align} 
where $x\equiv\left(1-\frac{\Phi_H^2}{N}\right)^{1/2}$.
\paragraph{Annulus solution.}
\begin{align}
    M&=\frac{ \Omega_{(n+1)}}{8  G} \left(\frac{n}{4\pi T}\right)^n\int_{r_{min}}^{r_{max}} drI_{r}\left(f(r)-\Omega^2 r^2+n+n \frac{\Phi^2_H}{4\pi^2 r^2}\frac{f(r)-\Omega^2 r^2}{f(r)-\frac{\Phi_H^2}{N\left(2\pi r\right)^2}} \right). \\
    J&=\frac{n  \Omega_{(n+1)}\Omega}{8  G} \left(\frac{n}{4\pi T}\right)^n\int_{r_{min}}^{r_{max}} dr r^{2}  I_{r}\ , \\
    S&=\frac{n \Omega_{(n+1)}}{8  G T} \left(\frac{n}{4\pi T}\right)^n\int_{r_{min}}^{r_{max}} drI_{r}\ , \\ 
    Q&=\frac{n \Omega_{(n+1)} \Phi_H}{16 \pi G } \left(\frac{n}{4\pi T}\right)^n\int_{r_{min}}^{r_{max}} dr r^{-2}  I_{r}\ , \\
    \mathcal{T}_{tot}&=-\frac{\Omega_(n+1)}{16\pi G} \int_{r_{min}}^{r_{max}} dr I_{r} \Big[(n+p)\Omega^2 r^2-pf(r)\\ \nonumber
    &\hspace{4cm}+n N\left(-f(r)+2\Omega^2 r^2-f(r)\Omega^2 r^2\frac{\frac{\Phi_H^2}{N(2\pi r)^2}}{f(r)-\frac{\Phi_H^2}{N(2\pi r)^2}}\right)\Big] ~~,
\end{align} 
where we have defined $I_{r}= r  \left(f(r)-\Omega^2 r^2\right)^\frac{n}{2} \left(1-\frac{\Phi_H^2}{Nf(r)\left(2\pi r\right)^2}\right)^\frac{n N}{2}$.

\subsection{Bound states in type II/M-theory}\label{physicalparameters}
Here we provide the explicit physical parameters for the solutions considered in type II/M-theory of section \ref{examplesmtypeii}.  The metrics and the local thermodynamics for the D0-D$p$ and F1-D$p$ bound states considered in this paper are given in \cite{Emparan:2011hg} and we will use the following shorthand notation \begin{equation}\label{shorthand}
    \sinh{\alpha_{i}}\equiv\sin{\theta}\sinh{\alpha}\ , \ \ \ \sinh{\alpha_{p}}\equiv\cos{\theta}\sinh{\alpha}\ ,
\end{equation} 
for the D0-D2 and F1-D$p$ bound states. Furthermore, we will not consider the single-charged cases explicitly, as these can easily be obtained from the properties of bound states. For example, setting $\cos{\theta}=0$ we recover the single-charged case with $\mathcal{Q}_{p}=0$, while setting $\sin{\theta}=0$, we recover set-ups with $\mathcal{Q}_{q}=0$. Finally, we will set 
\begin{equation}
    s_{i} \equiv \sinh{\alpha_{i}}\ , \ \ \ c \equiv \cosh{\alpha}\ ,
\end{equation} 
in order to avoid cluttering the various expressions for the thermodynamics.

\subsubsection{F1, D1, D3, F1-D3, M5}\label{dpf1dpm5}
We start with the case of $p=2k+1$ bound states as they are the most interesting since they couple to the background gauge field in the backgrounds \eqref{eq:AdSS5metric}. In the case of F1-D$p$ in type IIB, we have that $n=7-p$ and $\Tilde{L}=L$. Thus taking a look at \eqref{omegaflux}, \eqref{globalthermoq=p} and \eqref{mjsback}, we have that
\begin{align}\label{thermof1dp}
   \Omega^{2} &=\frac{f(R)}{R^{2}} \frac{p(1+(7-p)s^{2}_{p})+(7-p)s^{2}_{1}+\frac{R^{2}}{L^{2}}\Xi_{1,p}-4\frac{R(7-p)}{L}\sqrt{f(R)}s_{p}c}{7+p(7-p)s^{2}_{p}+(7-p)s^{2}_{1}+\frac{R^{2}}{L^{2}}\Xi_{1,p}-4\frac{R(7-p)}{L}\sqrt{f(R)}s_{p}c}\ , \\
   M & = \frac{\Omega_{8-p}V_{p}}{16\pi G}r_{0}^{7-p}\Big(f(R)^{3/2}\Xi_{1,p} - (7-p)\frac{R}{L}(4f(R)+1)s_{p}c\Big)~, \\ 
   J&=\frac{\Omega_{8-p}V_{p}R}{16\pi G}r_{0}^{7-p}\sqrt{p(1+(7-p)s^{2}_{p})+(7-p)s^{2}_{1}+\frac{R^{2}}{L^{2}}\Xi_{1,p}-4\frac{R(7-p)}{L}\sqrt{f(R)}s_{p}c}\ \times \nonumber \\ &\ \ \ \times
   \sqrt{7+p(7-p)s^{2}_{p}+(7-p)s^{2}_{1}+\frac{R^{2}}{L^{2}}\Xi_{1,p}-4\frac{R(7-p)}{L}\sqrt{f(R)}s_{p}c}~, \\
   S &= \frac{\Omega_{8-p}V_{p}c}{4G\sqrt{7-p}}r^{8-p}\sqrt{7+p(7-p)s^{2}_{p}+(7-p)s^{2}_{1}+\frac{R^{2}}{L^{2}}\Xi_{1,p}-4\frac{R(7-p)}{L}\sqrt{f(R)}s_{p}c}~, \\
   T &= \frac{1}{4\pi r_{0}c}\sqrt{\frac{f(R)(7-p)^{3}}{7+p(7-p)s^{2}_{p}+(7-p)s^{2}_{1}+\frac{R^{2}}{L^{2}}\Xi_{1,p}-4\frac{R(7-p)}{L}\sqrt{f(R)}s_{p}c}} \ , \\
   \mathcal{T}_{tot}&= - \frac{\Omega_{8-p}V_{p}}{16\pi G}\Big(\sqrt{f(R)}\frac{R^{2}}{L^{2}}\Xi_{1,p}+(7-p)\frac{R}{L}(4f(R)+1)s_{p}c \Big)\ ,
\end{align} 
ith $\Xi_{1,p}=8+(7-p)((1+p)s^{2}_{p}+2s^{2}_{1})$, while as seen in \eqref{globaltermophiq} the chemical potential and the global charge are not functions of the rapidity in this case, therefore their value can be easily read off from there. In the case of an M5-brane, the equations to consider are identical but in this case we have to set \begin{equation}
    \sin{\theta}=0\ , \ \ \ \tilde{L}=\frac{L}{2}\ , \ \ \ n=3 \ .
\end{equation} 
Finally, note that the F1-D3, D3 and M5 bound sate/branes can also be embedded as in \ref{productofoddspheres}. We will provide an example with $p=5$ in the next section. 

\subsubsection{D5, F1-D5}\label{d5f1d5}
Here we will consider the example of the $\mathbb{S}^{1}\times\mathbb{S}^{1}\times\mathbb{S}^{3}$ horizon topology. In particular, we will consider the D5 bound state embedded as in \eqref{embeds3s1s1}. Thus, we set $n=7-p=2$, $\tilde{L}=L$ and $\theta=2\pi$. Taking a look at \eqref{s1s1s3omega}, \eqref{globalthermoq=p} and \eqref{mjsback}, \begin{align}
     \Omega_{a}^{2} &=\frac{f(R)}{R_{a}^{2}} \frac{1+2s^{2}_{p}+\frac{R_{a}^{2}}{L^{2}}\Xi_{5}}{7+2(5s^{2}_{p}+s^{2}_{1})+\frac{R^{2}}{L^{2}}\Xi_{5}} \ , \\
    \Omega_{3}^{2} &=\frac{f(R)}{R_{3}^{2}} \frac{3(1+2s^{2}_{p})}{7+2(5s^{2}_{p}+s^{2}_{1})+\frac{R^{2}}{L^{2}}\Xi_{5}} \ , \\
    M & = \frac{\Omega_{3}V_{5}}{16\pi G}r_{0}^{2}\Big(f(R)^{3/2}\Xi_{1,5}\Big) ~, \\
    J_{a}&=\frac{\Omega_{3}V_{5}R_{a}}{16\pi G}r_{0}^{2}\sqrt{1+2s^{2}_{p}+\frac{R_{a}^{2}}{L^{2}}\Xi_{5}}
   \sqrt{7+2(5s^{2}_{p}+s^{2}_{1})+\frac{R^{2}}{L^{2}}\Xi_{5}}~, \\
   J_{3}&=\frac{\Omega_{3}V_{5}R_{3}}{16\pi G}r_{0}^{2}\sqrt{3(1+2s^{2}_{p})}
   \sqrt{7+2(5s^{2}_{p}+s^{2}_{1})+\frac{R^{2}}{L^{2}}\Xi_{5}}~, \\
   S &= \frac{\Omega_{3}V_{5}c r_{0}^{3}}{4G\sqrt{f(R)-\sum_{a}^{3}\Omega_{2}^{2}R_{a}^{2}}}~, \\
   T &= \frac{1}{4\pi r_{0}c}\sqrt{f(R)-\sum_{a}^{3}\Omega_{2}^{2}R_{a}^{2}} \ . \\
   \mathcal{T}_{tot}&= - \frac{\Omega_{3}V_{3}}{16\pi G}\sqrt{f(R)}\frac{R^{2}}{L^{2}}\Xi_{5}\ .
\end{align}

\subsubsection{M2}\label{m2}
Here we will consider the example of the $\mathbb{S}^{1}\times\mathbb{S}^{1}$ horizon topology. In particular, we will consider the M2 brane embedded as in \eqref{m2brane}. Thus, we set $n=8-p=6$, keep $\tilde{L}$ and $\theta=2\pi$. Taking a look at \eqref{s3s1s1omega}, \eqref{globalthermoq=p} and \eqref{mjsback}, \begin{align}
     \Omega_{a}^{2} &=\frac{f(R_{1})}{R_{a}^{2}} \frac{1+6s^{2}_{2}}{7+2s^{2}_{p}+\frac{R_{1}^{2}}{\tilde{L}^{2}}\Xi_{2}} \ . \\
    M & = \frac{\Omega_{7}V_{2}}{16\pi G}r_{0}^{7}\Big(f(R)^{3/2}\Xi_{2}\Big) ~, \\
    J_{a}&=\frac{\Omega_{7}V_{2}R_{a}}{16\pi G}r_{0}^{7}\sqrt{1+6s^{2}_{2}}
   \sqrt{7+2s^{2}_{p}+\frac{R_{1}^{2}}{\tilde{L}^{2}}\Xi_{2}}~, \\
   S &= \frac{\Omega_{7}V_{2}c r_{0}^{7}}{4G\sqrt{f(R)-\sum_{a}^{2}\Omega_{2}^{2}R_{a}^{2}}}~, \\
   T &= \frac{1}{4\pi r_{0}c}\sqrt{f(R)-\sum_{a}^{2}\Omega_{2}^{2}R_{a}^{2}} \ . \\
   \mathcal{T}_{tot}&= - \frac{\Omega_{7}V_{2}}{16\pi G}\sqrt{f(R)}\frac{R^{2}}{L^{2}}\Xi_{2}\ .
\end{align} 
This concludes the thermodynamic properties of the various solutions we considered. 

 \bibliographystyle{JHEP}

\begin{thebibliography}{10}
 	
 	\bibitem{Maldacena:1997re}
 	J.M.~Maldacena, \emph{{The Large N limit of superconformal field theories and
 			supergravity}}, \href{https://doi.org/10.4310/ATMP.1998.v2.n2.a1}{\emph{Adv.
 			Theor. Math. Phys.} {\bfseries 2} (1998) 231}
 	[\href{https://arxiv.org/abs/hep-th/9711200}{{\ttfamily hep-th/9711200}}].
 	
 	\bibitem{Emparan:2008eg}
 	R.~Emparan and H.S.~Reall, \emph{{Black Holes in Higher Dimensions}},
 	\href{https://doi.org/10.12942/lrr-2008-6}{\emph{Living Rev. Rel.} {\bfseries
 			11} (2008) 6} [\href{https://arxiv.org/abs/0801.3471}{{\ttfamily
 			0801.3471}}].
 	
 	\bibitem{witten1998antide}
 	E.~Witten, \emph{Anti-de sitter space, thermal phase transition, and
 		confinement in gauge theories},  1998.
 	
 	\bibitem{Hawking:1998kw}
 	S.W.~Hawking, C.J.~Hunter and M.~Taylor, \emph{{Rotation and the AdS / CFT
 			correspondence}},
 	\href{https://doi.org/10.1103/PhysRevD.59.064005}{\emph{Phys. Rev. D}
 		{\bfseries 59} (1999) 064005}
 	[\href{https://arxiv.org/abs/hep-th/9811056}{{\ttfamily hep-th/9811056}}].
 	
 	\bibitem{Gibbons:2004uw}
 	G.W.~Gibbons, H.~Lu, D.N.~Page and C.N.~Pope, \emph{{The General Kerr-de Sitter
 			metrics in all dimensions}},
 	\href{https://doi.org/10.1016/j.geomphys.2004.05.001}{\emph{J. Geom. Phys.}
 		{\bfseries 53} (2005) 49}
 	[\href{https://arxiv.org/abs/hep-th/0404008}{{\ttfamily hep-th/0404008}}].
 	
 	\bibitem{Konoplya:2007jv}
 	R.A.~Konoplya and A.~Zhidenko, \emph{{Stability of multidimensional black
 			holes: Complete numerical analysis}},
 	\href{https://doi.org/10.1016/j.nuclphysb.2007.04.016}{\emph{Nucl. Phys. B}
 		{\bfseries 777} (2007) 182}
 	[\href{https://arxiv.org/abs/hep-th/0703231}{{\ttfamily hep-th/0703231}}].
 	
 	\bibitem{Figueras:2014dta}
 	P.~Figueras and S.~Tunyasuvunakool, \emph{{Black rings in global anti-de Sitter
 			space}}, \href{https://doi.org/10.1007/JHEP03(2015)149}{\emph{JHEP}
 		{\bfseries 03} (2015) 149} [\href{https://arxiv.org/abs/1412.5680}{{\ttfamily
 			1412.5680}}].
 	
 	\bibitem{Caldarelli:2008pz}
 	M.M.~Caldarelli, R.~Emparan and M.J.~Rodriguez, \emph{{Black Rings in
 			(Anti)-deSitter space}},
 	\href{https://doi.org/10.1088/1126-6708/2008/11/011}{\emph{JHEP} {\bfseries
 			11} (2008) 011} [\href{https://arxiv.org/abs/0806.1954}{{\ttfamily
 			0806.1954}}].
 	
 	\bibitem{Armas:2010hz}
 	J.~Armas and N.A.~Obers, \emph{{Blackfolds in (Anti)-de Sitter Backgrounds}},
 	\href{https://doi.org/10.1103/PhysRevD.83.084039}{\emph{Phys. Rev. D}
 		{\bfseries 83} (2011) 084039}
 	[\href{https://arxiv.org/abs/1012.5081}{{\ttfamily 1012.5081}}].
 	
 	\bibitem{Armas:2015kra}
 	J.~Armas and M.~Blau, \emph{{Blackfolds, Plane Waves and Minimal Surfaces}},
 	\href{https://doi.org/10.1007/JHEP07(2015)156}{\emph{JHEP} {\bfseries 07}
 		(2015) 156} [\href{https://arxiv.org/abs/1503.08834}{{\ttfamily
 			1503.08834}}].
 	
 	\bibitem{Armas:2015qsv}
 	J.~Armas, N.A.~Obers and M.~Sanchioni, \emph{{Gravitational Tension, Spacetime
 			Pressure and Black Hole Volume}},
 	\href{https://doi.org/10.1007/JHEP09(2016)124}{\emph{JHEP} {\bfseries 09}
 		(2016) 124} [\href{https://arxiv.org/abs/1512.09106}{{\ttfamily
 			1512.09106}}].
 	
 	\bibitem{Chamblin:1999tk}
 	A.~Chamblin, R.~Emparan, C.V.~Johnson and R.C.~Myers, \emph{{Charged AdS black
 			holes and catastrophic holography}},
 	\href{https://doi.org/10.1103/PhysRevD.60.064018}{\emph{Phys. Rev. D}
 		{\bfseries 60} (1999) 064018}
 	[\href{https://arxiv.org/abs/hep-th/9902170}{{\ttfamily hep-th/9902170}}].
 	
 	\bibitem{Cvetic:2000nc}
 	M.~Cvetic, H.~Lu, C.N.~Pope, A.~Sadrzadeh and T.A.~Tran, \emph{{Consistent
 			SO(6) reduction of type IIB supergravity on S**5}},
 	\href{https://doi.org/10.1016/S0550-3213(00)00372-2}{\emph{Nucl. Phys. B}
 		{\bfseries 586} (2000) 275}
 	[\href{https://arxiv.org/abs/hep-th/0003103}{{\ttfamily hep-th/0003103}}].
 	
 	\bibitem{Cvetic:2004hs}
 	M.~Cvetic, H.~Lu and C.N.~Pope, \emph{{Charged Kerr-de Sitter black holes in
 			five dimensions}},
 	\href{https://doi.org/10.1016/j.physletb.2004.08.011}{\emph{Phys. Lett. B}
 		{\bfseries 598} (2004) 273}
 	[\href{https://arxiv.org/abs/hep-th/0406196}{{\ttfamily hep-th/0406196}}].
 	
 	\bibitem{Chong:2004na}
 	Z.W.~Chong, M.~Cvetic, H.~Lu and C.N.~Pope, \emph{{Charged rotating black holes
 			in four-dimensional gauged and ungauged supergravities}},
 	\href{https://doi.org/10.1016/j.nuclphysb.2005.03.034}{\emph{Nucl. Phys. B}
 		{\bfseries 717} (2005) 246}
 	[\href{https://arxiv.org/abs/hep-th/0411045}{{\ttfamily hep-th/0411045}}].
 	
 	\bibitem{Chong:2005hr}
 	Z.W.~Chong, M.~Cvetic, H.~Lu and C.N.~Pope, \emph{{General non-extremal
 			rotating black holes in minimal five-dimensional gauged supergravity}},
 	\href{https://doi.org/10.1103/PhysRevLett.95.161301}{\emph{Phys. Rev. Lett.}
 		{\bfseries 95} (2005) 161301}
 	[\href{https://arxiv.org/abs/hep-th/0506029}{{\ttfamily hep-th/0506029}}].
 	
 	\bibitem{Chong:2004dy}
 	Z.W.~Chong, M.~Cvetic, H.~Lu and C.N.~Pope, \emph{{Non-extremal charged
 			rotating black holes in seven-dimensional gauged supergravity}},
 	\href{https://doi.org/10.1016/j.physletb.2005.07.054}{\emph{Phys. Lett. B}
 		{\bfseries 626} (2005) 215}
 	[\href{https://arxiv.org/abs/hep-th/0412094}{{\ttfamily hep-th/0412094}}].
 	
 	\bibitem{Chong:2005da}
 	Z.W.~Chong, M.~Cvetic, H.~Lu and C.N.~Pope, \emph{{Five-dimensional gauged
 			supergravity black holes with independent rotation parameters}},
 	\href{https://doi.org/10.1103/PhysRevD.72.041901}{\emph{Phys. Rev. D}
 		{\bfseries 72} (2005) 041901}
 	[\href{https://arxiv.org/abs/hep-th/0505112}{{\ttfamily hep-th/0505112}}].
 	
 	\bibitem{Kunduri:2006ek}
 	H.K.~Kunduri, J.~Lucietti and H.S.~Reall, \emph{{Supersymmetric multi-charge
 			AdS(5) black holes}},
 	\href{https://doi.org/10.1088/1126-6708/2006/04/036}{\emph{JHEP} {\bfseries
 			04} (2006) 036} [\href{https://arxiv.org/abs/hep-th/0601156}{{\ttfamily
 			hep-th/0601156}}].
 	
 	\bibitem{Bobev:2023bxl}
 	N.~Bobev, M.~David, J.~Hong and R.~Mouland, \emph{{AdS$_{7}$ black holes from
 			rotating M5-branes}},
 	\href{https://doi.org/10.1007/JHEP09(2023)143}{\emph{JHEP} {\bfseries 09}
 		(2023) 143} [\href{https://arxiv.org/abs/2307.06364}{{\ttfamily
 			2307.06364}}].
 	
 	\bibitem{Cvetic:1999un}
 	M.~Cvetic, H.~Lu and C.N.~Pope, \emph{{Gauged six-dimensional supergravity from
 			massive type IIA}},
 	\href{https://doi.org/10.1103/PhysRevLett.83.5226}{\emph{Phys. Rev. Lett.}
 		{\bfseries 83} (1999) 5226}
 	[\href{https://arxiv.org/abs/hep-th/9906221}{{\ttfamily hep-th/9906221}}].
 	
 	\bibitem{Nastase:1999cb}
 	H.~Nastase, D.~Vaman and P.~van Nieuwenhuizen, \emph{{Consistent nonlinear K K
 			reduction of 11-d supergravity on AdS(7) x S(4) and selfduality in odd
 			dimensions}},
 	\href{https://doi.org/10.1016/S0370-2693(99)01266-6}{\emph{Phys. Lett. B}
 		{\bfseries 469} (1999) 96}
 	[\href{https://arxiv.org/abs/hep-th/9905075}{{\ttfamily hep-th/9905075}}].
 	
 	\bibitem{Nastase:1999kf}
 	H.~Nastase, D.~Vaman and P.~van Nieuwenhuizen, \emph{{Consistency of the AdS(7)
 			x S(4) reduction and the origin of selfduality in odd dimensions}},
 	\href{https://doi.org/10.1016/S0550-3213(00)00193-0}{\emph{Nucl. Phys. B}
 		{\bfseries 581} (2000) 179}
 	[\href{https://arxiv.org/abs/hep-th/9911238}{{\ttfamily hep-th/9911238}}].
 	
 	\bibitem{deWit:1986oxb}
 	B.~de~Wit and H.~Nicolai, \emph{{The Consistency of the S**7 Truncation in D=11
 			Supergravity}},
 	\href{https://doi.org/10.1016/0550-3213(87)90253-7}{\emph{Nucl. Phys. B}
 		{\bfseries 281} (1987) 211}.
 	
 	\bibitem{Cvetic:1999xp}
 	M.~Cvetic, M.J.~Duff, P.~Hoxha, J.T.~Liu, H.~Lu, J.X.~Lu et~al.,
 	\emph{{Embedding AdS black holes in ten-dimensions and eleven-dimensions}},
 	\href{https://doi.org/10.1016/S0550-3213(99)00419-8}{\emph{Nucl. Phys. B}
 		{\bfseries 558} (1999) 96}
 	[\href{https://arxiv.org/abs/hep-th/9903214}{{\ttfamily hep-th/9903214}}].
 	
 	\bibitem{Bhattacharyya:2010yg}
 	S.~Bhattacharyya, S.~Minwalla and K.~Papadodimas, \emph{{Small Hairy Black
 			Holes in $AdS_5 x S^5$}},
 	\href{https://doi.org/10.1007/JHEP11(2011)035}{\emph{JHEP} {\bfseries 11}
 		(2011) 035} [\href{https://arxiv.org/abs/1005.1287}{{\ttfamily 1005.1287}}].
 	
 	\bibitem{Liu:1999ai}
 	J.T.~Liu and R.~Minasian, \emph{{Black holes and membranes in AdS(7)}},
 	\href{https://doi.org/10.1016/S0370-2693(99)00500-6}{\emph{Phys. Lett. B}
 		{\bfseries 457} (1999) 39}
 	[\href{https://arxiv.org/abs/hep-th/9903269}{{\ttfamily hep-th/9903269}}].
 	
 	\bibitem{Dias:2015pda}
 	O.J.C.~Dias, J.E.~Santos and B.~Way, \emph{{Lumpy AdS$_{5}$\texttimes{} S$^{5}$
 			black holes and black belts}},
 	\href{https://doi.org/10.1007/JHEP04(2015)060}{\emph{JHEP} {\bfseries 04}
 		(2015) 060} [\href{https://arxiv.org/abs/1501.06574}{{\ttfamily
 			1501.06574}}].
 	
 	\bibitem{Dias:2016eto}
 	O.J.C.~Dias, J.E.~Santos and B.~Way, \emph{{Localised $AdS_5\times S^5$ Black
 			Holes}}, \href{https://doi.org/10.1103/PhysRevLett.117.151101}{\emph{Phys.
 			Rev. Lett.} {\bfseries 117} (2016) 151101}
 	[\href{https://arxiv.org/abs/1605.04911}{{\ttfamily 1605.04911}}].
 	
 	\bibitem{Cardona:2020unx}
 	B.~Cardona and P.~Figueras, \emph{{Critical lumpy black holes in AdS${}_p\times
 			S^q$}}, \href{https://doi.org/10.1007/JHEP05(2021)265}{\emph{JHEP} {\bfseries
 			21} (2020) 265} [\href{https://arxiv.org/abs/2103.06932}{{\ttfamily
 			2103.06932}}].
 	
 	\bibitem{Dias:2022eyq}
 	O.J.C.~Dias, P.~Mitra and J.E.~Santos, \emph{{New phases of $ \mathcal{N} $ = 4
 			SYM at finite chemical potential}},
 	\href{https://doi.org/10.1007/JHEP05(2023)053}{\emph{JHEP} {\bfseries 05}
 		(2023) 053} [\href{https://arxiv.org/abs/2207.07134}{{\ttfamily
 			2207.07134}}].
 	
 	\bibitem{Armas:2012bk}
 	J.~Armas, T.~Harmark, N.A.~Obers, M.~Orselli and A.V.~Pedersen, \emph{{Thermal
 			Giant Gravitons}}, \href{https://doi.org/10.1007/JHEP11(2012)123}{\emph{JHEP}
 		{\bfseries 11} (2012) 123} [\href{https://arxiv.org/abs/1207.2789}{{\ttfamily
 			1207.2789}}].
 	
 	\bibitem{Armas:2013ota}
 	J.~Armas, N.A.~Obers and A.V.~Pedersen, \emph{{Null-Wave Giant Gravitons from
 			Thermal Spinning Brane Probes}},
 	\href{https://doi.org/10.1007/JHEP10(2013)109}{\emph{JHEP} {\bfseries 10}
 		(2013) 109} [\href{https://arxiv.org/abs/1306.2633}{{\ttfamily 1306.2633}}].
 	
 	\bibitem{Emparan:2011br}
 	R.~Emparan, \emph{{Blackfolds}},  in \emph{{Black holes in higher dimensions}},
 	G.T.~Horowitz, ed., pp.~180--212 (2012)
 	[\href{https://arxiv.org/abs/1106.2021}{{\ttfamily 1106.2021}}].
 	
 	\bibitem{Emparan:2011hg}
 	R.~Emparan, T.~Harmark, V.~Niarchos and N.A.~Obers, \emph{{Blackfolds in
 			Supergravity and String Theory}},
 	\href{https://doi.org/10.1007/JHEP08(2011)154}{\emph{JHEP} {\bfseries 08}
 		(2011) 154} [\href{https://arxiv.org/abs/1106.4428}{{\ttfamily 1106.4428}}].
 	
 	\bibitem{Caldarelli:2010xz}
 	M.M.~Caldarelli, R.~Emparan and B.~Van~Pol, \emph{{Higher-dimensional Rotating
 			Charged Black Holes}},
 	\href{https://doi.org/10.1007/JHEP04(2011)013}{\emph{JHEP} {\bfseries 04}
 		(2011) 013} [\href{https://arxiv.org/abs/1012.4517}{{\ttfamily 1012.4517}}].
 	
 	\bibitem{Grignani:2010xm}
 	G.~Grignani, T.~Harmark, A.~Marini, N.A.~Obers and M.~Orselli, \emph{{Heating
 			up the BIon}}, \href{https://doi.org/10.1007/JHEP06(2011)058}{\emph{JHEP}
 		{\bfseries 06} (2011) 058} [\href{https://arxiv.org/abs/1012.1494}{{\ttfamily
 			1012.1494}}].
 	
 	\bibitem{Grignani:2011mr}
 	G.~Grignani, T.~Harmark, A.~Marini, N.A.~Obers and M.~Orselli,
 	\emph{{Thermodynamics of the hot BIon}},
 	\href{https://doi.org/10.1016/j.nuclphysb.2011.06.002}{\emph{Nucl. Phys. B}
 		{\bfseries 851} (2011) 462}
 	[\href{https://arxiv.org/abs/1101.1297}{{\ttfamily 1101.1297}}].
 	
 	\bibitem{Armas:2015nea}
 	J.~Armas and M.~Blau, \emph{{New Geometries for Black Hole Horizons}},
 	\href{https://doi.org/10.1007/JHEP07(2015)048}{\emph{JHEP} {\bfseries 07}
 		(2015) 048} [\href{https://arxiv.org/abs/1504.01393}{{\ttfamily
 			1504.01393}}].
 	
 	\bibitem{Armas:2017myl}
 	J.~Armas, T.~Harmark and N.A.~Obers, \emph{{Extremal Black Hole Horizons}},
 	\href{https://doi.org/10.1007/JHEP03(2018)099}{\emph{JHEP} {\bfseries 03}
 		(2018) 099} [\href{https://arxiv.org/abs/1712.09364}{{\ttfamily
 			1712.09364}}].
 	
 	\bibitem{Grignani:2012iw}
 	G.~Grignani, T.~Harmark, A.~Marini, N.A.~Obers and M.~Orselli, \emph{{Thermal
 			string probes in AdS and finite temperature Wilson loops}},
 	\href{https://doi.org/10.1007/JHEP06(2012)144}{\emph{JHEP} {\bfseries 06}
 		(2012) 144} [\href{https://arxiv.org/abs/1201.4862}{{\ttfamily 1201.4862}}].
 	
 	\bibitem{Armas:2022bkh}
 	J.~Armas, G.~Batzios and J.P.~van~der Schaar, \emph{{Holographic duals of the $
 			\mathcal{N} $ = 1* gauge theory}},
 	\href{https://doi.org/10.1007/JHEP04(2023)021}{\emph{JHEP} {\bfseries 04}
 		(2023) 021} [\href{https://arxiv.org/abs/2212.02777}{{\ttfamily
 			2212.02777}}].
 	
 	\bibitem{Armas:2018rsy}
 	J.~Armas, N.~Nguyen, V.~Niarchos, N.A.~Obers and T.~Van~Riet,
 	\emph{{Meta-stable non-extremal anti-branes}},
 	\href{https://doi.org/10.1103/PhysRevLett.122.181601}{\emph{Phys. Rev. Lett.}
 		{\bfseries 122} (2019) 181601}
 	[\href{https://arxiv.org/abs/1812.01067}{{\ttfamily 1812.01067}}].
 	
 	\bibitem{Armas:2019asf}
 	J.~Armas, N.~Nguyen, V.~Niarchos and N.A.~Obers, \emph{{Thermal transitions of
 			metastable M-branes}},
 	\href{https://doi.org/10.1007/JHEP08(2019)128}{\emph{JHEP} {\bfseries 08}
 		(2019) 128} [\href{https://arxiv.org/abs/1904.13283}{{\ttfamily
 			1904.13283}}].
 	
 	\bibitem{Emparan:2009at}
 	R.~Emparan, T.~Harmark, V.~Niarchos and N.A.~Obers, \emph{{Essentials of
 			Blackfold Dynamics}},
 	\href{https://doi.org/10.1007/JHEP03(2010)063}{\emph{JHEP} {\bfseries 03}
 		(2010) 063} [\href{https://arxiv.org/abs/0910.1601}{{\ttfamily 0910.1601}}].
 	
 	\bibitem{Armas:2016mes}
 	J.~Armas, J.~Gath, V.~Niarchos, N.A.~Obers and A.V.~Pedersen, \emph{{Forced
 			Fluid Dynamics from Blackfolds in General Supergravity Backgrounds}},
 	\href{https://doi.org/10.1007/JHEP10(2016)154}{\emph{JHEP} {\bfseries 10}
 		(2016) 154} [\href{https://arxiv.org/abs/1606.09644}{{\ttfamily
 			1606.09644}}].
 	
 	\bibitem{Armas:2018ibg}
 	J.~Armas, J.~Gath, A.~Jain and A.V.~Pedersen, \emph{{Dissipative hydrodynamics
 			with higher-form symmetry}},
 	\href{https://doi.org/10.1007/JHEP05(2018)192}{\emph{JHEP} {\bfseries 05}
 		(2018) 192} [\href{https://arxiv.org/abs/1803.00991}{{\ttfamily
 			1803.00991}}].
 	
 	\bibitem{Camps:2012hw}
 	J.~Camps and R.~Emparan, \emph{{Derivation of the blackfold effective theory}},
 	\href{https://doi.org/10.1007/JHEP03(2012)038}{\emph{JHEP} {\bfseries 03}
 		(2012) 038} [\href{https://arxiv.org/abs/1201.3506}{{\ttfamily 1201.3506}}].
 	
 	\bibitem{DiDato:2015dia}
 	A.~Di~Dato, J.~Gath and A.V.~Pedersen, \emph{{Probing the Hydrodynamic Limit of
 			(Super)gravity}}, \href{https://doi.org/10.1007/JHEP04(2015)171}{\emph{JHEP}
 		{\bfseries 04} (2015) 171}
 	[\href{https://arxiv.org/abs/1501.05441}{{\ttfamily 1501.05441}}].
 	
 	\bibitem{Emparan:2007wm}
 	R.~Emparan, T.~Harmark, V.~Niarchos, N.A.~Obers and M.J.~Rodriguez, \emph{{The
 			Phase Structure of Higher-Dimensional Black Rings and Black Holes}},
 	\href{https://doi.org/10.1088/1126-6708/2007/10/110}{\emph{JHEP} {\bfseries
 			10} (2007) 110} [\href{https://arxiv.org/abs/0708.2181}{{\ttfamily
 			0708.2181}}].
 	
 	\bibitem{Nguyen:2021srl}
 	N.~Nguyen and V.~Niarchos, \emph{{On matched asymptotic expansions of
 			backreacting metastable anti-branes}},
 	\href{https://doi.org/10.1007/JHEP06(2022)055}{\emph{JHEP} {\bfseries 06}
 		(2022) 055} [\href{https://arxiv.org/abs/2112.04514}{{\ttfamily
 			2112.04514}}].
 	
 	\bibitem{becker_becker_schwarz_2006}
 	K.~Becker, M.~Becker and J.H.~Schwarz, \emph{String Theory and M-Theory: A
 		Modern Introduction}, Cambridge University Press (2006),
 	\href{https://doi.org/10.1017/CBO9780511816086}{10.1017/CBO9780511816086}.
 	
 	\bibitem{Brown:1992br}
 	J.D.~Brown and J.W.~York, Jr., \emph{{Quasilocal energy and conserved charges
 			derived from the gravitational action}},
 	\href{https://doi.org/10.1103/PhysRevD.47.1407}{\emph{Phys. Rev. D}
 		{\bfseries 47} (1993) 1407}
 	[\href{https://arxiv.org/abs/gr-qc/9209012}{{\ttfamily gr-qc/9209012}}].
 	
 	\bibitem{Armas:2023tyx}
 	J.~Armas and A.~Jain, \emph{{Approximate higher-form symmetries, topological
 			defects, and dynamical phase transitions}},
 	\href{https://arxiv.org/abs/2301.09628}{{\ttfamily 2301.09628}}.
 	
 	\bibitem{Caldarelli:2008mv}
 	M.M.~Caldarelli, O.J.C.~Dias, R.~Emparan and D.~Klemm, \emph{{Black Holes as
 			Lumps of Fluid}},
 	\href{https://doi.org/10.1088/1126-6708/2009/04/024}{\emph{JHEP} {\bfseries
 			04} (2009) 024} [\href{https://arxiv.org/abs/0811.2381}{{\ttfamily
 			0811.2381}}].
 	
 	\bibitem{Armas:2018atq}
 	J.~Armas and A.~Jain, \emph{{Magnetohydrodynamics as superfluidity}},
 	\href{https://doi.org/10.1103/PhysRevLett.122.141603}{\emph{Phys. Rev. Lett.}
 		{\bfseries 122} (2019) 141603}
 	[\href{https://arxiv.org/abs/1808.01939}{{\ttfamily 1808.01939}}].
 	
 	\bibitem{Armas:2018zbe}
 	J.~Armas and A.~Jain, \emph{{One-form superfluids \& magnetohydrodynamics}},
 	\href{https://doi.org/10.1007/JHEP01(2020)041}{\emph{JHEP} {\bfseries 01}
 		(2020) 041} [\href{https://arxiv.org/abs/1811.04913}{{\ttfamily
 			1811.04913}}].
 	
 	\bibitem{Armas:2024caa}
 	J.~Armas, G.~Batzios and A.~Jain, \emph{{Higher-group global symmetry and the
 			bosonic M5 brane}},  \href{https://arxiv.org/abs/2402.19458}{{\ttfamily
 			2402.19458}}.
 	
 	\bibitem{Kubiznak:2014zwa}
 	D.~Kubiznak and R.B.~Mann, \emph{{Black hole chemistry}},
 	\href{https://doi.org/10.1139/cjp-2014-0465}{\emph{Can. J. Phys.} {\bfseries
 			93} (2015) 999} [\href{https://arxiv.org/abs/1404.2126}{{\ttfamily
 			1404.2126}}].
 	
 	\bibitem{Karch:2015rpa}
 	A.~Karch and B.~Robinson, \emph{{Holographic Black Hole Chemistry}},
 	\href{https://doi.org/10.1007/JHEP12(2015)073}{\emph{JHEP} {\bfseries 12}
 		(2015) 073} [\href{https://arxiv.org/abs/1510.02472}{{\ttfamily
 			1510.02472}}].
 	
 	\bibitem{Emparan:2008qn}
 	R.~Emparan, \emph{{Exact Microscopic Entropy of Non-Supersymmetric Extremal
 			Black Rings}},
 	\href{https://doi.org/10.1088/0264-9381/25/17/175005}{\emph{Class. Quant.
 			Grav.} {\bfseries 25} (2008) 175005}
 	[\href{https://arxiv.org/abs/0803.1801}{{\ttfamily 0803.1801}}].
 	
 	\bibitem{Blanco-Pillado:2007eit}
 	J.J.~Blanco-Pillado, R.~Emparan and A.~Iglesias, \emph{{Fundamental Plasmid
 			Strings and Black Rings}},
 	\href{https://doi.org/10.1088/1126-6708/2008/01/014}{\emph{JHEP} {\bfseries
 			01} (2008) 014} [\href{https://arxiv.org/abs/0712.0611}{{\ttfamily
 			0712.0611}}].
 	
 	\bibitem{Camps:2008hb}
 	J.~Camps, R.~Emparan, P.~Figueras, S.~Giusto and A.~Saxena, \emph{{Black Rings
 			in Taub-NUT and D0-D6 interactions}},
 	\href{https://doi.org/10.1088/1126-6708/2009/02/021}{\emph{JHEP} {\bfseries
 			02} (2009) 021} [\href{https://arxiv.org/abs/0811.2088}{{\ttfamily
 			0811.2088}}].
 	
 	\bibitem{Emparan:2009vd}
 	R.~Emparan, T.~Harmark, V.~Niarchos and N.A.~Obers, \emph{{New Horizons for
 			Black Holes and Branes}},
 	\href{https://doi.org/10.1007/JHEP04(2010)046}{\emph{JHEP} {\bfseries 04}
 		(2010) 046} [\href{https://arxiv.org/abs/0912.2352}{{\ttfamily 0912.2352}}].
 	
 	\bibitem{Armas:2011uf}
 	J.~Armas, J.~Camps, T.~Harmark and N.A.~Obers, \emph{{The Young Modulus of
 			Black Strings and the Fine Structure of Blackfolds}},
 	\href{https://doi.org/10.1007/JHEP02(2012)110}{\emph{JHEP} {\bfseries 02}
 		(2012) 110} [\href{https://arxiv.org/abs/1110.4835}{{\ttfamily 1110.4835}}].
 	
 	\bibitem{Armas:2014bia}
 	J.~Armas and T.~Harmark, \emph{{Black Holes and Biophysical (Mem)-branes}},
 	\href{https://doi.org/10.1103/PhysRevD.90.124022}{\emph{Phys. Rev. D}
 		{\bfseries 90} (2014) 124022}
 	[\href{https://arxiv.org/abs/1402.6330}{{\ttfamily 1402.6330}}].
 	
 	\bibitem{Armas:2014rva}
 	J.~Armas and T.~Harmark, \emph{{Constraints on the effective fluid theory of
 			stationary branes}},
 	\href{https://doi.org/10.1007/JHEP10(2014)063}{\emph{JHEP} {\bfseries 10}
 		(2014) 063} [\href{https://arxiv.org/abs/1406.7813}{{\ttfamily 1406.7813}}].
 	
 	\bibitem{Armas_2013}
 	J.~Armas, \emph{How fluids bend: the elastic expansion for higher-dimensional
 		black holes}, \href{https://doi.org/10.1007/jhep09(2013)073}{\emph{Journal of
 			High Energy Physics} {\bfseries 2013} (2013) }.
 	
 	\bibitem{Armas:2013aka}
 	J.~Armas, J.~Gath and N.A.~Obers, \emph{{Electroelasticity of Charged Black
 			Branes}}, \href{https://doi.org/10.1007/JHEP10(2013)035}{\emph{JHEP}
 		{\bfseries 10} (2013) 035} [\href{https://arxiv.org/abs/1307.0504}{{\ttfamily
 			1307.0504}}].
 	
 	\bibitem{Armas:2012jg}
 	J.~Armas and N.A.~Obers, \emph{{Relativistic Elasticity of Stationary Fluid
 			Branes}}, \href{https://doi.org/10.1103/PhysRevD.87.044058}{\emph{Phys. Rev.
 			D} {\bfseries 87} (2013) 044058}
 	[\href{https://arxiv.org/abs/1210.5197}{{\ttfamily 1210.5197}}].
 	
 	\bibitem{Armas:2012ac}
 	J.~Armas, J.~Gath and N.A.~Obers, \emph{{Black Branes as Piezoelectrics}},
 	\href{https://doi.org/10.1103/PhysRevLett.109.241101}{\emph{Phys. Rev. Lett.}
 		{\bfseries 109} (2012) 241101}
 	[\href{https://arxiv.org/abs/1209.2127}{{\ttfamily 1209.2127}}].
 	
 	\bibitem{Caldarelli:2012hy}
 	M.M.~Caldarelli, J.~Camps, B.~Gout\'eraux and K.~Skenderis,
 	\emph{{AdS/Ricci-flat correspondence and the Gregory-Laflamme instability}},
 	\href{https://doi.org/10.1103/PhysRevD.87.061502}{\emph{Phys. Rev. D}
 		{\bfseries 87} (2013) 061502}
 	[\href{https://arxiv.org/abs/1211.2815}{{\ttfamily 1211.2815}}].
 	
 	\bibitem{Gath:2013qya}
 	J.~Gath and A.V.~Pedersen, \emph{{Viscous asymptotically flat
 			Reissner-Nordstr\"om black branes}},
 	\href{https://doi.org/10.1007/JHEP03(2014)059}{\emph{JHEP} {\bfseries 03}
 		(2014) 059} [\href{https://arxiv.org/abs/1302.5480}{{\ttfamily 1302.5480}}].
 	
 	\bibitem{Armas:2019iqs}
 	J.~Armas and E.~Parisini, \emph{{Instabilities of Thin Black Rings: Closing the
 			Gap}}, \href{https://doi.org/10.1007/JHEP04(2019)169}{\emph{JHEP} {\bfseries
 			04} (2019) 169} [\href{https://arxiv.org/abs/1901.09369}{{\ttfamily
 			1901.09369}}].
 	
 	\bibitem{Nguyen:2019syc}
 	N.~Nguyen, \emph{{Comments on the stability of the KPV state}},
 	\href{https://doi.org/10.1007/JHEP11(2020)055}{\emph{JHEP} {\bfseries 11}
 		(2020) 055} [\href{https://arxiv.org/abs/1912.04646}{{\ttfamily
 			1912.04646}}].
 	
 	\bibitem{Cvetic:1999ne}
 	M.~Cvetic and S.S.~Gubser, \emph{{Phases of R charged black holes, spinning
 			branes and strongly coupled gauge theories}},
 	\href{https://doi.org/10.1088/1126-6708/1999/04/024}{\emph{JHEP} {\bfseries
 			04} (1999) 024} [\href{https://arxiv.org/abs/hep-th/9902195}{{\ttfamily
 			hep-th/9902195}}].
 	
 	\bibitem{Harmark:1999xt}
 	T.~Harmark and N.A.~Obers, \emph{{Thermodynamics of spinning branes and their
 			dual field theories}},
 	\href{https://doi.org/10.1088/1126-6708/2000/01/008}{\emph{JHEP} {\bfseries
 			01} (2000) 008} [\href{https://arxiv.org/abs/hep-th/9910036}{{\ttfamily
 			hep-th/9910036}}].
 	
 \end{thebibliography}

 \providecommand{\href}[2]{#2}\begingroup\raggedright\endgroup


\end{document}